\documentclass[a4paper,fleqn,longmktitle]{cas-sc}

\usepackage[T1]{fontenc}
\usepackage[utf8]{inputenc}

\makeatletter
\def\@seccntformat#1{\@ifundefined{#1@cntformat}%
   {\csname the#1\endcsname\quad}
   {\csname #1@cntformat\endcsname}
}
\makeatother

\usepackage[numbers, square, sort&compress]{natbib}
\usepackage{rotating}
\usepackage{gensymb}
\usepackage[section]{placeins}
\usepackage{nomencl} 
\usepackage{framed} 
\usepackage{multicol} 
\usepackage{tabularx}
\usepackage{pdflscape}
\usepackage{longtable}

\usepackage{nomencl} 
\usepackage[boxruled]{algorithm2e} 
\makenomenclature
\renewcommand*\nompreamble{\begin{multicols}{2}}
\renewcommand*\nompostamble{\end{multicols}}

\renewcommand\nomgroup[1]{%
  \item[\bfseries
  \ifstrequal{#1}{A}{Abbreviations}{%
  \ifstrequal{#1}{V}{Variables}{%
  \ifstrequal{#1}{P}{Parameter}{%
  \ifstrequal{#1}{S}{Sets}}}}
]}

\begin{document}
\let\WriteBookmarks\relax
\def\floatpagepagefraction{1}
\def\textpagefraction{.001}
\shorttitle{Nuclear power and renewables}
\shortauthors{Göke et al.}

\title [mode = title]{Flexible nuclear power and fluctuating renewables? - A techno-economic analysis for decarbonized energy systems
}                  

\author[1]{Leonard Göke}[orcid=0000-0002-3219-7587]
\author[2,3]{Alexander Wimmers}[orcid=0000-0003-3686-0793]
\cormark[1]
\ead{awi@wip.tu-berlin.de}
\author[2,3]{Christian von Hirschhausen}[orcid=0000-0002-0814-8654]

\address[1]{Energy and Process Systems Engineering, Department of Mechanical and Process Engineering, ETH Zürich, Tannenstrasse 3, Zürich 8092, Switzerland}
\address[2]{Workgroup for Infrastructure Policy (WIP), Technische Universität Berlin, Straße des 17. Juni 135, 10623 Berlin, Germany}
\address[3]{Energy, Transportation, Environment Department, German Institute for Economic Research (DIW Berlin), Mohrenstraße 58, 10117 Berlin, Germany}

\cortext[cor1]{Corresponding author: \href{mailto:awi@wip.tu-berlin.de}{awi@wip.tu-berlin.de}}

\begin{abstract}
Many governments are considering constructing new nuclear power plants to support the decarbonization of the energy system. On the one hand, dispatchable nuclear plants can complement fluctuating generation from wind and PV. On the other hand, escalating construction costs and times raise economic concerns. In this paper, we extensively review construction costs and times. On this basis, we apply a detailed multi-vector energy model to analyze the cost-efficient share of nuclear power in fully decarbonized energy systems, i.e., energy systems that do not utilize any fossil fuels. Our analysis finds that even if, reversing the historical trend, overnight construction costs of nuclear half to 4,000 US-\$\textsubscript{2018} per kW and construction times remain below ten years, the cost-efficient share of nuclear power in European electricity generation is only around 10\%. The analysis still omits the social costs of nuclear power, such as the risk of accidents or waste management. Nuclear plants must operate inflexibly and at capacity factors close to 90\% to recover their investment costs, implying that operational flexibility---even if technically possible---is not economically viable. As a result, grid infrastructure, flexible demand in multi-energy systems, and storage are more efficient options for integrating fluctuating wind and photovoltaic generation.

\end{abstract}

\begin{highlights}
\item Integrated energy model to assess the potential of nuclear power for decarbonization
\item Cost analysis for nuclear reveals large difference between current and projected data 
\item Even if overnight costs halve, the share of nuclear in Europe is only around 10\%
\item Operation of nuclear adapting to wind and photovoltaic is not economically viable
\end{highlights}


\maketitle

\section{Introduction}\label{introduction}

In the energy crisis ensued by the Russian invasion of Ukraine in 2022, surging prices for fossil fuels added to the debate on alternatives to gas, coal, and oil for energy generation. In addition to ongoing and planned efforts to expand renewable energy sources, many governments have recently declared or emphasized their intention to invest in new nuclear power plants: In addition to restarting many of its shuttered reactors, Japan is considering investing in new reactors for the first time since the meltdown of the Fukushima Daiichi plant \citep{ft}, France recently pledged to construct up to 14 new plants \citep{nyt} and the United Kingdom announced plans for new plants to pursue their target of 24 GW nuclear capacity by 2050 \citep{bb}, followed by an announcement at the COP28 to triple installed nuclear power capacities by 2050 despite ongoing industry challenges and aging fleets \citep{bose_questioning_2024}.

\subsection{The case for and against nuclear}

Currently, surging prices for fossil fuels only encourage existing plans for nuclear power that pre-date the crisis and originate from government strategies to mitigate carbon emissions and combat climate change \citep{Goldstein2019}. On the one hand, nuclear power offers two technical advantages: First, nuclear plants are dispatchable. Currently, only a few plants operate in load-following mode due to added costs and material stress \citep{ramana_small_2021, schneider_world_2023, monitoring_analytics_llc_state_2023}. However, research suggests that new reactor concepts and current high-capacity light-water reactors are eligible for efficient load-following operations \citep{Jenkins2018, Lynch2022, mit_future_2018,pistner_analyse_2024}. Therefore, nuclear power could complement the fluctuating supply from renewables and ensure the security of supply in decarbonized energy systems. Second, nuclear energy has significant energy potential. Nuclear power is far less constrained by land availability or meteorological conditions than renewables \citep{landUse}. At the same time, decarbonization in heating, industry, and transport requires electricity as a primary energy source, either directly or indirectly, e.g., by using synthetic fuels produced from electricity \citep{Luderer2021,Bogdanov2021}. In conclusion, nuclear power could substantially contribute to fulfilling the increasing demand for electricity.

On the other hand, there are economic arguments against nuclear power. New plants are capital-intensive, especially since actual construction cost and time frequently exceed forecasts \citep{Lovins2022,Rothwell2022}. For instance, in France, the official construction costs of Flamanville 3 have so far quadrupled to 12,600 €/kW. At the same time, project completion is delayed by more than a decade \citep{edf_universal_2022,Rothwell2022}. Similarly, the costs of Olkiluoto 3 in Finland tripled over a construction time of 17 years \citep{dw}. Costs of the Hinkley Point C project in the United Kingdom have increased by 2.6, from 18 to 47.9 bn GBP\textsubscript{2015}. EDF has repeatedly moved back its start date and doubts that the reactor will be operational before 2030 -- 14 years after construction started \citep{wnn_improved_2024}. Whether this trend is reversible is debatable: Drawing on the historic expansion of pressurized water reactors in France, \citet{Berthelemy2015} suggest that reinforced investments and standardization could lower construction costs and time. \citet{Grubler2010} states that the expansion of pressurized water reactors never achieved positive learning effects in the first place. In addition, there are other challenges, such as the maturity of new reactor concepts, aging fleets prone to early closures, and limited industry capabilities \citep{bose_questioning_2024}. As a result, high investment costs can render nuclear power more expensive even compared to renewables in unfavorable sites, like offshore wind turbines in deep waters or photovoltaic (PV) in locations with little sunshine \citep{mckenna2014}.

\subsection{Flexibility in renewable systems}

Besides flexible nuclear plants, many technological alternatives are conceivable to balance wind and photovoltaic (PV). One option is battery storage, commonly suggested as a device for short-term storage over several hours or days, in particular complementing PV generation, often from decentralized prosumers \citep{Child2019, Schill2020j}. Lithium-ion batteries are the dominant technology, but sodium-sulfide and vanadium redox flow batteries are also promising \citep{Deguenon2023, Lund2015}. Another option for short-term flexibility is to adapt consumption to fluctuating supply, referred to as demand-side management. Although this option can reduce peak-load by 7\% to 26\%, this shift is constrained to one or two hours, significantly limiting the added flexibility \citep{Gils2014}.

However, beyond the shifting of conventional demand, future electrification of transport and heating adds new options for demand-side management. Research consistently finds that flexbility in transport and heating is cost-efficient and capable of substituting battery storage \citep{Brown2018}. Electrification of transport can add substantial flexibility and support renewable integration if battery electric vehicle (BEV) charging adapts to renewable supply \citep{Schuller2015,Verzijlbergh2014}. The benefits slightly increase if, in addition, bi-directional loading feed electricity back into the grid \citep{Gunkel2020,Nagel2024}. For heating, the flexibility potential significantly differs between individual and district heating. In individual heating, costs of small-scale heat storage and operational constraints hinder flexibility \citep{Bloess2019}, and electric heating might rather increase than reduce flexibility demands \citep{Schill2020h}. A promising option is utilizing the thermal inertia of buildings as passive thermal storage \citep{Bloess2018}. For district heating, operational flexibility and economies of scale in energy storage can provide substantial flexibility \citep{Schuller2015}. Of particular interest is long-duration heat storage that, unlike all options discussed above, does not offer short-term flexibility but balances seasonal mismatches of renewable supply and demand \citep{Brown2018,Schill2020j}.

Another option for system flexibility is transmission infrastructure. Imports from other regions can cover demand if local supply is low, and vice versa; exports can reduce excess supply. Research consistently shows that electricity transmission benefits the integration of wind, reducing the need for both short- and long-term storage \citep{Schlachtberger2017,Roth2023}. The second transmission infrastructure considered in the literature are hydrogen grids. Hydrogen grids enable flexible electricity consumption from electrolyzers producing hydrogen, particularly with hydrogen storage \citep{Neumann2023}. Given the high expected demand for hydrogen in decarbonized energy systems, several studies identify flexible hydrogen production as a significant, in some cases even the most significant, lever for renewable integration \citep{Wang2018, Gils2021, Guerra2023}. Finally, hydrogen and other synthetic fuels, for instance, produced from biomass, enable dispatchable thermal generation in decarbonized systems, a pricey but indispensable flexibility option for long periods with low renewable supply that short-term storage options, such as batteries, cannot cover.

\subsection{Contribution and structure of this work}

Against this background, we investigate the economic efficiency of nuclear power in two steps: First, we review projected and actual costs of new nuclear capacity to compute the conceivable range of levelized costs of energy (LCOE), discuss plausibility, and compare the outcome to renewables. The cost review focuses on OECD countries due to the discrepancy of nuclear development between OECD and non-OECD countries observed by \citet{Rothwell2022}. In contrast to most previous research, we also review construction times and their impact on capital costs \citep{wealer_investing_2021}.

Second, we investigate the efficient share of nuclear power for previously obtained cost ranges in a decarbonized European energy system with a comprehensive energy planning model \citep{Goeke2020a}. This analysis reflects the best practices for modelling nuclear power in decarbonized energy systems recommended by \citet{Bistline2023}: Building on the previous section, we test a broad range of cost parameters based on publicly available data. The model applies a high temporal resolution using a full year of hourly time-series and considers different national circumstances thanks to a large spatial scope. Furthermore, the model includes all competing options for flexibility and firm capacity discussed above---most notably, long-duration energy storage. In addition, it covers the electricity sector and a multi-vector energy system that reflects additional demand for decarbonizing heat, industry, and transport and the flexibility these sectors can add.

Careful consideration of the best practices recommended distinguishes our system analysis from the existing research. Most importantly, existing studies on nuclear power omit pivotal options to balance fluctuating supply, most notably only considering the electricity sector rather than a multi-vector energy system. As a result, they cannot capture key characteristics of demand and flexibility in future decarbonized systems \citep{Shirizadeh2021, Price2023, mit2022}. In addition, several previous studies omit long-duration energy storage as well or limit their scope to a single country or region \citep{Duan2022,Baik2021,fattahi_analyzing_2022}.

The inevitable simplifications of the techno-economic model are made in a way that favors investment in nuclear power. Therefore, the computed share of nuclear power in the energy mix is not an accurate estimate but rather an upper bound of what might occur if nuclear power achieved significant cost reductions. In particular, our techno-economic analysis omits external social costs of nuclear power associated with the risk of accidents \citep{wheatley_reassessing_2016}, waste management \citep{krall_nuclear_2022}, decommissioning \citep{lordan-perret_ex-ante_2023}, and proliferation \citep{lehtveer_nuclear_2015}. Although their robust quantification is difficult, external costs for nuclear typically exceed estimates for renewable energies \citep{stirling_limits_1997}.

The remainder of this paper is structured as follows: The next section \ref{methods} describes the methodology to compute LCOEs and introduces the applied energy planning model. Section \ref{main} presents the results for the LCOE analysis and the planning model. Section \ref{diss} discusses these results, and section \ref{conclusion} concludes.

\section{Methods} \label{methods}

The first subsection \ref{lcoe} describes the computation of LCOEs for nuclear with an emphasis on the review of construction costs and financing costs during construction. Then, we introduce the energy planning model for a more detailed evaluation of nuclear power. The applied model uses the open AnyMOD.jl modeling framework described in greater detail in previous publications \citep{Goeke2020a, goke_how_2023}.

\subsection{Computation of LCOEs} \label{lcoe}

Whether nuclear power is an economically efficient option for decarbonization strongly depends on its costs. Operational costs for fuel or maintenance constitute only a minor share while construction and financing account for 80\% of the total costs of nuclear power \citep{haas_historical_2019, mackerron_nuclear_1992}. At the same time, the costs of construction and financing are also subject to the greatest uncertainty. Therefore, we provide an in-depth review of construction and financing costs and vary them across the observed range in the subsequent model analysis. For all other parameters, section \ref{cost_analysis} of the appendix provides a review. Table \ref{tab:assumptions} lists the assumptions used for the analysis. For these parameters, we use optimistic assumptions corresponding to at least the 25th percentile of the reviewed data. If not stated otherwise, all monetary values in this paper are inflation-adjusted and given in US-\$\textsubscript{2018}. Section \ref{lcoeSup} of the appendix provides the corresponding conversion factors.

To assess the range of overnight construction costs, we reviewed academic publications and industry reports providing 88 individual future projections or reported costs for different reactor concepts and countries \citep{nrel_2021_2021,oecd_full_2018,shirvan_overnight_2022, Rothwell2022, barkatullah_current_2017, lazard2021, wealer_investing_2021, international_atomic_energy_agency_advances_2020, green_smr_2019, rothwell_economics_2016, Duan2022, Grubler2010, stein_advancing_2022, mit_future_2018, boldon_small_2014, stewart_capital_2021, nrel_2021_2021, iea_projected_2020, tolley_economic_2004, dixon_advanced_2017}. We focus on light-water reactors (LWR), the most common reactor concept\footnote{Table \ref{tab:OCC_results} in the appendix shows that considering a broader range of reactor types does not give lower cost projections.}, and OECD countries since \citet{Rothwell2022} observes non-OECD estimates are not comparable. Cost projections are often provided as n-th-of-a-kind cost, presenting assumptions of sometimes significant cost reductions through standardization and learning effects, e.g., \citet{shirvan_overnight_2022}. These projections, however, are not given for a specific year or cumulated investment, complicating the comparison with cost projections for renewables.

\begin{figure}
    \centering
    \includegraphics[scale=0.15]{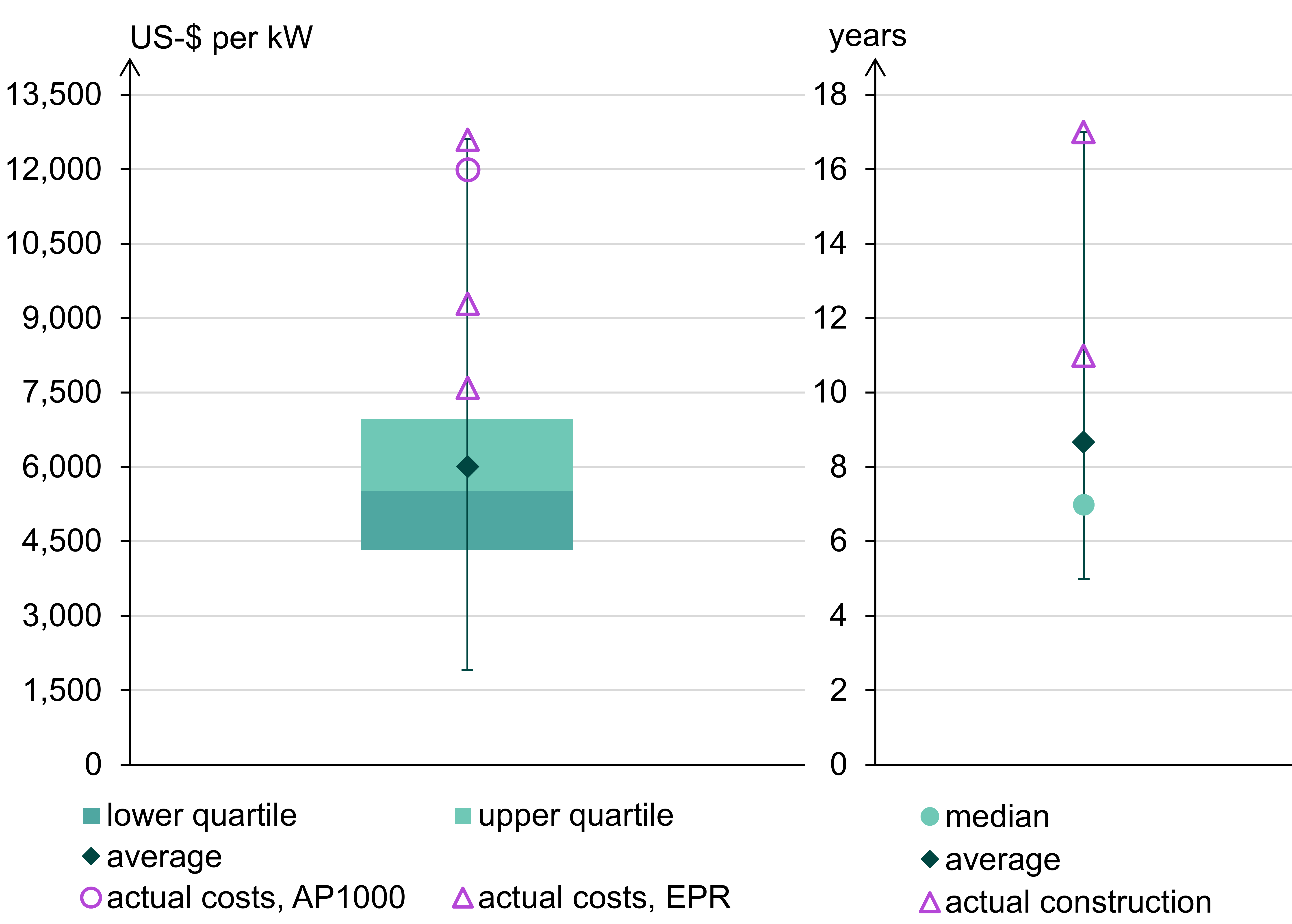}
    \caption{Boxplots for overnight construction costs and construction times}
    \label{fig:OCC_boxplot}
\end{figure}

The left boxplot in Fig. \ref{fig:OCC_boxplot} summarizes the distribution of the 45 remaining estimates. Overall, overnight construction cost projections range from 1,914 US-\$  per kW projected by \citet{oecd_full_2018} to 12,600 US-\$ per kW, the actual costs of the Flamanville-3 project as of 2020 \citep{Rothwell2022}. The median is at 5,430 US-\$ per kW; 50\% of estimates are between 4,328 and 9,966 US-\$ per kW. Future projections are generally much lower than recent observations of actual costs for the European Pressurized Reactor (EPR) and the Westinghouse AP1000. Cost observations are still preliminary for Hinkley Point C, Flamanville, and Voglte-4. They could further increase as construction or operations testing is ongoing. At the time of data collection, Olkiluoto-3 and Voglte-3 were still undergoing pre-operation procedures and had only recently begun operations throughout 2023 \citep{iaea_power_2024}. \footnote{For the Shin Hanul 1 reactor that recently started operation in South Korea, final costs were not publicly disclosed. KEPCO's financial statements suggest costs as low as 3,120 US-\$ per kW \citep{kepco_interim_2022}, which is very low considering the construction time of approximately 12 years.} Some studies suggest that historically, learning rates of 5 to 10\% were achieved in the nuclear industry, which would coincide with the mentioned projections \citep{Berthelemy2015, mignacca_economics_2020, lloyd_expanding_2020}. However, the discrepancy between projected and actual costs is in line with previous analyses on the actual development of nuclear construction costs in Western countries since the 1970s \citep{Grubler2010, davis_prospects_2012, koomey_reactor-level_2007, escobar_rangel_revisiting_2015, harding_economics_2007}, placing doubt on supposed learning rates \citep{steigerwald_uncertainties_2023}.

The right boxplot in Fig. \ref{fig:OCC_boxplot} presents a similar image for construction time. Future projections range from 5 to 9 years, but recent and ongoing projects take 10 to 17 years. These overruns drive up financing costs and can also threaten energy security.

\begin{table}
\caption{Parameters for nuclear power} \label{tab:assumptions}
\centering
\begin{tabular}{p{3cm}|p{2cm}|p{2cm}}
    \hline
    Parameter                           &Unit       &Value / Range\\
    \hline
    Overnight construction cost   &US-\$/kW     &1,914 - 12,600\\
    Annual fixed operational cost       &US-\$/kW     &88.81\\
    Variable operational cost  (incl. fuel)          &US-\$/MWh    &10.96\\
    Capacity factor                     &\%         &95\\
    Construction time             &years      &4 - 10\\
    Depreciation period        &years      &40\\
    \hline
\end{tabular}
\end{table}

Together, construction and financing costs constitute total construction costs (TCC). Financing costs reflect the paid interest and depend on the depreciation time and interest rate. Following the methodology for nuclear power investments introduced by \citet{rothwell_economics_2016}, our computation of financing costs factors in the construction time. We compute the financing costs during construction $f$ using Equation \ref{idc}. In the equation, $i$ is the interest rate, $t$ is the construction time, and $c$ is the overnight construction costs.

\begin{equation} \label{idc}
f = (\frac{i}{2}\cdot t + \frac{i^2}{6}\cdot t^2) \cdot c
\end{equation}

To compare LCOEs of different technologies and as input for the later analysis with the energy system model, we compute the annuity $a$ for the sum of financing and overnight construction costs according to Equation \ref{ann}. In the formula, $d$ is the depreciation period.

\begin{equation} \label{ann}
a = (f + c) \cdot \frac{(1+i)^d \cdot i}{(1+i)^d -1}
\end{equation}

Finally, the LCOE $l$ for a single year corresponds to the sum of annuity and variable costs $v$ divided by the yearly generation. The yearly generation corresponds to the product of the capacity factor $u$ and the number of hours in a year.

\begin{equation}
l = \frac{a + v \cdot u \cdot 8760}{u \cdot 8760}
\end{equation}

LCOEs depend on the depreciation period, which corresponds to the assumed operational lifetime of a plant in our case. Our literature analysis shows that most studies assume operational lifetimes of 40 years; for details, see \ref{tab:operational_lifetime} in the appendix. For reference, the average age of the closed global fleet is less than 30 years \citep{schneider_world_2023}, though EPRs currently under construction in Europe are planned to operate for 60 years \citep{thomas_epr_2010}. An extended operational lifetime might also require additional investments in later years or increase maintenance costs, but these effects are difficult to quantify from the available literature.

\subsection{Energy system model} \label{sec:model}

To assess the cost-efficient share of nuclear power, we apply an energy planning model that minimizes total system costs. This cost minimization takes a strictly techno-economic social planner's perspective, not considering the different agents within the system, such as producers, consumers, or governments, and their respective behavior. Formulated as a linear optimization problem, the model decides on the optimal strategy to satisfy an exogenous final demand for energy services. More specifically, the model determines two kinds of variables: First, the capacity investment into different technologies for converting, storing, generating, and transporting energy, and second, the operation of the invested capacities to satisfy the final demand. Technology and demand data are based on projections for the year 2040. 

\begin{figure}
	\centering
		\includegraphics[scale=0.8]{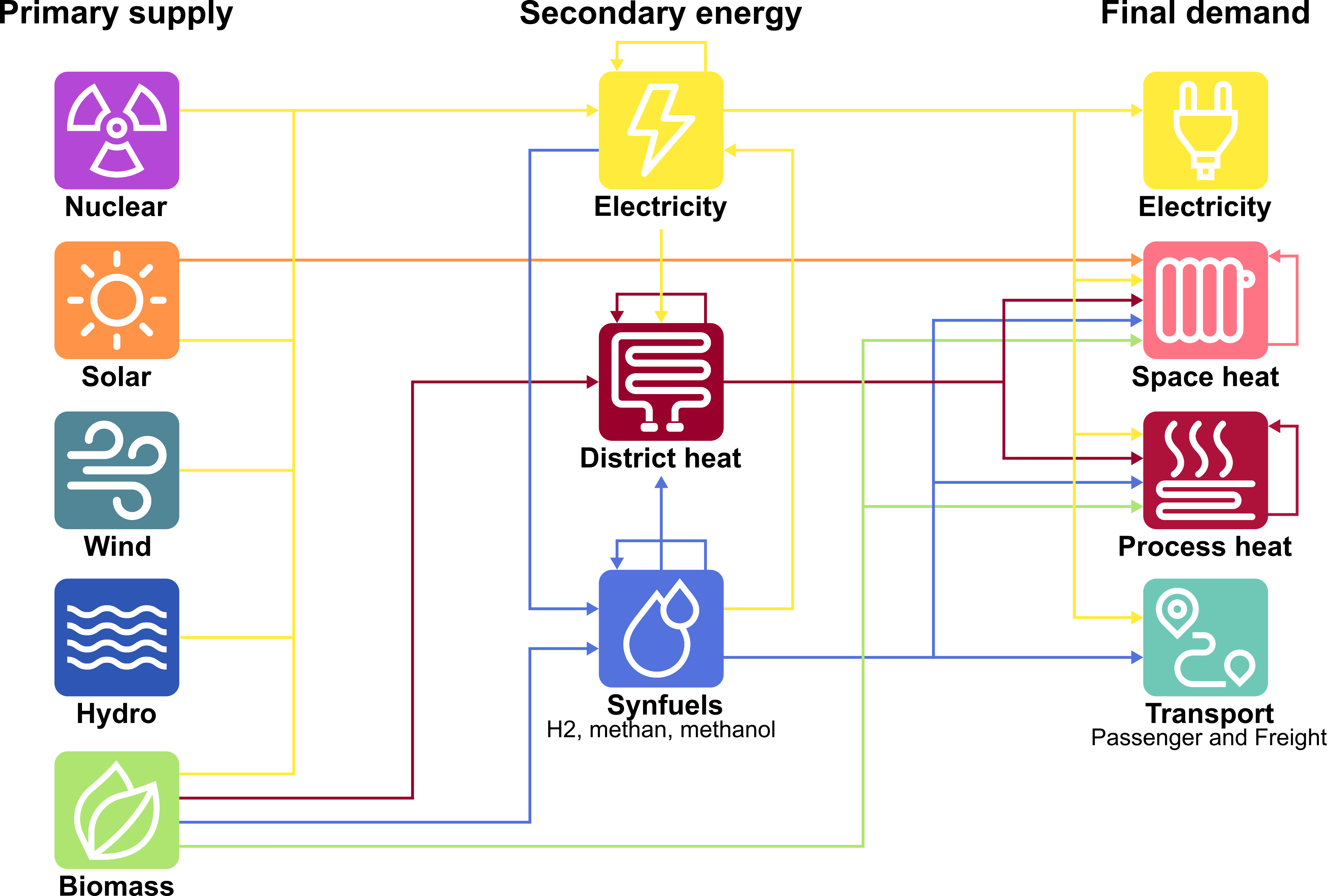}
	\caption{Overview of model structure}
	\label{fig:overMod}
\end{figure}

Fig. \ref{fig:overMod} gives a high-level overview of the considered sources for primary energy, secondary energy carriers, and final demands. The following paragraphs will comprehensively summarize the modeled energy carriers and the technologies connecting them. The appendix \ref{graphSec} provides an exhaustive account of all 22 energy carriers and 125 technologies. Full documentation of all technologies here is neither informative nor valuable; for instance, the model already includes ten different combined heat and power (CHP) turbines that differ in terms of input (hydrogen or synthetic gas), output (district heat or process heat on different temperature levels) and technology (backpressure or extraction). The choice and parametrization of all non-nuclear technologies except vehicles build on reports of the \citet{DEA}; vehicle data from \citet{Robinius2020}. Table \ref{tab:assumptions} gives all assumptions regarding nuclear power.

Since our analysis focuses on a decarbonized energy system, the primary supply on the left side of Fig. \ref{fig:overMod} precludes fossil fuels. In this analysis, as further discussed in section \ref{diss}, nuclear is considered exclusively as a source of electricity. For electricity generation from PV, the model differentiates photovoltaic in open spaces, on roofs of residential buildings, and roofs of industrial buildings; for electricity generation from wind, the model differentiates on- and offshore. All wind and PV technologies fluctuate, and their supply is contingent on an exogenous time-series of capacity factors. For hydro supply, the model considers run-of-river and reservoirs. Run-of-river fluctuates, like wind and PV. Reservoirs have exogenous and fluctuating inflows, but supply is dispatchable as long as the water in the reservoir is sufficient. For biomass, the most versatile energy source, the model distinguishes raw biogas, solid biomass, and non-solid biomass. Boilers or CHP plants can directly burn biomass to generate electricity and district heat, space heat, or process heat up to 500°C. Alternatively, various routes are available to convert biomass into synthetic fuels, like the gasification of solid biomass or upgrading of raw biogas to synthetic methane, and the liquefaction of solid and non-solid biomass to synthetic methanol. Since renewables sources are the key competitor of nuclear power, section \ref{renewSup} further elaborates on their representation.

The middle column of Fig. \ref{fig:overMod} shows the secondary energy, the intermediate step from primary supply to final demand. Heat-pumps and electric boilers convert electricity into district heat; equally, each synthetic fuel, hydrogen, methane, or methanol, can fuel a different boiler technology to provide district heat. Alkali, solid-oxide, or proton exchange membrane electrolyzers utilize electricity to generate hydrogen while feeding waste heat into district heating networks. Vice versa, synthetic fuels can fuel different engines or turbines to generate electricity and district heat. In addition, there are several conversion pathways among the synthetic fuels: methane pyrolysis creates hydrogen from methane, methanation synthesizes methane from hydrogen (and raw biogas), and hydrogen-to-methanol conversion is possible but requires a carbon input supplied by direct air capture. 

Furthermore, the model can invest in storage systems for secondary energy carriers, as indicated by the arrows starting and ending at the same carrier. Electricity storage encompasses lithium-ion batteries, redox flow batteries, and pumped hydro storage. Pumped hydro storages are the most mature and cost-efficient systems but have little growth potential, at least in Europe; lithium-ion batteries are the most mature and efficient battery technology but exceed redox flow batteries in investment costs \citep{mit2022}. Short-term storage of district heat requires investment in water tanks and long-term storage in pit thermal storage. Caverns can store hydrogen and synthetic gas but depend on geological conditions \citep{mit2022}. For gas, the potential corresponds to today's gas storages; the potential for hydrogen caverns builds on \citet{Caglayan2020}. In addition, investment into tanks for hydrogen storage is possible without any restrictions on potential, but investment costs per energy are substantially higher \citep{mit2022}.

The right column of Fig. \ref{fig:overMod} lists the final energy demands the model considers. Section \ref{demFlex} details how the model represents the flexibility of final demand. The final electricity demand corresponds to residential, service, and industry appliances. 

Next, there is a final demand for space and process heating. The model further splits process heat into three temperature levels: low temperature up to 100°C, medium temperature from 100°C to 500°C, and high temperature above 500°C. To cover the demand for space heat, the model can invest in solar thermal heating, boilers and heat-pumps fuelled by electricity, the connection to a district heating network, and boilers fuelled by biomass or synthetic fuels. The same options are available for low temperature heat, except for solar thermal heating; medium and high temperatures preclude heat-pumps and district heating. Even if a technology can provide a particular heat type, its capacity can still be constrained. The potential for district heating and ground-source heat-pumps in space heating reflects settlement structures. District heating can only supply heat in cities, towns, and suburbs;  ground-source heat-pumps in rural areas and half the demand in towns and suburbs. In process heating, today's shares of district heating and electric boilers are an upper limit since it is indeterminate whether their temperature level is sufficient to supply a larger share of demand. Technologies for space heat and low-temperature heat utilize heat storage, as section \ref{demFlex} elaborates.

Finally, the model encompasses the demand for transport services. Passenger transport includes private road, public road, and public rail transport; freight includes heavy road, light road, and rail transport. Vehicle options for road transport include BEVs, fuel cells, and internal combustion engines using synthetic fuels. In addition, compressed natural gas vehicles are available for private passenger road transport and electric overhead lines for heavy freight transport. Rail transport can run on electricity, diesel, or fuel cell engines. Finally, there is a fixed demand for methanol for aviation and navigation.

The linear model formulation disregards operational restrictions, e.g., ramping rates or start-up times, for nuclear power or any other technology. Previous studies agree that operational detail has little impact on results if models include options for short-term flexibility, like batteries or demand-side response \citep{Poncelet2016,Poncelet2020}. In addition, our analysis excludes most technologies subject to significant operational restrictions, such as coal plants. As a result, the omission only creates a bias in favor of nuclear power, which aligns with our intention to compute an upper bound on its deployment.

The temporal scope of the model consists of a single year. For the hourly profiles of electricity demand, heat demand, and renewable capacity factors, the model uses historical data from 2008, an average climatic year regarding these inputs. Limiting planning to one year of climatic data is a common but perfectible practice. Previous research finds that supply and demand vary substantially by climate year, and a single year is insufficient to plan a secure system \citep{Bloomfield2016,Ruhnau2022}; considering multiple---ideally smartly selected---climatic years improves accuracy but is challenging since it linearly increases model size which entails an exponential increase of computation time \citep{Ullmark2024,Goeke2024}. In addition to climate conditions, a single year of data also discounts the risk of coincident nuclear outages, for instance, during the Winter of 2022/2023 in France \citep{schneider_world_2023}.

\subsubsection{Spatial resolution and starting grid} \label{map}

The spatial scope includes the European Union (excluding isolated island states), the United Kingdom, Switzerland, Norway, and the Balkan region.

\begin{figure}
	\centering
		\includegraphics[scale=0.25]{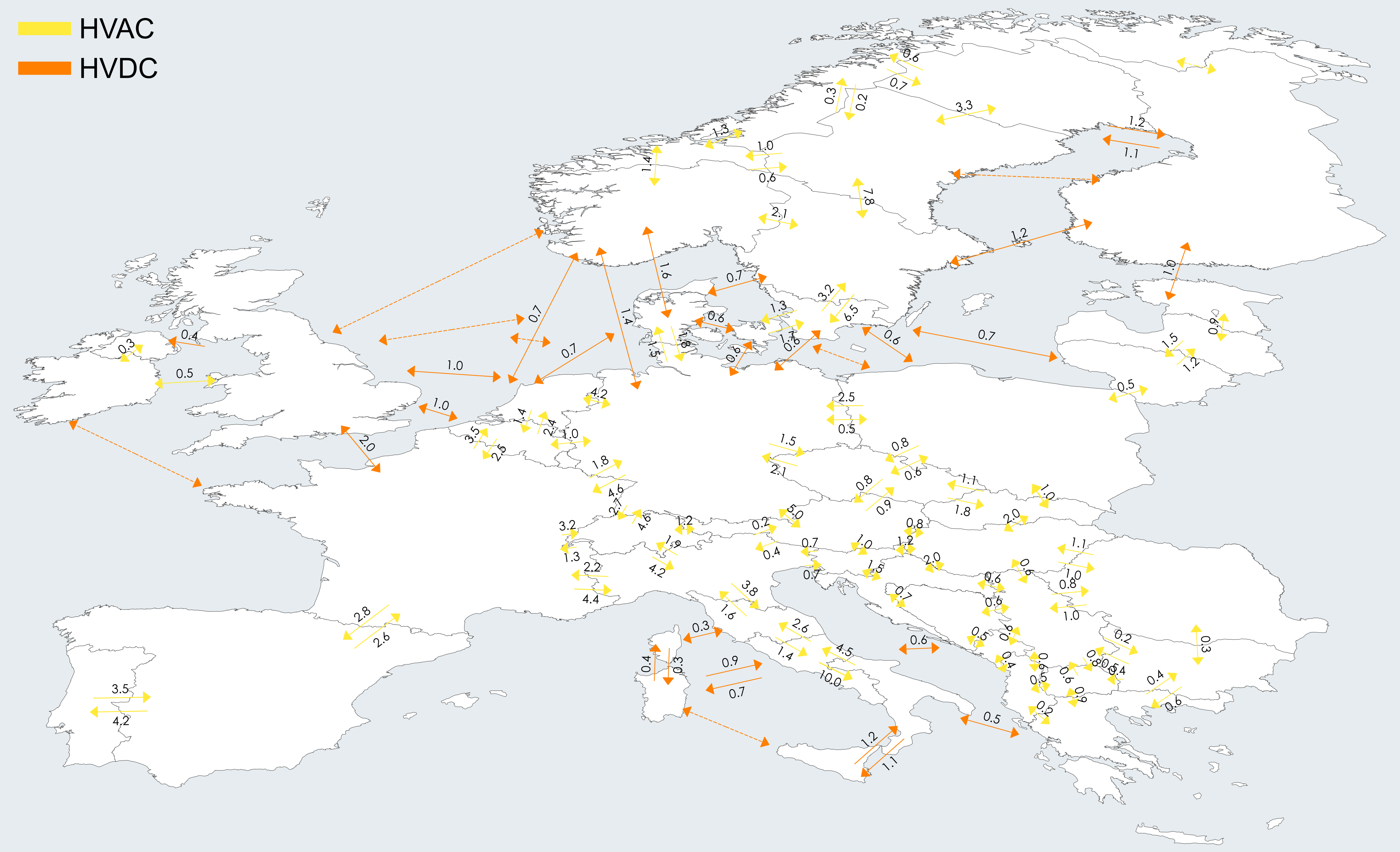}
	\caption{Spatial resolution and grid connections for electricity}
	\label{fig:startGrid1}
\end{figure}

The spatial resolution for electricity corresponds to the electricity market zones as shown in Fig. \ref{fig:startGrid1}. The exchange of electricity between zones requires HVAC (high-voltage alternating current) or HVDC (high-voltage direct current) lines, as indicated by the arrows in the figure. Whether HVAC or HVDC transmission connects two zones depends on the information provided by the European grid operator, but generally, HVDC is limited to long distances over sea \citet{entsoe2020}. The model can invest in new transmission lines; today's lines are available without investment, only incurring maintenance costs to reflect their long technical lifetime. The costs of new lines also include maintenance costs. The numbers in Fig. \ref{fig:startGrid1} specify today's capacities. The dotted arrows represent potential connections without any pre-existing capacity.

\begin{figure}[!htbp]
	\centering
		\includegraphics[scale=0.13]{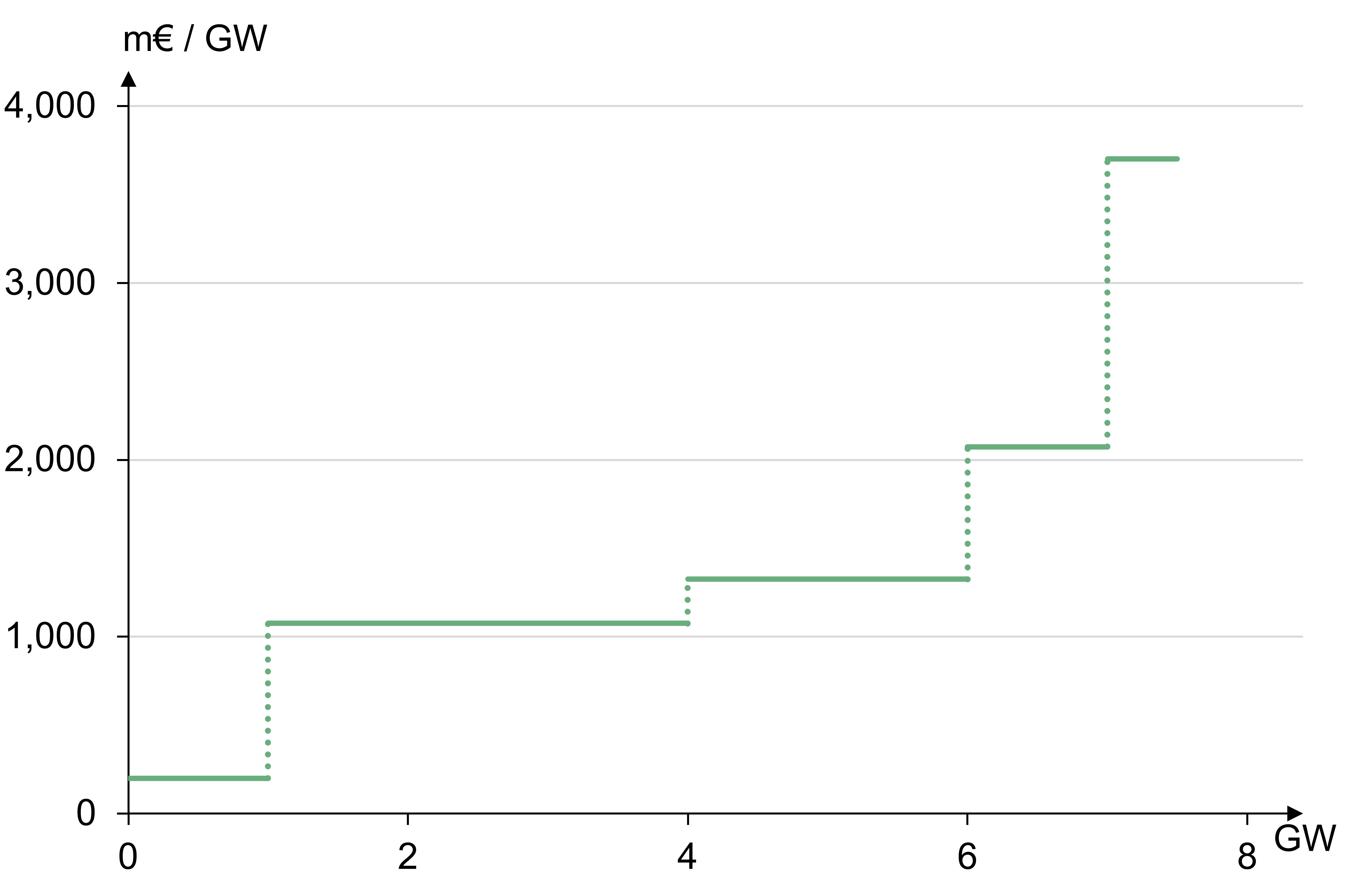}
	\caption{Potential-cost curve for expanding the electricity grid between Germany and the Netherlands}
	\label{fig:ntcExp}
\end{figure}

Electricity exchange within a market zone is unrestricted. This "copper plate" assumption neglects interzonal congestion but it allows building on specific ENTSO-E data listing potential expansion projects in the European electricity grid \citet{entsoe2020}. From this data, we obtain a potential-cost curve for each connection with increasing investment costs and an upper limit on expansion. For example, Fig. \ref{fig:ntcExp} shows this curve for the connection between Germany and the Netherlands. In this case, the specific investment costs rise from 200 to 3,700 million € per GW, and the total expansion limit is 7.5 GW. A more detailed spatial resolution cannot utilize this data and has to rely on potential and cost data that is not connection-specific and subject to substantial uncertainty; for instance, literature values for expansion costs vary by a factor of 5 for underground cables \citep{acer,nep}.

We use a transport instead of flow formulation to model the grid operation since previous research found this simplification to be sufficiently accurate \citep[][cited by \citealp{goke_how_2023}]{Neumann2020a}. In line with the same source, transmission losses amount to 5\% and 3\% per 1,000 km for HVAC and HVDC grids, respectively. The distance used is the distance between the geographic centers of each zone.

\begin{figure}
	\centering
		\includegraphics[scale=0.25]{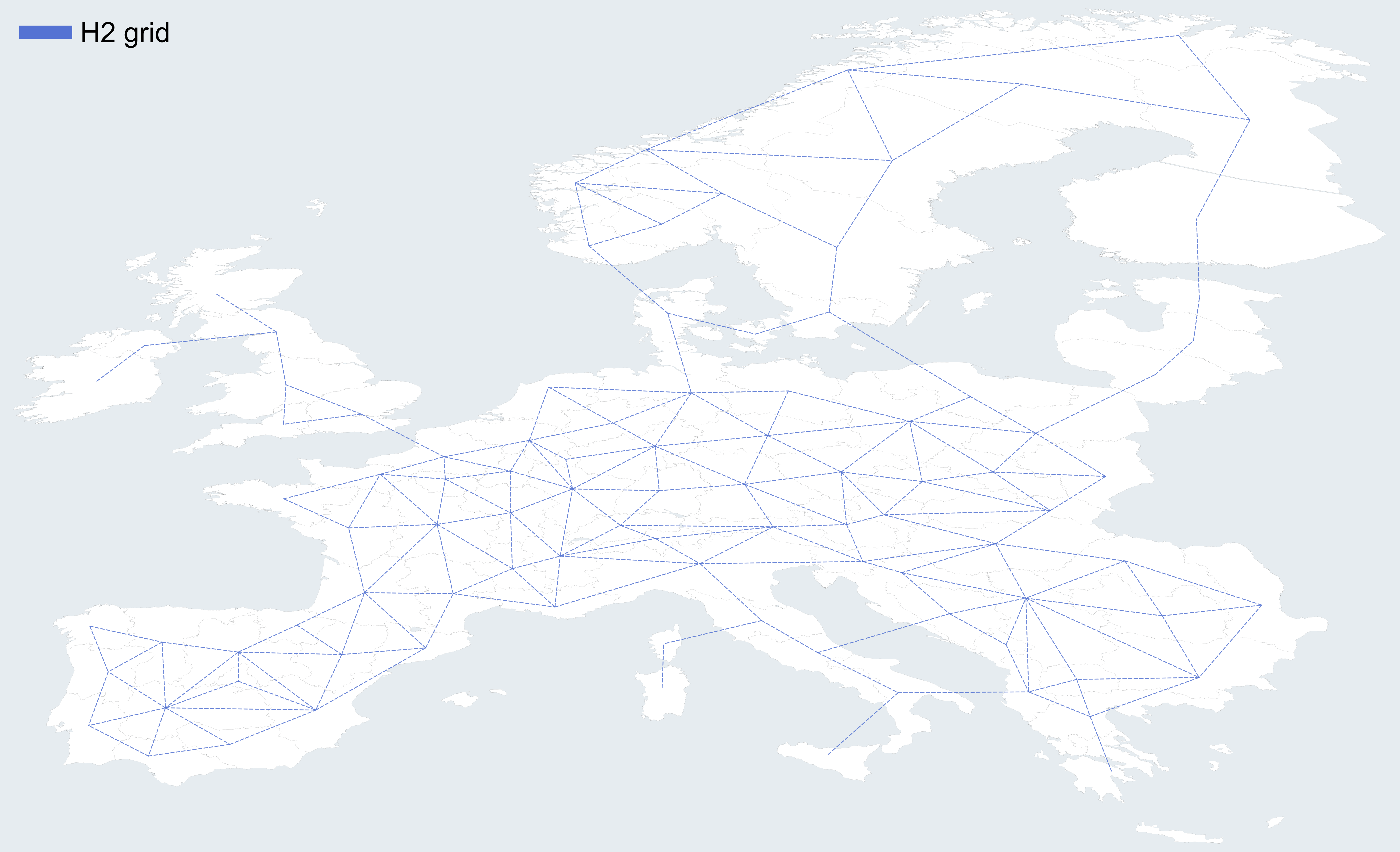}
	\caption{Spatial resolution and grid connections for hydrogen}
	\label{fig:startGrid2}
\end{figure}

For hydrogen, the model uses a different, more detailed spatial resolution consisting of the 96 clusters shown in Fig. \ref{fig:startGrid2}. For hydrogen exchange between clusters, the model can invest in pipelines indicated by the dotted lines in Fig. \ref{fig:startGrid2}. These pipelines are subject to costs of 0.4 million EUR per GW and km and energy losses of 2.44\% per 1,000 km \citep{DEA}. The distance between the geographic center of clusters serves as an estimate for pipeline length.

The model also deploys the 96-cluster resolution for heat and synthetic gases; passenger and freight transport uses the zonal resolution; and methanol and biomass use a national resolution. The exchange of methanol and biomass between countries is possible but does not require grid infrastructure. Instead, it uses trucks and only incurs variable costs based on \citet{jrc2015}. 

\subsubsection{Flexibility and electricity demand} \label{demFlex}

Nuclear power has a key advantage over renewable wind, or PV---it is dispatchable. In our analysis, nuclear plants are considered fully flexible and not subject to operational restrictions. To what extent these characteristics drive investment strongly depends on whether and how a model includes alternatives to nuclear for flexibility on the one hand but also accurately accounts for inflexibilities in the system on the other. One method in the applied model to control flexibility is varying the temporal resolution by energy carrier in the same way as the spatial resolution \citep{Goeke2020b}. An added benefit of this method is that it reduces problem size and, thus, frees computational resources that can be applied where more detail is essential. As a result, the model achieves an hourly time series for the whole year for electricity---the carrier pivotal for the analysis and most sensitive to fluctuations in supply and demand.

For space and process heat, the model applies a four-hour resolution capturing the thermal inertia of buildings and the possibility of rescheduling processes \citep[][cited by \citealp{goke_how_2023}]{Heinen2017}. In the model, this translates into an energy balance for each four-hour step, unlike the energy balance for electricity that enforces supply to match demand in each hour individually. In other regards, the four-hour resolution does not affect technology representation. For instance, the consumption of an electric heat pump remains an hourly variable subject to an hourly capacity constraint and hourly fluctuations of the coefficient of performance. As a result, even with a four-hour-resolution, substantial operational constraints remain. Especially in the case of process heat, the flexibility is still severely limited due to high capacity factors averaging 97\%. For instance, when the demand profile prescribes a utilization of 97\% for a four-hour time-step; the process can, at most, reduce production to 88\% in one hour and increase to 100\% in the remaining three since it must still meet the total demand.

In contrast to the lower resolutions that add flexibility, the model also adds constraints to reflect that the consumer's demand profiles predetermine the operation of space and process heating systems. Unlike technologies supplying electricity or district heat, there is no encompassing network, thus, enabling some systems to reduce and others to increase supply. To reflect this, we set the generation of space and process heating in each time-step proportional to the installed capacities. The model can invest in local heat storage for space and process heating to add operational flexibility. Regarding storage, a heat-pump or boiler can increase production above the predetermined level and store the excess generation to reduce production later and instead discharge stored energy. Since the heating systems are disconnected, each technology requires a separate dedicated storage system.

District heating uses a four-hour resolution to account for the inertia of the network \citep[][cited by \citealp{goke_how_2023}]{Triebs2022}. Besides the operational flexibility, the model can invest in water tanks for short-term storage and in pits for seasonal storage.

Demand from BEVs is flexible within limits: First, an hourly profile restricts charging and reflects when BEVs connect to the grid. Second, the sum of charged electricity must match consumption for driving each day, assuming vehicle batteries are sufficient to balance supply and demand within a single day. Accordingly, private passenger transport in the model uses a daily resolution. We assume a charging capacity of 5 kW for private passenger vehicles and apply a 75\% safety margin for all charging profiles \citep{entsoe2021}.

Overall, our representation of BEV flexibility is a middle ground of the wide range of assumptions in the literature. For instance, \citet{Heinisch2021} conclude that 85\% of demand from BEVs will be flexible, while \citet{Mangipinto2022} state that only 44\% of BEVs adopting smart charging is plausible. Our implementation restricts the share of smart charging only implicitly but makes conservative assumptions on the numeric parameters. The charging capacity of 5 kW is very low compared to other literature values. \citet{Mangipinto2022} assume an average of 17.52 kW, including fast chargers, or still 6.1 kW when excluding fast chargers; in a technology-optimistic scenario, capacities even increase to 84.2 kW and 54 kW, respectively. In addition, they apply an 80\% instead of a 75\% safety margin. \citet{Strobel2022} assume a charging capacity of 4.7 kW at home and 11 kW at work. Assuming that the vehicle's battery can balance supply and demand within each day is plausible, considering the average daily demand is 4 kWh and the average energy storage capacity of contemporary BEVs is 71.6 kWh \citep{evDatabase}.

Finally, we do not consider bidirectional loading, also known as vehicle-to-grid, where BEVs can operate like regular batteries and supply energy back to the grid \citep{Hannan2022}. Other electricity consumption in the transport sector, such as rail transport, is inflexible and uses an hourly resolution.

Lastly, power-to-x and synthetic fuels are flexible elements in the modeled energy system. As section \ref{sec:model} explains, the model includes various pathways, either utilizing biomass or electricity for generating synthetic fuels, most prominently hydrogen. 
The produced fuels can again generate electricity or cover final demand in other sectors. In some cases, like high-temperature heat or aviation and navigation, they are even the only options. Consequently, in the model, power-to-x is used for both demand-side management and long-term storage. The general option to invest in storage for synthetic fuels adds to the potential flexibility.

For hydrogen and methane, the model uses a daily resolution to reflect the carriers' inertia and the pipelines' storage capabilities. In contrast, the standard practice in the literature on integrated energy systems is a uniform hourly resolution since few models can vary temporal resolution by energy carrier. A notable exception are \citet{Wang2022}, who also deploy an hourly resolution for electricity but a daily resolution for hydrogen. However, although other research does not account for the storage capability of pipelines, it acknowledges it \citep{Guerra2023,He2021}. In addition, daily resolutions are standard practice in research exclusively modeling hydrogen \citep{Parolin2022,Arpino2024}. Furthermore, the daily clearing of gas markets suggests that a daily resolution captures the system's inherent flexibility.

\subsubsection{Renewable supply}  \label{renewSup}

In the model, the level of investment into nuclear power depends on the potential and costs of renewable energy. Therefore, this section further details the used assumptions regarding renewable supply.

The model considers the technical potential of renewables, i.e., wind and PV, and neglects sociological and political implications, such as local opposition to wind farms. For nuclear power (and transmission infrastructure), we equally abstract from limited public acceptance. Thus, our study uses data from \citet{Auer2020} on the technical potential of wind and PV. To put these numbers into context, we compare them against the results of a meta-analysis on renewable potential in Europe in Tab. \ref{tab:ee_pot2} and Germany in Tab. \ref{tab:ee_pot} \citep{Risch2022,Tour2023}. The material demonstrates that the assumed potential aligns with the meta-analyses' results. For Germany, capacities are close to the reference, assuming current legislation except for PV; here, we are more conservative. On a European level, our assumptions are close to or lower than the median of literature values.

\begin{table}
\caption{Comparison of renewable potential for Europe (excl. Norway and Balkan) in GW.} \label{tab:ee_pot2}
\centering
\begin{tabular}{lcccc}
    \hline
            & Own assumptions     &  \multicolumn{3}{c}{Literature values in} \\
                                & based on  &   \multicolumn{3}{c}{ \citet{Tour2023}}\\
                    & \citet{Auer2020} &  First quartile &  Median    & Third quartile\\
    \hline
    Wind, onshore   & 4,508   & 1,518   & 3,421   & 7,489 \\
    Wind, offshore  & 1,678    & 1,628    & 4,844   & 8,704\\
    PV, openspace   & 2,793   & 1,252  & 4,235   & 8,625 \\
    PV, rooftop     & 1,558   & 610   & 952   & 1,665\\
    \hline
\end{tabular}
\end{table}

\begin{table}
\caption{Comparison of renewable potential for Germany in GW.} \label{tab:ee_pot}
\centering
\begin{tabular}{lcccc}
    \hline
            & Own assumption     & Results for  & \multicolumn{2}{c}{Literature values} \\
                                & based on  & current legislation &   \multicolumn{2}{c}{in \citet{Risch2022}}\\
                    & \citet{Auer2020} & in \citet{Risch2022} & Lowest  & Highest\\
    \hline
    Wind, onshore   & 386   & 385   & 68   & 1,188 \\
    Wind, offshore  & 84    & 79   & 34   & 99.6\\
    PV, openspace   & 301   & 456   & 90   & 1,285\\
    PV, rooftop     & 177   & 492  & 43   & 746\\
    \hline
\end{tabular}
\end{table}

To reflect that sites for renewables differ in quality, wind and PV technologies are further subdivided. Onshore wind and open space PV have three sub-technologies each; rooftop PV has two. Offshore wind is subdivided into two technologies with identical full-load hours, but different investment costs representing shallow and deep waters. Figures in section \ref{supEE} of the appendix sort the entire potential of different renewables by full-load hours and demonstrate how increased deployment results in smaller yields. Although electricity uses the zonal resolution, the potential and time series of wind and PV are cluster-specific to accurately describe location-dependent fluctuations. In the case of hydropower, we assume today's capacities are available without any investment due to their long technical lifetime but prohibit any expansion, assuming the available potential is fully utilized.

The use of biomass in each country is subject to an upper energy limit of 1,081 TWh for the entire model, including waste for the production of biomass \citep{jrc2015}. In addition to domestic production, the model can import hydrogen by ship at costs of 131.8 US-\$ per MWh and by pipeline from Morocco or Egypt at 90.7 and 86.8 US-\$ per MWh, respectively \citep{Hampp2021}.

\section{Results} \label{main}

This section \ref{main} first compares the computed range of LCOEs for nuclear power with renewable sources. Afterwards, we present the results of the more elaborate analysis applying the energy planning model.

\subsection{Levelized cost comparison}

\begin{figure}[!htbp]
	\centering
		\includegraphics[scale=0.1]{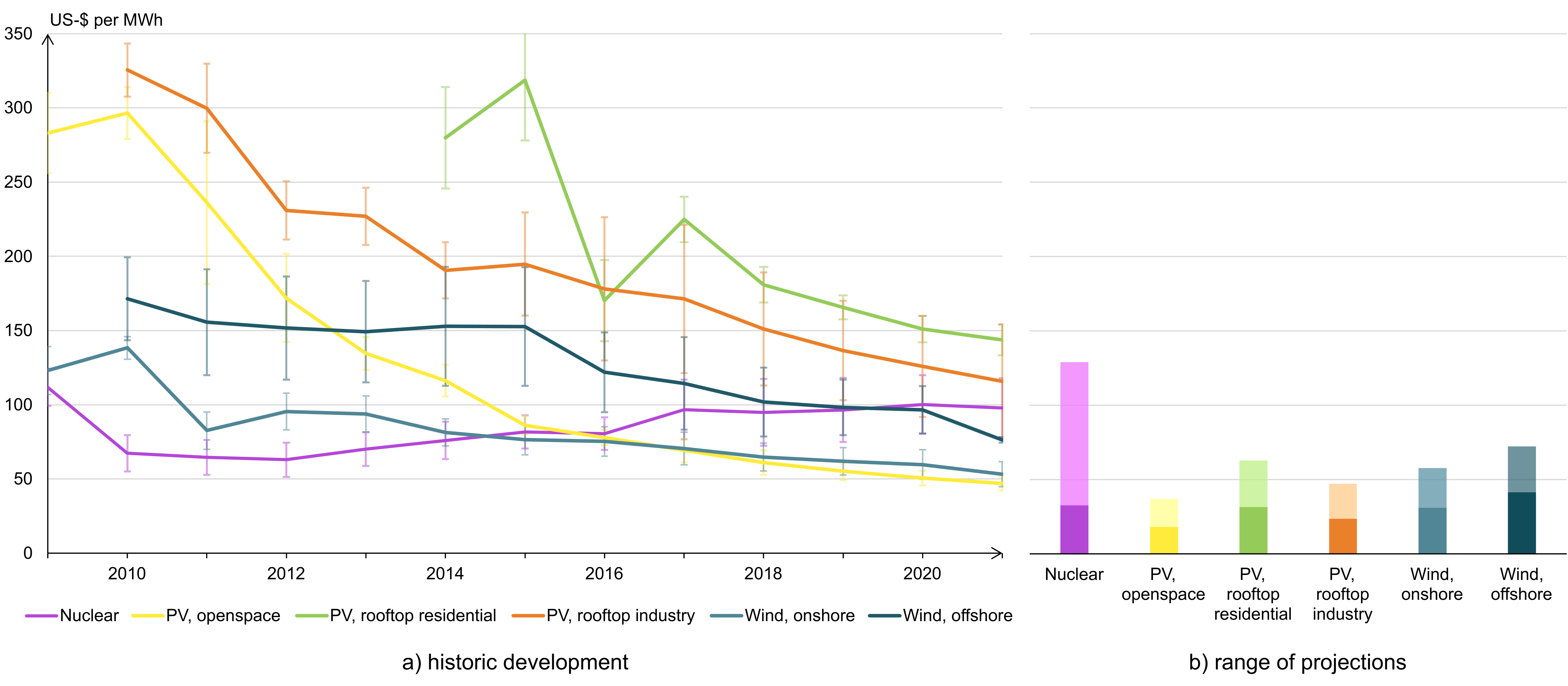}
	\caption{Comparison of LCOEs for generation technologies}
	\label{fig:lcoe}
\end{figure}

Based on the review of construction projections, we compute a conceivable range for LCOE for nuclear power using the formula presented in section \ref{lcoe}. Fig. \ref{fig:lcoe} compares this range against historic LCOEs on the left side (a) and against cost projections for renewable energy on the right (b). Historic LCOEs for all energy sources build on the corresponding reports by \citeauthor{lazard2021} from 2009 to 2021.\footnote{To exclude subsidies and harmonize interest rates, we re-computed LCOEs based on the reported techno-economic assumptions instead of directly copying the values in the Lazard reports \citep{lazard2009,lazard2010,lazard2011,lazard2012,lazard2013,lazard2014,lazard2015,lazard2016,lazard2017,lazard2018,lazard2019,lazard2020,lazard2021}.} The projection for nuclear results from the parameters in Table \ref{tab:assumptions}, projections for renewable sources use 2040 data from \citet{DEA} for cost assumptions and the 10th to 90th percentile of respective energy yields in Europe for full-load hours. For full documentation of assumptions on costs and renewable potential, see the respective sections \ref{cost_analysis} and \ref{lcoeSup} in the appendix.

The comparison of historic LCOEs shows how, in the past, costs have increased for nuclear power but decliend for renewable energy, particularly PV. Current costs for nuclear power reported by Lazard are at the upper end of projections. For wind, current costs are already within the projected range; for PV, current costs are above projections, implying a continuous degression of costs. The comparison of projected LCOEs suggests that nuclear can only compete with renewables if the industry can reverse the historical trend and achieve substantial learning effects and economies of scale to reduce project costs.

In addition to the LCOE comparison shown here, that assume a uniform interest rate of 5\%, the appendix \ref{diffInter} shows projections for 0\% and 10\% interest rates, respectively. This assumption changes the absolute values significantly, but the ratios among technologies do not change substantially since all considered technologies are capital-intensive. Therefore, we do not expect a substantial impact of interest rates on our key result, the cost-efficient share of nuclear, and will proceed assuming a 5\% interest rate for the subsequent analyses.

Finally, as mentioned in section \ref{lcoe}, the planned lifetime for some reactors under construction in Europe is 60 years instead of the 40 years we assume. With the reference interest rate of 5\%, the range of LCOEs for nuclear decreases from 32.6 US-\$ per kW to 96.4 US-\$ per kW for 40 years to 31.3 US-\$ per kW to 87.3 US-\$ for 60 years. The impact on the lower end is small since operational costs that scale with lifetime dominate here. The effect is more prominent at the upper end but still limited since discounting dampens the impact of longer lifetimes. Accordingly, the effect is more pronounced for a discount rate of 0\%, and the range moves from 24.8 US-\$ to 41 US-\$ for 40 years to 22.9 US-\$ to 27.1 US-\$ for 60 years. When increasing the lifetime of renewables by the same factor of 50\%, they achieve a similar reduction in LCOEs. In conclusion, longer lifetimes do not significantly move the lower ends of LCOEs for nuclear; the upper-end only changes substantially if interest rates are close to zero.

\subsection{Planning model analysis}

However, comparing LCOEs for nuclear and renewables is of limited insight because nuclear and renewables ultimately provide different goods: nuclear plants are dispatchable and can, within limits, adapt supply to demand; renewable energy sources like wind and PV cannot. Therefore, we elaborate our comparison and deploy the techno-economic energy system planning model described in section \ref{sec:model}. First, section \ref{sankey} describes the overall system design for the highest observed nuclear share of 52\%, and a fully renewable system. These results provide the context for analyzing nuclear power and flexibility in the subsequent section \ref{nuPoFlex}.

\subsubsection{Overall system design} \label{sankey}

\begin{sidewaysfigure}[!htbp]
	\centering
		\includegraphics[scale=0.5]{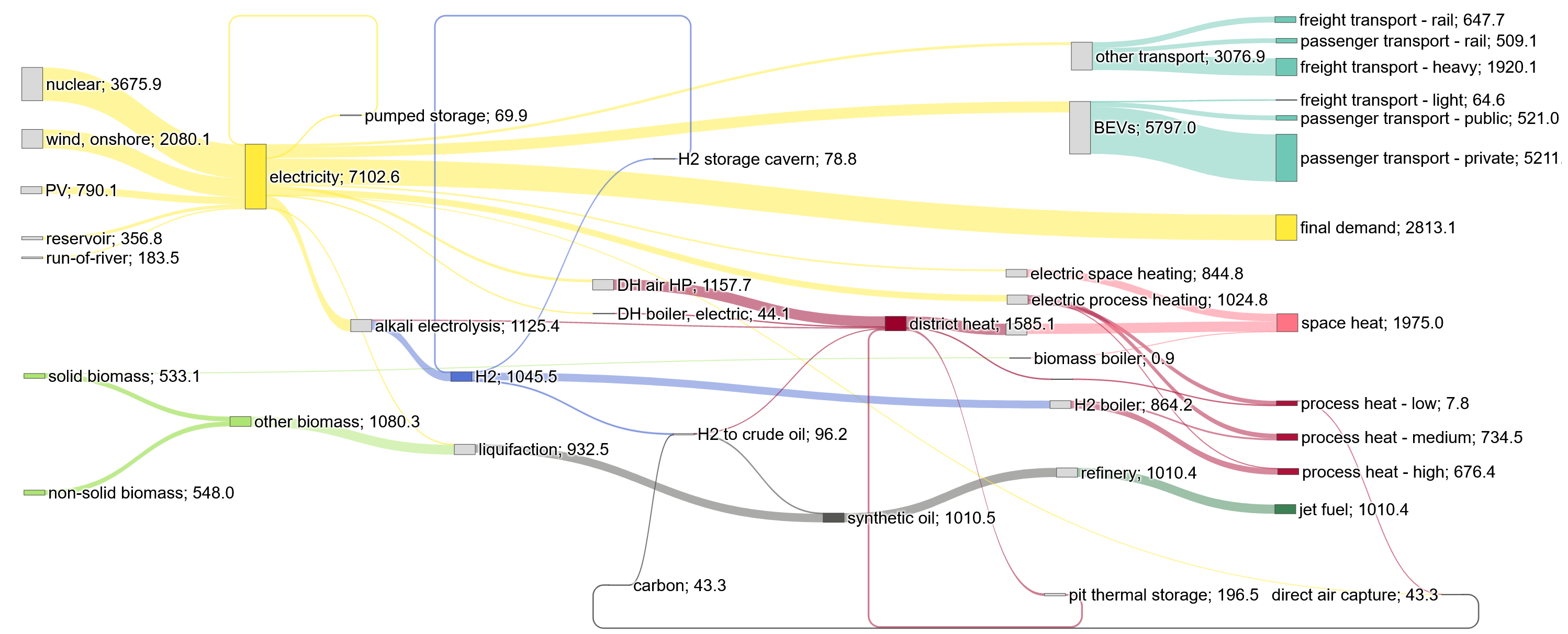}
	\caption{Sankey diagram for the 52\%-nuclear system, in TWh/Gpkm/Gtkm}
	\label{fig:sankNu}
\end{sidewaysfigure}

\begin{sidewaysfigure}[!htbp]
	\centering
		\includegraphics[scale=0.5]{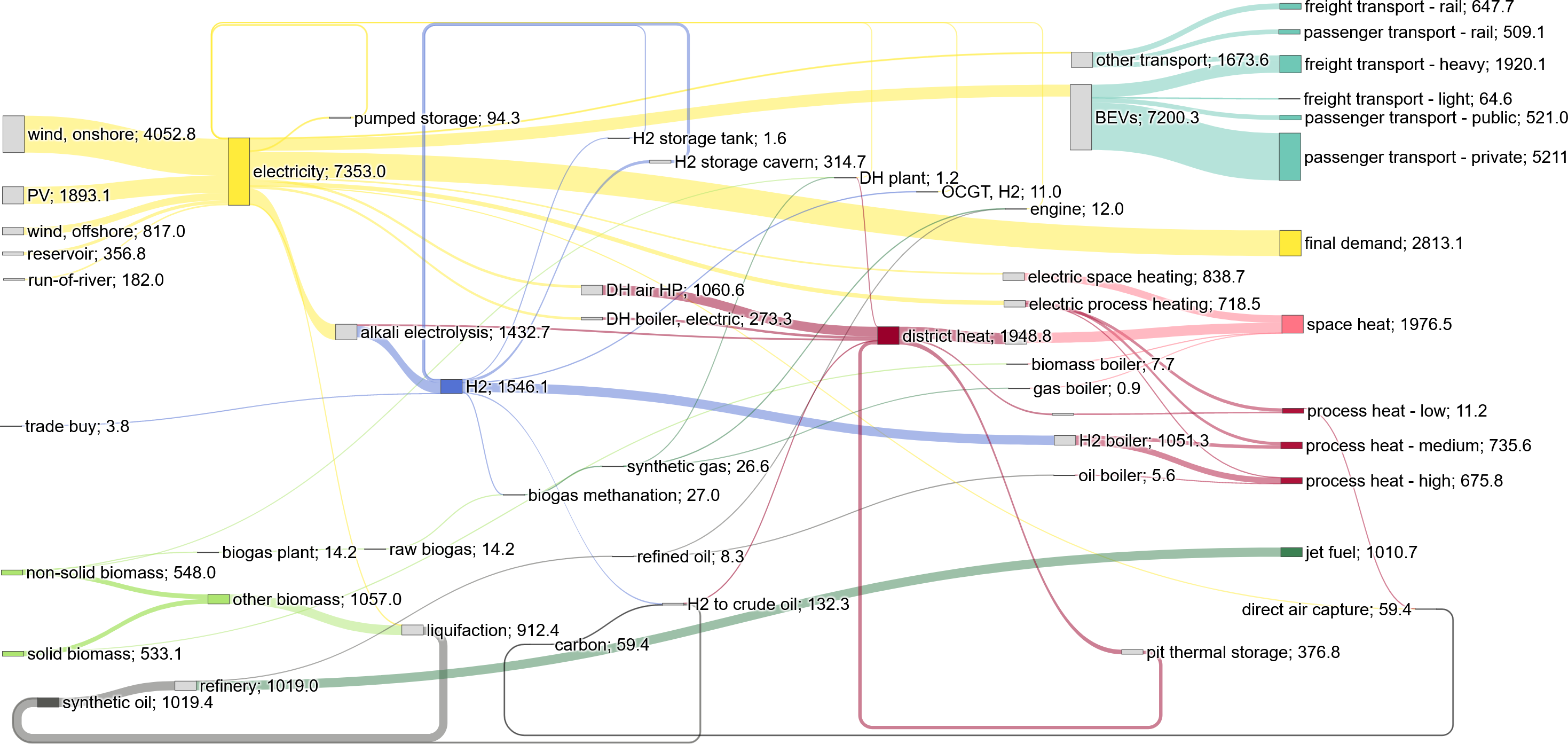}
	\caption{Sankey diagram for the system without nuclear, in TWh/Gpkm/Gtkm}
	\label{fig:sankNoNu}
\end{sidewaysfigure}

The Sankey diagrams in Figs. \ref{fig:sankNu} and \ref{fig:sankNoNu} show the energy flows and conversion processes in the solved model for two extreme cases, one with a 52\% nuclear share, the highest share we observed, and a renewable system without any nuclear. The diagrams are the quantitative counterpart to Fig \ref{fig:overMod}, showing primary energy sources on the left, secondary energy carriers in the middle, and final demand on the right. Colored nodes represent energy carriers; grey nodes conversion or storage technologies. Accordingly, the ratio of flows entering and leaving a technology reflects its average yearly efficiency. Many technology nodes aggregate multiple individual technologies in the model; for instance, electric space heating includes air and ground heat pumps. All energy flows in the diagram are in TWh; the final demand for passenger transport is in Gpkm, and the final demand for freight transport is in Gtkm. The presented quantities are the sum across the entire model scope according to section \ref{map}.

In both cases, electrification dominates the energy system. The 52\%-nuclear system supplies 7,100 TWh of electricity with a 29\% share of onshore wind and an 11\% share of PV. In the case without nuclear energy, the supply amounts to 7350 TWh with 55\% onshore wind, 26\% PV, and, unlike the nuclear system, 11\% offshore wind. Hydro supply is identical for both cases at 540 TWh since we did not allow its expansion. 

In both cases, the model decides on full transport sector electrification, resulting in 1,500 TWh of consumption. In space heating, both cases fully utilize the available potential for district heating and cover the remaining demand with residential heat pumps, resulting in a consumption of 800 TWh. Biomass utilization is also very similar in both cases. Since for methanol, or synthetic fuel, production via liquefaction is the most cost-efficient process to meet the final demand for synthetic fuels, this process utilizes most of the available biomass potential.

However, the biomass conversion also shows the first significant difference between the 52\%-nuclear and the renewable system. In the renewable system, 3.6\% do not go towards liquefication but instead fuel peak-load boilers and plants to supply electricity and heat when renewable generation is low. Due to dispatchable nuclear plants, this does not occur in the 52\%-nuclear system, and liquefication can utilize the entire biomass potential.

Heat supply reveals further differences between the two cases. In the renewable case, process heating has a higher share of hydrogen boilers to reduce the inflexible and almost constant consumption from electric boilers. Peak-load plants add to the hydrogen demand, resulting in 250 TWh of additional production by electrolyzers compared to the 52\%-nuclear system. Since this production depends on weather conditions, storage of hydrogen increases by 236 TWh as well. In district heating, the stored heat in the renewable case almost doubles to 380 TWh. Correspondingly, the heat supply from two sources increases: First, the inflexible access heat provided by electrolyzers, and second, the heat generation of electric boilers. The additional boilers are presumably not installed as peak-load technologies but to utilize rare peaks of excess renewable generation and low costs.

Overall, the subtle but clear differences between 52\%-nuclear and the renewable system prove the ability of the model to reflect the effect of dispatchable or fluctuating electricity supply on the overall energy system.

\subsubsection{Role of nuclear power and flexibility} \label{nuPoFlex}

\begin{figure}[!htbp]
	\centering
		\includegraphics[scale=0.15]{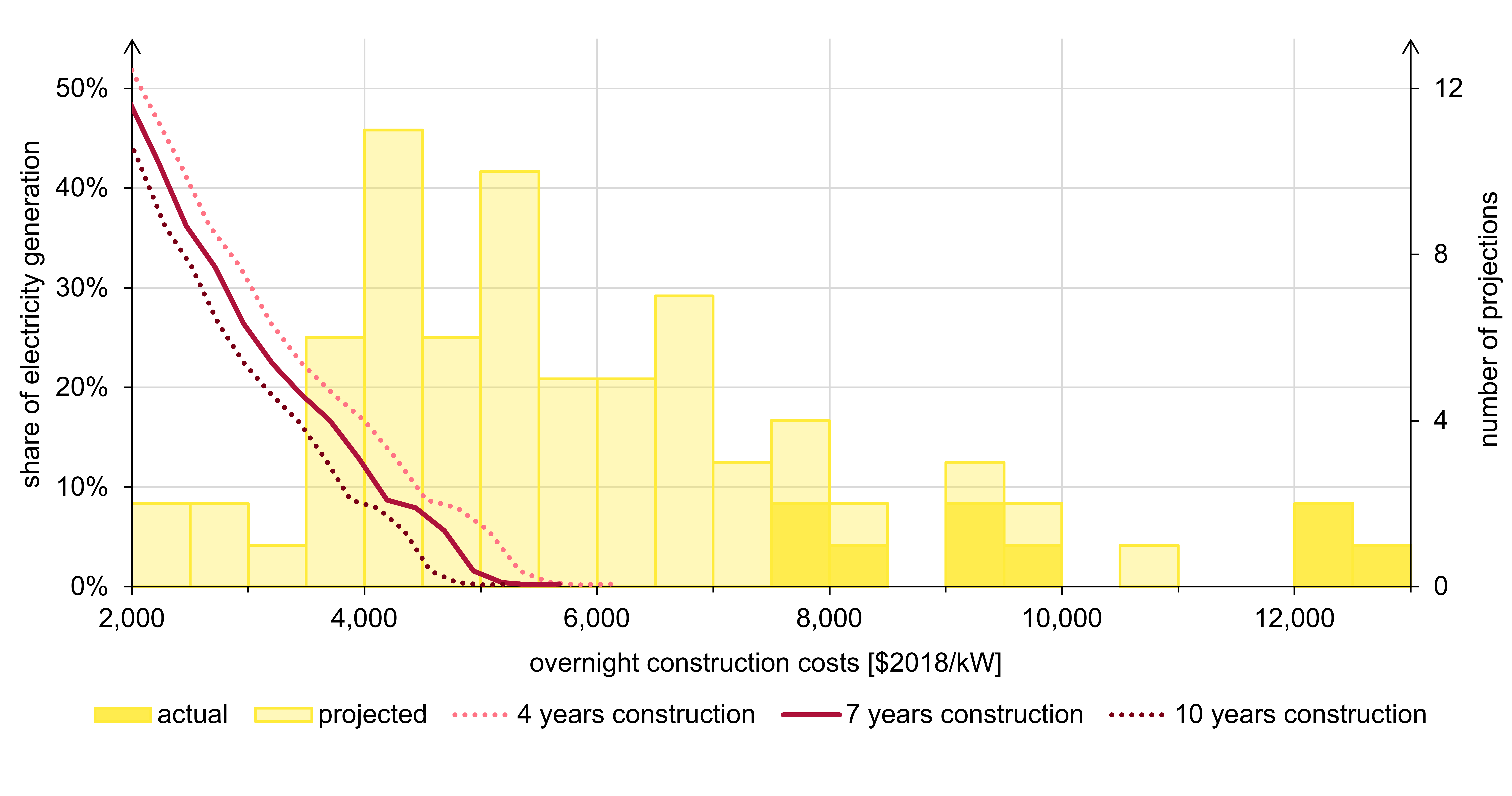}
	\caption{Share of nuclear power in electricity generation and distribution of overnight construction cost projections}
	\label{fig:costShare}
\end{figure}

Fig. \ref{fig:costShare} presents the key results of the energy system model analysis. On the left axis, the graph shows the share of nuclear power in total electricity generation depending on the overnight construction costs and time. The right axis provides the corresponding distribution of cost projections in the literature and actual costs reported by recent projects.

The results confirm the tentative conclusion from the LCOE analysis. If construction costs and time drop substantially, nuclear power provides a significant share of electricity. For example, for nuclear to provide 10\% of total electricity generation, roughly the current global share \cite{schneider_world_2023}, overnight construction costs must drop to 4,120 US-\$ per kW at 7 years construction time, and to around 4,450 US-\$ per kW at 4 years construction time, allwhile only 20\% of overnight construction cost projections lie below 4,120 US-\$ per kW, and actual costs of recent projects lie well above 7,500 US-\$ per kW. At the lowest construction costs, projected by \citet{oecd_full_2018}, nuclear power provides up to 52 \% of total electricity generation as shown in Fig. \ref{fig:sankNu}.

Despite its assumed flexibility, nuclear power substitutes rather than complements fluctuating supply from wind and PV generation. In the most extreme case, with a 52 \% generation share, nuclear power plants operate at an average capacity factor of 94.4\%, close to the 95\% limit. The capacity factor changes slightly with smaller nuclear capacities and, consequently, higher renewable shares. At a nuclear share of 7.9\%, the capacity factor still amounts to 88.7\%, which is above the average of European nuclear plants operating today \citep{wealer_ten_2021}. This observation implies that nuclear power is, even if technically capable, economically unsuited to provide system flexibility due to its high investment costs.

\begin{figure}[!htbp]
	\centering
		\includegraphics[scale=0.15]{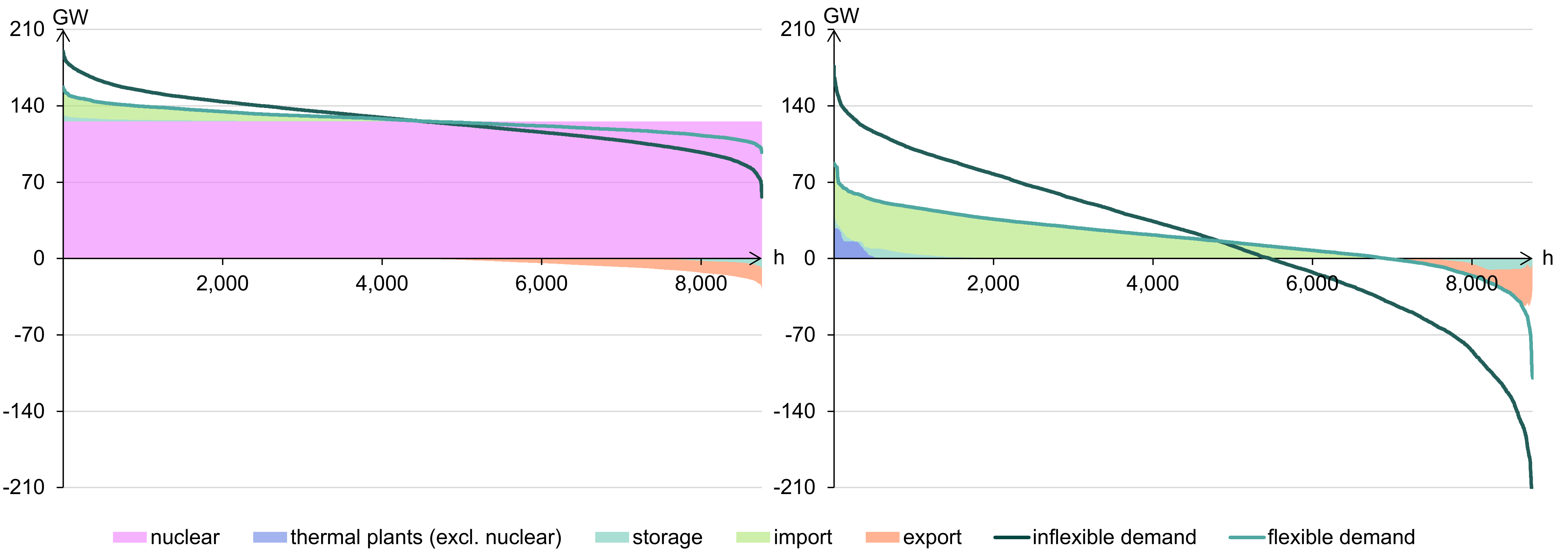}
	\caption{Residual load curve for Germany in a system with a nuclear share of 52\% (left) and no nuclear at all (right)}
	\label{fig:durCur}
\end{figure}

To highlight the system's renewable integration, Fig. \ref{fig:durCur} exemplarily compares the residual load curves of Germany for a system with the highest nuclear share of 52\% and a fully renewable system. The residual load is the remaining electricity demand after subtracting the supply of fluctuating renewables from the total demand. Plotted in descending order, the curve shows the residual peak load for each hour on the y-axis, the amount of residual demand as the area above the x-axis, and excess energy as the area below the x-axis. Appendix \ref{timeSeries} shows an extract of the hourly model results from which the duration curves were computed.

The graphs each show two residual load curves: 'Flexible demand' shows the actual residual load in the model; 'inflexible demand' is computed ex-post, assuming demand cannot adapt to supply. To what extent demand is flexible depends on its use. The conventional final demand for electricity is inflexible but only constitutes around 39\% of total demand, with exact shares differing from case to case. Additionally, creating synthetic fuels takes up 24\%, electrified transportation 20\%, and heating 17\% of total electricity demand. This demand can be flexible within specific restrictions detailed in section \ref{methods}. Some of this flexibility comes without additional cost, like flexible charging of BEVs. However, most options require the model to invest, for instance, in heat or hydrogen storage. In the results, process, and space heating are only responsible for 1\% of the shift from the inflexible to the flexible demand curve. Investment into local heat storage for space or district heat does not occur. Flexible charging of BEVs accounts for 14\%, flexible district heating for 24\%, and finally, the greatest source of flexibility is the flexible production of hydrogen at 61\%. Here, the inflexible reference is operating electrolyzers at a constant capacity for the entire year.

In the 52\%-nuclear system, wind and PV only account for 10\% of generation in Germany, and the need for flexibility is small. The residual load is flat and does not substantially adapt to supply. For the system without nuclear, the results are contrary. Demand substantially adapts to supply ro reduce the amount of residual demand and excess energy. For instance, electrolyzers operate more flexibly, causing their full-load hours to decrease to 3,343 compared to 5,979 in the 52\%-nuclear system while the capacity doubles from 35.9 to 77.1 GW. Investments in heat and hydrogen storage provide additional flexibility. Storage capacities for district heating increase to 40.9 TWh compared to 18.2 TWh in the 52\%-nuclear system; for hydrogen, even to 12.4 from 1.7 TWh.

Colored areas in the graph indicate how the system covers the residual load. In the 52\%-nuclear system, 132 GW of German nuclear capacity operates at close to full capacity and covers the predominant share of residual load. Additional imports of 86.9 TWh from neighboring countries cover peak loads. In this scenario, Germany exports 78.5 TWh of excess supply. In the system without nuclear, the residual load is smaller. However, imports are higher and cover 244 TWh of demand---the predominant share of residual load. Dispatchable thermal plants fueled by hydrogen and biogas only provide 7.3 TWh of generation. In neither scenario we see additional investment in electricity storage, and both systems only deploy the existing pumped storage and hydro reservoirs.

Although integrating renewables requires additional investments, nuclear power is only a cost-efficient alternative at very low construction costs. Otherwise, the investment costs of nuclear significantly outweigh the costs of renewables and corresponding flexibility options. For instance, the system without nuclear shows 9.7 billion US-\$ of additional annualized investment into demand side flexibility compared to a system with a nuclear share of 52\%. Adding 8.2 billion US-\$ for additional grid infrastructure and 3.0 billion US-\$ for thermal plants, the total costs of renewable integration add up to 20.9 billion US-\$. For comparison, we assume overnight construction costs of nuclear drop to 5,430 US-\$ per kW, the median of projections in Fig. \ref{fig:costShare}: According to \ref{fig:costShare}, the cost-optimal share of nuclear at these costs is still 0\%. Nonetheless assuming a system with a 52\% share of nuclear, one would save 20.9 billion US-\$ of integration costs, but generation costs increase by 67.8 billion US-\$ compared to the renewable scenarios. This thought experiment demonstrates how flexible nuclear does incur smaller integration costs than wind and PV, but these savings cannot outweigh higher generation costs. 

\begin{figure}[!htbp]
	\centering
		\includegraphics[scale=0.65]{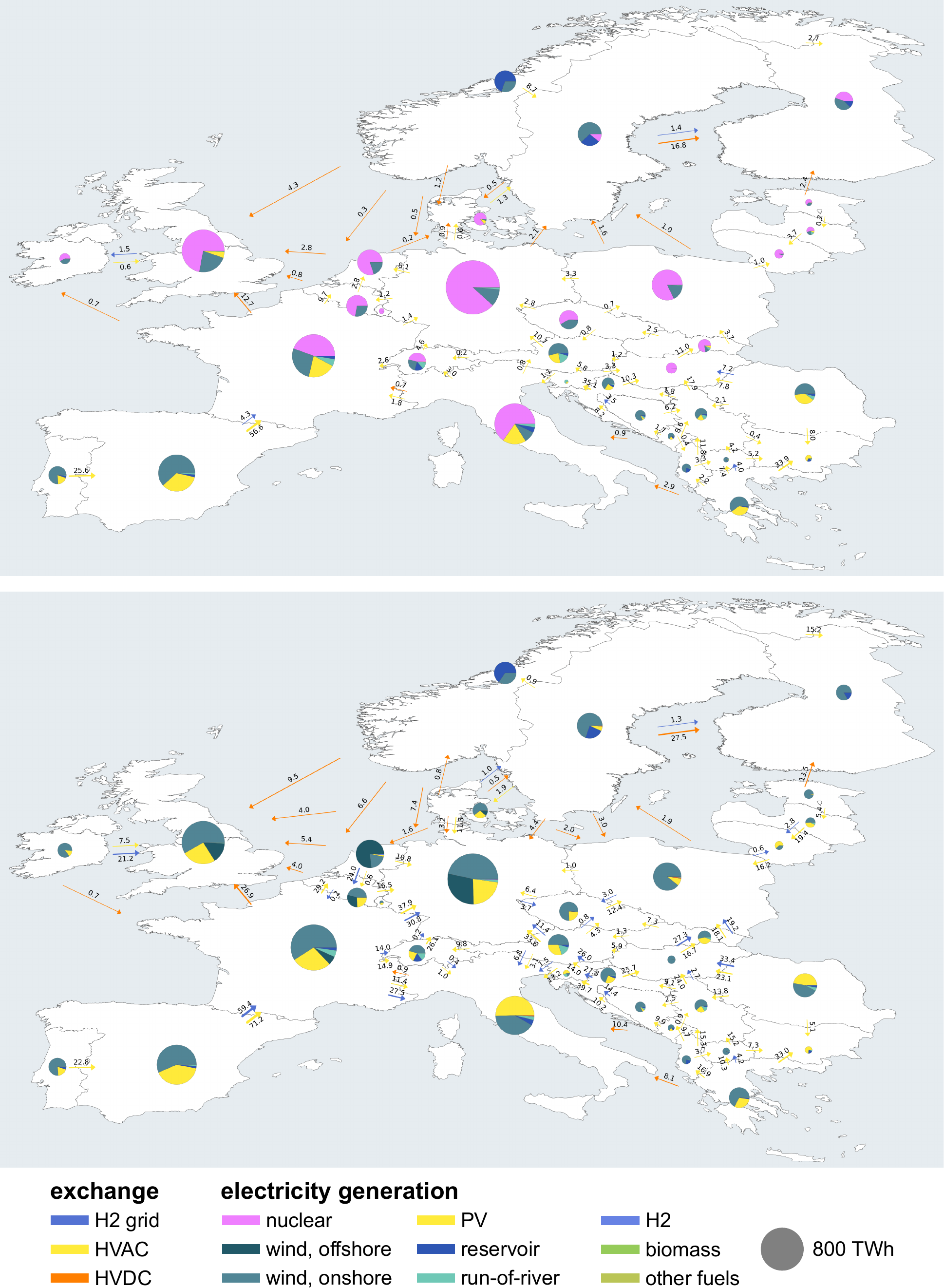}
	\caption{Electricity generation and exchange in the 52\%-nuclear system (top) and the system without nuclear (bottom). Arrows indicate net transfer of electricity between countries.}
	\label{fig:mapRes}
\end{figure}

Fig. \ref{fig:mapRes} further investigates the role of import and export for both systems. It shows each country's electricity generation and net exchange of hydrogen and electricity via high-voltage alternating current (HVAC) or direct current (HVDC) grids. The exchanges of hydrogen and electricity increase substantially in the system without nuclear, reflecting the 8.2 billion US-\$ rise in costs for grid infrastructure. The maps show that this infrastructure not only serves renewable integration but also reduces generation costs because it allows the utilization of renewables where full-load hours are highest, for instance, in southern Europe for PV or the North Sea for offshore wind.

\section{Discussion} \label{diss}

Although building on an extensive review of cost projections and a detailed planning model, our analysis faces some inevitable limitations. 
Wherever possible, we choose our simplifying assumptions in a way that imposes a positive bias on nuclear power, so results are instead an upper bound rather than reliable estimates for the cost-efficient share of nuclear. This section discusses these limitations, including those that do not impose an obvious positive bias on nuclear power.

First, some limitations result from the general fundamental method. We consider the construction cost of nuclear (and all other technologies) to be exogenous. However, costs can depend on the level of investment due to learning effects. Therefore, cases in Fig. \ref{fig:costShare} with low nuclear shares, but construction costs dropping below 5,000 US-\$ per kW, are implausible since achieving such a cost reduction through learning effects would require a more substantial investment into nuclear. Furthermore, the model only considers the single year 2040, but decarbonizing the energy system takes decades, and going further into the future will increasingly favor renewables due to expected cost reductions \citep{DEA}. Therefore, we choose 2040 as the earliest date until substantial cost reductions for nuclear power are also conceivable, an optimistic assumption considering the construction times of recent European nuclear projects.

Second, the model makes several assumptions on demand-side flexibility, as explained in \ref{demFlex}. In particular, the representation of flexible BEVs, space heating, and process heating relied on a reduced temporal resolution instead of explicit modeling. The model results show that BEVs shift 48 TWh of consumption when the share of nuclear is zero and 26 TWh for the highest nuclear share compared to a total demand of 194 TWh. Overall, the share of flexible consumption never exceeds 25\%, which is well below the lower end of literature values of 44\%. Furthermore, the flexibility from space and process heating in the results is negligible. Therefore, our representation of flexibility does not significantly bias results. 

Next, section \ref{sec:model} explains that the temporal scope of the model consists of a single climatic year, which is still a necessary simplification due to computational limits. However, using a single average year missing extreme periods, with low renewable supply or high nuclear outages, does not fundamentally question our results. The previous section finds that nuclear plants operate as base-load power plants even with substantial cost reductions. As a result, neither nuclear power nor fluctuating renewables are viable options to back up the system against rare and extreme periods. Instead of backup capacity, both technologies compete to supply utilizable electricity, which a single average year at a high temporal resolution can capture reasonably well.

As explained in section \ref{map}, computational limits and data availability necessitate abstracting from interzonal congestions and limiting the grid representation to the connections of zones. Research shows that further increasing spatial resolution can significantly improve the accuracy of energy planning models \citep{Frysztacki2021}; yet nothing indicates that lower resolutions impose a bias on nuclear power or renewables. In principle, it is possible to build plants close to demand, while renewables depend on local potential. But nuclear power plants are typically located in remote areas, and research shows that it is possible to place renewables, particularly PV, close to demand to reduce grid expansion \citep{Kendziorski2022}.

Finally, we omit unit commitment constraints, as already discussed in section \ref{sec:model}. Of all considered technologies, these operational restrictions are most pronounced for nuclear power.

Beyond modeling details, the analysis only considers nuclear power as a source of electricity. Nevertheless, the model can still convert this electricity to process heat or hydrogen, but we do not consider integrated reactor concepts. Since nuclear reactors are essentially generating heat, there are concepts for integrated generation of electricity and process heat or hydrogen \citep{iaea_opportunities_2017}. These concepts are typically non-light water reactors in an early development stage, and their commercial availability by 2040 is hard to predict \citep{committee2023,oecd_beyond_2022,kupitz_small_2001}. The low technology readiness level also limits operating experiences and prevents a reasonable quantification of costs \citep{pistner_analyse_2024}. Therefore, we limit our analysis to light-water reactor concepts for electricity generation. This focus also aligns with government plans for nuclear expansion outlined in the introduction motivating our work \cite{nyt,bb}.

As discussed in section \ref{introduction}, our techno-economic analysis also omits costs for nuclear waste management, plant decommissioning, potential accidents, and proliferation risks. Since the model heavily expands transmission infrastructure to enable the integration of renewables, results can differ not only if the public does not accept the expansion of renewables but also if the public does not accept the expansion of electricity or hydrogen grids---but does accept more nuclear power.

\section{Conclusion} \label{conclusion}

Our least-cost analysis shows that even a 10\% share of nuclear power in electricity generation in a fully decarbonized European energy system is only cost-efficient if construction costs drop substantially from current levels to around 4,000 US-\$\textsubscript{2018} per kW. However, in recent years, costs and construction times for nuclear power in the last ten years have increased rather than decreased. For other characteristics of nuclear, like availability and operational costs, we apply optimistic assumptions and limit the variation to the main cost component, capital cost. The analysis does not consider social costs, like the risk of accidents or waste management, since these externalities escape robust quantification. Neither does it consider limited public acceptance of nuclear power, renewables, or electricity and hydrogen grids.  

A crucial detail to understand these results is that apart from grid infrastructure, most investments into system flexibility occur outside of the electricity sector. Instead of battery storage or conventional demand-side management, new electricity demand induced by the decarbonization of heating, transport, and industry adds substantial flexibility. For instance, hydrogen production, accounting for 25\% of electricity demand in our model, supports renewable integration by adapting to renewable supply and using storage to match production with demand.

Results do not indicate that fluctuating renewables and flexibly operated nuclear plants are economic complements. Even when operational restrictions are omitted, nuclear power runs at capacity factors around 90\%. From a cost perspective, investing in system flexibility rather than running capital-intensive nuclear plants at low capacity factors is preferable.

Our analysis puts previous least-cost analyses on nuclear power limited to the electricity sector and a single country into perspective. As we showed, this setup omits two key options for system integration of fluctuating renewables: First, grid infrastructure connecting regions with different generation profiles, and second, the potential flexibility of electricity demand added to decarbonize the heating, transport, and industry sector. As a result, previous analyses overestimated the value of dispatchable nuclear power plants. Not considering other types of firm generation, like hydrogen or biogas-fueled plants, adds to this bias.

Our results suggest that efforts to advance nuclear technology---if undertaken at all---should focus on reducing costs and construction time. Placing too much effort and investments into nuclear power and shifting money from renewables could put European energy security at risk as nuclear plant projects are prone to cost increases and substantial delays that could lead to a lack of capacity. Concepts for more flexible operation are not promising because even fully flexible plants almost exclusively operate at full capacity for economic reasons. Since the share of nuclear also affects investments into grid infrastructure and energy storage, path dependencies in system development arise. Anticipating cost-competitive nuclear can result in a system that cannot integrate a large share of renewables. Therefore, the risk of nuclear power not achieving substantial cost reductions should thus be accounted for in long-term planning. This paper does not consider potential non-electrical applications for nuclear power, such as high-temperature provision for industrial processes, which should be addressed in future research.

Planning with an increased spatial resolution and multiple climatic years could further refine our results, though computational limits in state-of-the-art energy planning prevent even more detailed analyses. Finally, future research should investigate reactor concepts for the integrated generation of electricity and process heat or hydrogen as soon as meaningful cost and performance metrics are available.

\section*{Supplementary material}

All data and running scripts used in this paper are openly available here: \url{https://github.com/leonardgoeke/EuSysMod/tree/greenfield_nu} Due to their size, model outputs are not part of the repository but are available upon request.

The applied model uses the open AnyMOD.jl modeling framework \citep{Goeke2020a}. The applied version of the tool is openly available under this link. \url{https://github.com/leonardgoeke/AnyMOD.jl/releases/tag/flexibleElectrificationWorkingPaper}

\section*{Acknowledgments}

The research leading to these results has received funding from the Deutsche Forschungsgemeinschaft (DFG) under project number 423336886.

\section*{Declaration of interest}

The authors declare that they have no known competing financial interests or personal relationships that could have appeared to influence the work reported in this paper.

\section*{Credit Author Statements}

Leonard Göke: Conceptualization, Methodology, Validation, Visualization, Writing- Original draft preparation Alexander Wimmers: Methodology, Data curation, Writing- Reviewing and Editing Christian von Hirschhausen: Supervision, Funding acquisition

\printcredits

\bibliographystyle{unsrtnat}
\bibliography{cas-refs}

\appendix

\section{Review of other parameters} \label{cost_analysis}

Table \ref{tab:assumption_real_occ} show how future projections for overnight construction costs are substantially below recently reported costs. Table \ref{tab:OCC_results} compares the costs of different reactor types. Table \ref{tab:inflation} provides the factors for inflation adjustments normalized to 2018 following inflationtool.com.

\begin{figure}
    \centering
    \includegraphics[scale=0.1]{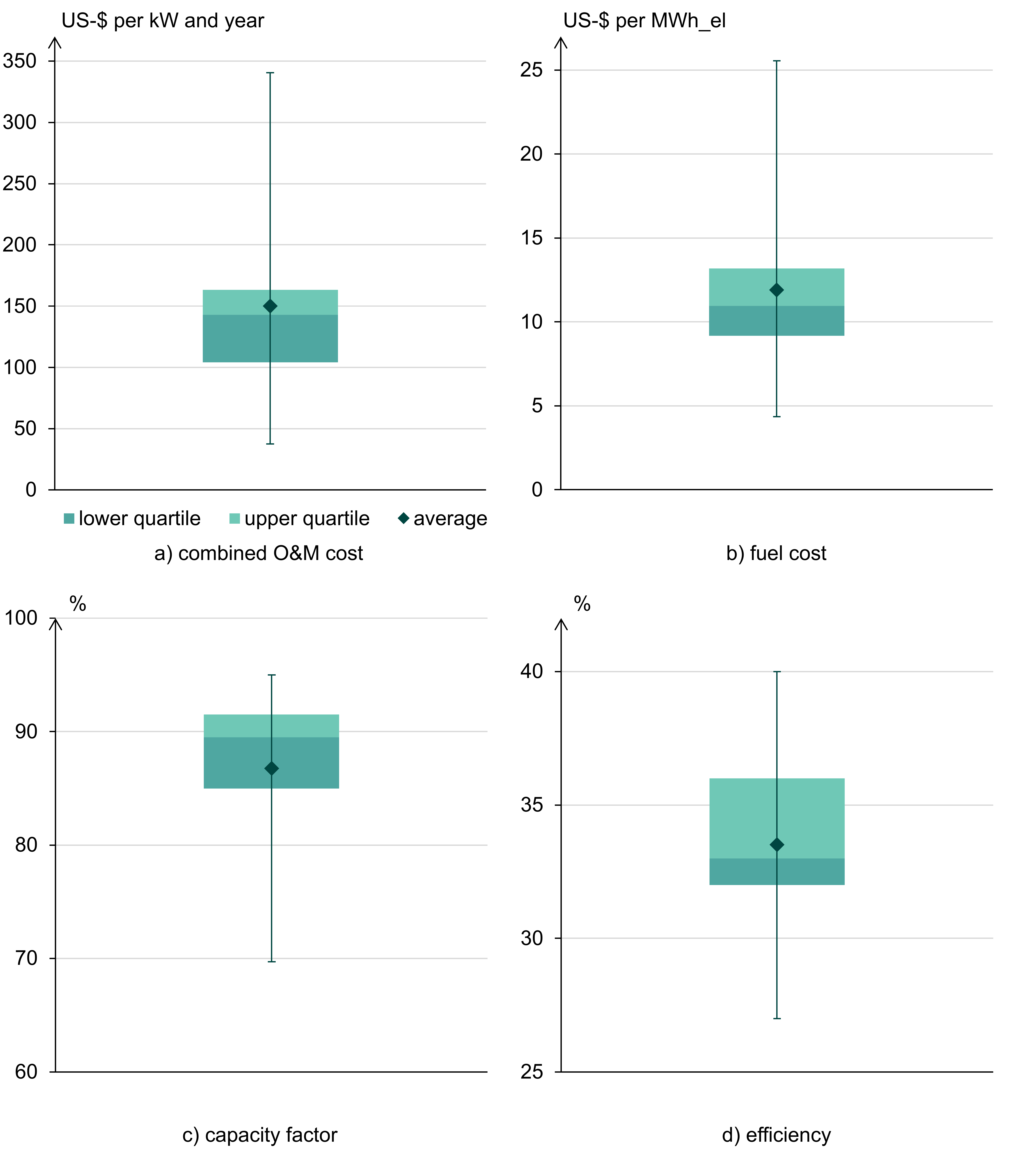}
    \caption{Boxplot diagram for for capacity factors, construction time, fuel cost, operational lifetime, efficiency and combined O\&M cost}
    \label{other_boxplot}
\end{figure}

Analysis of nuclear power in the paper builds on several techno-economic assumptions but only the most pivotal ones, construction costs and time, are discussed in depth in the paper. The following material provides a comprehensive overview of the entire reviewed material. First, Fig. \ref{other_boxplot} shows boxplots illustrating the distribution of values for different parameters. Afterwards, several tables list the underlying data for all parameters. These include overnight construction cost (see Tables \ref{tab:occ_unfiltered}, \ref{tab:occ_real_lwr_oecd}, \ref{tab:occ_assumed_lwr_oecd}), O\&M cost (Tables \ref{tab:opex_unfiltered}, \ref{tab:opex_combination_results}, \ref{tab:opex_combined}), nuclear fuel cost (see Table \ref{tab:fuel_cost}), capacity factor (see Table \ref{tab:capacity_factors}), construction time (see Table \ref{tab:construction_time}), operational lifetime (see Table \ref{tab:operational_lifetime}), and thermal efficiency (see Table \ref{tab:thermal efficiency}).

In literature, operation \& maintenance (O\&M) costs are often provided as two components. The first component, fixed O\&M costs, does not depend on yearly generation quantities, but the second component, variable O\&M costs, does and is given in US-\$/MWh. Some references however only provide one component, e.g., \citet{budi_fuel_2019, ingersoll_cost_2020, international_atomic_energy_agency_advances_2020}. For further details see Table \ref{tab:opex_unfiltered}. 

To make literature O\&M cost comparable, we combine fixed and variable costs to a single value in US-\$/kW. For this, we assume an optimistic 95\% capacity factor to determine nuclear full load hours and are thus able to determine combined O\&M costs for all references. Hereby, we also caclulate the average ratio of fixed and variable costs to then determine the average fixed and variable O\&M costs from the 25\%-quantile of the combined O\&M cost value, see Tables \ref{tab:opex_combination_results} and \ref{tab:opex_combined}. This results in the input values given in Table \ref{tab:assumptions}.

\begin{table}
\caption{Comparison of future projections and recently reported costs for LWR costs in US-\$ per kW.} \label{tab:assumption_real_occ}
\centering
\begin{tabular}{p{2cm}|p{2cm}|p{2cm}}
    \hline
    Measure         &Assumptions  &Real \\
    \hline
    Mean            &5,144.10   &9108,35   \\
    Median          &5,122.00   &9250.00   \\
    25\%-Quantile   &4,174.44   &7867,00   \\
    75\%-Quantile   &6,218.11   &10134,00   \\
    \hline
\end{tabular}
\end{table}

\begin{table}
\caption{Overnight construction costs by reactor type in US-\$ per kW.} \label{tab:OCC_results}
\centering
\begin{tabular}{p{2cm}|p{2cm}|p{2cm}|p{2cm}|p{2cm}}
    \hline
    Measure         &All types  &Standard LWR  &SMR    &Non-LWR\\
    \hline
    Mean            &6,279.41   &6,008.85   &7,774.17   &5,030.55\\
    Median          &5,353.64   &5,515.00   &6,270.29    &5,311.51\\
    25\%-Quantile   &4,319.80   &4,328.00   &4,472.66   &4,453.58\\
    75\%-Quantile   &7,146.76   &6,965.92   &7,978.75   &5,487.23\\
    \hline
\end{tabular}
\end{table}

\begin{table}[]
\centering
\caption{Inflation taken from inflationtool.com.} \label{tab:inflation}
\begin{tabular}{l|l}    
Year & Inflation until 2018 \\ \hline
1995          & 64.79\%                       \\
1996          & 60.09\%                       \\
1997          & 56.44\%                       \\
1998          & 54.05\%                       \\
1999          & 50.75\%                       \\
2000          & 45.82\%                       \\
2001          & 41.81\%                       \\
2002          & 39.60\%                       \\
2003          & 36.50\%                       \\
2004          & 32.94\%                       \\
2005          & 28.58\%                       \\
2006          & 24.56\%                       \\
2007          & 21.11\%                       \\
2008          & 16.63\%                       \\
2009          & 17.05\%                       \\
2010          & 15.16\%                       \\
2011          & 11.63\%                       \\
2012          & 9.37\%                        \\
2013          & 7.79\%                        \\
2014          & 6.07\%                        \\
2015          & 5.94\%                        \\
2016          & 4.62\%                        \\
2017          & 2.44\%                        \\
2018          & 0.00\%                        \\
2019          & -1.81\%                       \\
2020          & -3.07\%                       \\
2021          & -7.91\%                       \\
2022          & -17.61\%                     
\end{tabular}
\end{table}

\begin{landscape}
\footnotesize
    \begin{longtable} {p{2cm}|p{1cm}|p{2.5cm}|p{2.5cm}|p{1.3cm}|p{1.25cm}|p{1.25cm}|p{2cm}}
    \caption{Unfiltered overnight construction costs for nuclear plants.} \label{tab:occ_unfiltered}\\
    \hline
    Reference & Page & Cost Description & Reactor & Reference Year & Capacity in kW & OCC & Unit\\
    \hline
    \endfirsthead
    \caption[]{(continued)}\\
    \hline
    Reference & Page & Cost Description & Reactor & Reference Year & Capacity in kW & OCC & Unit\\
    \hline
    \endhead
    \hline
    \multicolumn{8}{l}{Continued on next page.}\\
    \hline
    \endfoot
    \hline
    \endlastfoot
    
        \citet{barkatullah_current_2017}& 132 & Asssumed parameters & n.a. & 2017 & n.a. & 4646 & US-\$(2018) per kWe \\ 
        \citet{boldon_small_2014}& 19 & Input Water-Cooled SMR & SMR & 2014 & 1260 & 3767 & US-\$(2011) per kW \\
        \citet{stein_advancing_2022}& 30 & Lower Cost, High Learning (Best Case) // FOAK CAPEX & Traditional Nuclear & 2020 & n.a. & 4783 & US-\$(2020) per kWe \\
        \citet{stein_advancing_2022}& 30 & Lower Cost, High Learning (Best Case) // FOAK CAPEX & SMR & 2020 & n.a. & 5108 & US-\$(2020) per kWe \\
        \citet{stein_advancing_2022}& 30 & Lower Cost, High Learning (Best Case) // FOAK CAPEX & HTGR & 2020 & n.a. & 5518 & US-\$(2020) per kWe \\
        \citet{stein_advancing_2022}& 30 & Lower Cost, High Learning (Best Case) // FOAK CAPEX & Advanced reactor with thermal storage (ARTES) & 2020 & n.a. & 4000 & US-\$(2020) per kWe \\
        \citet{stein_advancing_2022}& 30 & Upper Cost, Low Learning (Worst Case) // FOAK CAPEX & Traditional Nuclear & 2020 & n.a. & 6338 & US-\$(2020) per kWe \\
        \citet{stein_advancing_2022}& 30 & Upper Cost, Low Learning (Worst Case) // FOAK CAPEX & SMR & 2020 & n.a. & 6974 & US-\$(2020) per kWe \\
        \citet{stein_advancing_2022}& 30 & Upper Cost, Low Learning (Worst Case) // FOAK CAPEX & HTGR & 2020 & n.a. & 7500 & US-\$(2020) per kWe \\
        \citet{stein_advancing_2022}& 30 & Upper Cost, Low Learning (Worst Case) // FOAK CAPEX & Advanced reactor with thermal storage (ARTES) & 2020 & n.a. & 6220 & US-\$(2020) per kWe \\
        \citet{dixon_advanced_2017}& xii & Mean Costs & Thermal LWR Reactor & 2017 & n.a. & 4300 & US-\$(2017) per kW \\
        \citet{dixon_advanced_2017}& xii & Mean Costs & Fast Reactors & 2017 & n.a. & 4700 & US-\$(2017) per kW \\ 
        \citet{dixon_advanced_2017}& xii & Mean Costs & Gas-cooled reactors & 2017 & n.a. & 5170 & US-\$(2017) per kW \\
        \citet{dixon_advanced_2017}& xii & Mean Costs & PHWR & 2017 & n.a. & 4230 & US-\$(2017) per kW \\ 
        \citet{Duan2022}& 2 & Capital Costs & US nuclear & n.a. & n.a. & 4000 &US-\$ per kW \\ 
        \citet{Duan2022}& 2 & Capital Costs & US nuclear & n.a. & n.a. & 6317 &US-\$ per kW \\ 
        \citet{green_smr_2019}& 18 & HTGR & SMR & 2016 & 250 & 5000 &US-\$ per kW \\ 
        \citet{green_smr_2019}& 18 & CAREM25 & SMR & 2017 & 25 & 21900 &US-\$ per kW \\ \
        \citet{green_smr_2019}& 21 & Median (Low) & SMR & 2013 & 45 & 4000 &US-\$ per kW \\ 
        \citet{green_smr_2019}& 21 & Median (High) & SMR & 2013 & 45 & 16300 &US-\$ per kW \\ 
        \citet{green_smr_2019}& 21 & Median (Low) & SMR & 2013 & 225 & 3200 &US-\$ per kW \\ 
        \citet{green_smr_2019}& 21 & Median (High) & SMR & 2013 & 225 & 7100 &US-\$ per kW \\ 
        \citet{Grubler2010}& 5179 & Capital Costs France with no cost increase since 1998 & French PWR & 1998 & n.a. & 2600 & US-\$(2008) per kWe \\ 
        \citet{Grubler2010}& 5179 & Capital Costs France with no cost increase since 1998 & French PWR & 1998 & n.a. & 2100 & US-\$(2008) per kWe \\ 
        \citet{international_atomic_energy_agency_advances_2020}& 56 & SMART (FOAK) & SMR & 2020 & 30 & 10000 & US-\$ per kW \\ 
        \citet{iea_projected_2020}& 49 & Nuclear New Build - France & EPR & 2019 & 1650 & 4013 & US-\$ per kW \\ 
        \citet{iea_projected_2020}& 49 & Nuclear New Build - Japan & ALWR & 2019 & 1152 & 3963 & US-\$ per kW \\ 
        \citet{iea_projected_2020}& 49 & Nuclear New Build - Korea & ALWR & 2019 & 1377 & 2157 & US-\$ per kW \\ 
        \citet{iea_projected_2020}& 49 & Nuclear New Build - Russia & VVER & 2019 & 1122 & 2271 & US-\$ per kW \\ 
        \citet{iea_projected_2020}& 49 & Nuclear New Build - Slovak Republic & Other nuclear & 2019 & 1004 & 6920 & US-\$ per kW \\
        \citet{iea_projected_2020}& 49 & Nuclear New Build - United States & LWR & 2019 & 1100 & 4250 & US-\$ per kW \\
        \citet{iea_projected_2020}& 49 & Nuclear New Build - China & LWR & 2019 & 950 & 2500 & US-\$ per kW \\ 
        \citet{iea_projected_2020}& 49 & Nuclear New Build - India & LWR & 2019 & 950 & 2778 & US-\$ per kW \\ 
        \citet{lazard2021}& 18 & Nuclear (New Build, Low Case) & n.a. & 2021 & 2200 & 7800 & US-\$(2022) per kWe \\ 
        \citet{lazard2021}& 18 & Nuclear (New Build, High Case) & n.a. & 2021 & 2200 & 12800 & US-\$(2022) per kWe \\ 
        \citet{mit_future_2018}& 236 & Overnight Costs (Standard LWR) & LWR & 2018 & n.a. & 5000 & US-\$(2017) per kWe \\ 
        \citet{mit_future_2018}& 70 & Total Overnight Costs & Molten Salt Reactor & 2018 & 2275 (MWth) & 5400 & US-\$(2017) per kW \\
        \citet{mit_future_2018}& 236 & FOAK CAPEX & HTGR & 2017 & n.a. & 5200 & US-\$(2017) per kW \\ 
        \citet{mit_future_2018}& 159, 220 & Benchmark Scenario Cost & Nuclear & 2018 & n.a. & 4900 & US-\$(2017) per kW \\
        \citet{mit_future_2018}& 220 & Barakah OCC & APR1400 & 2010 & n.a. & 4000 & US-\$(2017) per kW \\ 
        \citet{mit_future_2018}& 149 & US (high) & Nuclear & 2018 & 1000 & 6880 & US-\$(2017) per kW \\ 
        \citet{mit_future_2018}& 149 & US (nominal) & Nuclear & 2018 & 1000 & 5500 & US-\$(2017) per kW \\ 
        \citet{mit_future_2018}& 149 & US (low) & Nuclear & 2018 & 1000 & 4100 & US-\$(2017) per kW \\
        \citet{mit_future_2018}& 149 & US (very low) & Nuclear & 2018 & 1000 & 2750 & US-\$(2017) per kW \\ 
        \citet{mit_future_2018}& 150 & China (nominal) & Nuclear & 2018 & 1000 & 2800 & US-\$(2017) per kW \\ 
        \citet{mit_future_2018}& 150 & China (low) & Nuclear & 2018 & 1000 & 2080 & US-\$(2017) per kW \\ 
        \citet{mit_future_2018}& 151 & France (nominal) & Nuclear & 2018 & 1000 & 6800 & US-\$(2017) per kW \\ 
        \citet{mit_future_2018}& 151 & France (low) & Nuclear & 2018 & 1000 & 5070 & US-\$(2017) per kW \\ 
        \citet{mit_future_2018}& 150 & UK (nominal) & Nuclear & 2018 & 1000 & 8140 & US-\$(2017) per kW \\ 
        \citet{mit_future_2018}& 150 & UK (low) & Nuclear & 2018 & 1000 & 6070 & US-\$(2017) per kW \\ 
        \citet{national_nuclear_laboratory_smr_2016}& 46 & LCOE build-up for GWe LWR & LWR & n.a. & n.a. & 26.99 & US-\$ per MWh \\ 
        \citet{nrel_2021_2021}&Xls Sheet "Nuclear" & Nuclear-Moderate & Nuclear & 2019 & n.a. & 6297 & US-\$(2019) per kWe \\ 
        \citet{oecd_full_2018}& 68 & Nuclear (min) & n.a. & 2020 & 535 & 1807 & US-\$(2015) per kWe \\ 
        \citet{oecd_full_2018}& 68 & Nuclear (mean) & n.a. & 2020 & 1343 & 4249 & US-\$(2015) per kWe \\ 
        \citet{oecd_full_2018}& 68 & Nuclear (median) & n.a. & 2020 & 1300 & 4896 & US-\$(2015) per kWe \\ 
        \citet{oecd_full_2018}& 68 & Nuclear (max) & n.a. & 2020 & 3300 & 6215 & US-\$(2015) per kWe \\         
        \citet{rothwell_economics_2016}& 106 & FOAK, no sec & SMR & 2013 & 360 & 7343 &US-\$ per kW \\ 
        \citet{rothwell_economics_2016}& 106 & NOAK, no sec & SMR & 2013 & 360 & 5357 &US-\$ per kW \\ 
        \citet{rothwell_economics_2016}& 106 & FOAK, sec = 0.1\% & SMR & 2013 & 360 & 6722 &US-\$ per kW \\ 
        \citet{rothwell_economics_2016}& 106 & NOAK, sec = 0.1\% & SMR & 2013 & 360 & 4885 &US-\$ per kW \\ 
        \citet{Rothwell2022}& 2 & Taishin 1 \& 2, China & EPR & 2019 & 1600 & 3200 & US-\$(2018) per kWe \\ 
        \citet{Rothwell2022}& 2 & Sanmen, China & AP1000 & 2018 & 1150 & 3270 & US-\$(2018) per kWe \\ 
        \citet{Rothwell2022}& 6 & NuScale NOAK & SMR & 2022 & n.a. & 3600 & US-\$(2018) per kWe \\ 
        \citet{Rothwell2022}& 4 & Belgium & EPR & 2015 & 1600 & 5515 & US-\$(2018) per kWe \\ 
        \citet{Rothwell2022}& 4 & UK & EPR & 2015 & 1600 & 6588 & US-\$(2018) per kWe \\ 
        \citet{Rothwell2022}& 4 & Hungary & VVER1200 & 2015 & 1200 & 6745 & US-\$(2018) per kWe \\ 
        \citet{Rothwell2022}& 2 & Olkiluoto, Finland & EPR & 2020 & 1600 & 7600 & US-\$(2018) per kWe \\
        \citet{Rothwell2022}& 6 & NuScale FOAK & SMR & 2022 & n.a. & 8000 & US-\$(2018) per kWe \\ 
        \citet{Rothwell2022}& 2 & Hinkley Point, Uk & EPR & 2019 & 1600 & 9300 & US-\$(2018) per kWe \\ 
        \citet{Rothwell2022}& 2 & Vogtle Station, US & AP1000 & 2021 & 1150 & 12000 & US-\$(2018) per kWe \\ 
        \citet{Rothwell2022}& 6 & TerraPower FOAK & SMR & 2022 & n.a. & 12000 & US-\$(2018) per kWe \\
        \citet{Rothwell2022}& 2 & Flamanville, France & EPR & 2020 & 1600 & 12600 & US-\$(2018) per kWe \\ 
        \citet{shirvan_overnight_2022}& 7 & 10th unit AP1000 (Should cost) & AP1000 & 2019 & 1150 & 2900 & US-\$(2018) per kWe \\ 
        \citet{shirvan_overnight_2022}& 7 & Next AP1000 (Should cost) & AP1000 & 2019 & 1150 & 4300 & US-\$(2018) per kWe \\
        \citet{shirvan_overnight_2022}& 7 & 10th unit AP1000 (Post-COVID high estimate) & AP1000 & 2022 & 1150 & 4500 & US-\$(2022) per kWe \\ 
        \citet{shirvan_overnight_2022}& 7 & Historic (Pre-Three-Mile-Island) & PWR & 1979 & 1200 & 4700 & US-\$(2018) per kWe \\
        \citet{shirvan_overnight_2022}& 7 & Next AP1000 (Post-COVID high estimate) & AP1000 & 2022 & 1150 & 6800 & US-\$(2022) per kWe \\ 
        \citet{shirvan_overnight_2022}& 7 & Vogtle Station, US (Official) & AP1000 & 2021 & 1150 & 7956 & US-\$(2018) per kWe \\
        \citet{shirvan_overnight_2022}& 7 & Vogtle Station, US (Independent) & AP1000 & 2022 & 1150 & 9200 & US-\$(2018) per kWe \\ 
        \citet{shirvan_overnight_2022}& 7 & Historic (Post-Three-Mile-Island) & PWR & >1979 & 1200 & 9512 & US-\$(2018) per kWe \\
        \citet{stewart_capital_2021}& 13 & FOAK (US) & Large Passive Safety PWR & 2018 & n.a. & 4328 & US-\$(2018) per kWe \\ 
        \citet{stewart_capital_2021}& 13 & FOAK (China) (min) & Large Passive Safety PWR & 2018 & n.a. & 1700 & US-\$(2018) per kWe \\ 
        \citet{stewart_capital_2021}& 13 & FOAK (China) (max) & Large Passive Safety PWR & 2018 & n.a. & 2200 & US-\$(2018) per kWe \\ 
        \citet{stewart_capital_2021}& 13 & First-4 unit average & Large Active Safety PWR & 2018 & n.a. & 5337 & US-\$(2018) per kWe \\ 
        \citet{stewart_capital_2021}& 13 & APR1400 4-unit average - Barakah & APR1400 & 2018 & n.a. & 4358 & US-\$(2018) per kWe \\ 
        \citet{stewart_capital_2021}& 13 & 10th of a kind & SMR & 2018 & n.a. & 3856 & US-\$(2018) per kWe \\
        \citet{tolley_economic_2004}& 7 & No-Policy Nuclear LCOE Installations (Max OCC) & Advanced Nuclear Reactor & 2003 & 1000 & 1800 & US-\$(2003) per kW \\ 
        \citet{wealer_investing_2021}& 5 & Inputs for MC & n.a. & 2018 & 1600 & 4000-9000 & US-\$(2018) per kWe \\\hline
    \end{longtable}
    \normalsize
\end{landscape}

\begin{table}[!ht]
    \caption{Overnight construction costs reported for LWR in OECD countries.} \label{tab:occ_real_lwr_oecd}
    \centering
    \begin{tabular}{l|l|l}
    \hline
        Reference & Cost Description & Cost in US-\$(2018) per kW \\ \hline
        \citet{lazard2021}& Nuclear (New Build, High Case) & 10883.43 \\ 
        \citet{lazard2021}& Nuclear (New Build, Low Case) & 6632.09 \\
        \citet{Rothwell2022}& Flamanville, France & 12600 \\ 
        \citet{Rothwell2022}& Hinkley Point, Uk & 9300 \\
        \citet{Rothwell2022}& Olkiluoto, Finland & 7600 \\ 
        \citet{Rothwell2022}& Vogtle Station, US & 12000 \\ 
        \citet{shirvan_overnight_2022}& Historic (Post-Three-Mile-Island) & 9512 \\ 
        \citet{shirvan_overnight_2022}& Historic (Pre-Three-Mile-Island) & 4700 \\ 
        \citet{shirvan_overnight_2022}& Vogtle Station, US (Independent) & 9200 \\ 
        \citet{shirvan_overnight_2022}& Vogtle Station, US (Official) & 7956 \\ \hline
    \end{tabular}
    
\end{table}

\footnotesize
\begin{longtable} {p{3cm}|p{4cm}|p{4cm}}
    \caption{Overnight construction costs projected for LWR in OECD countries.} \label{tab:occ_assumed_lwr_oecd}\\
    \hline
    Reference & Cost Description & Cost in US-\$(2018) per kW \\ \hline
    \hline
    \endfirsthead
    \caption[]{(continued)}\\
    \hline
    Reference & Cost Description & Cost in US-\$(2018) per kW \\ \hline
    \hline
    \endhead
    
    \multicolumn{3}{l}{Continued on next page.}\\
    \hline
    \endfoot
    \hline
    \endlastfoot
        \citet{barkatullah_current_2017}& Asssumed parameters & 4646 \\ 
        \citet{stein_advancing_2022}& Lower Cost, High Learning (Best Case) // FOAK CAPEX & 4640.54 \\ 
        \citet{stein_advancing_2022}& Upper Cost, Low Learning (Worst Case) // FOAK CAPEX & 6149.22 \\ 
        \citet{dixon_advanced_2017}& Mean Costs & 4404.92 \\ 
        \citet{Duan2022}& Capital Costs 1 & 4000 \\ 
        \citet{Duan2022}& Capital Costs 2 & 6317 \\ 
        \citet{iea_projected_2020}& Nuclear New Build - France & 3941.66 \\ 
        \citet{iea_projected_2020}& Nuclear New Build - Japan & 3892.54 \\ 
        \citet{iea_projected_2020}& Nuclear New Build - Korea & 2118.65 \\ 
        \citet{iea_projected_2020}& Nuclear New Build - Slovak Republic & 6796.97 \\ 
        \citet{iea_projected_2020}& Nuclear New Build - United States & 4174.44 \\ 
        \citet{mit_future_2018}& Benchmark Scenario Cost & 5019.56 \\
        \citet{mit_future_2018}& France (low) & 5193.71 \\ 
        \citet{mit_future_2018}& France (nominal) & 6965.92 \\ 
        \citet{mit_future_2018}& Overnight Costs (Standard LWR) & 5122 \\ 
        \citet{mit_future_2018}& UK (low) & 6218.11 \\ 
        \citet{mit_future_2018}& UK (nominal) & 8338.62 \\ 
        \citet{mit_future_2018}& US (high) & 7047.87 \\
        \citet{mit_future_2018}& US (low) & 4200.04 \\ 
        \citet{mit_future_2018}& US (nominal) & 5634.2 \\
        \citet{mit_future_2018}& US (very low) & 2817.1 \\
        \citet{nrel_2021_2021}& Nuclear-Moderate & 6185.05 \\ 
        \citet{oecd_full_2018}& Nuclear (max) & 6584.17 \\ 
        \citet{oecd_full_2018}& Nuclear (mean) & 4501.39 \\ 
        \citet{oecd_full_2018}& Nuclear (median) & 5186.82 \\ 
        \citet{oecd_full_2018}& Nuclear (min) & 1914.34 \\ 
        \citet{Rothwell2022}& Belgium & 5515 \\ 
        \citet{Rothwell2022}& Hungary & 6745 \\ 
        \citet{Rothwell2022}& UK & 6588 \\ 
        \citet{shirvan_overnight_2022}& 10th unit AP1000 (Post-COVID high estimate) & 3826.21 \\ 
        \citet{shirvan_overnight_2022}& 10th unit AP1000 (Should cost) & 2900 \\ 
        \citet{shirvan_overnight_2022}& Next AP1000 (Post-COVID high estimate) & 5781.82 \\
        \citet{shirvan_overnight_2022}& Next AP1000 (Should cost) & 4300 \\ 
        \citet{stewart_capital_2021}& First-4 unit average & 5337 \\ 
        \citet{stewart_capital_2021}& FOAK (US) & 4328 \\ 
        \citet{wealer_investing_2021}& Inputs for MC MAX & 9000 \\ 
        \citet{wealer_investing_2021}& Inputs for MC MIN & 4000 \\
\end{longtable}
\normalsize

\begin{landscape}
\footnotesize
    \begin{longtable} {p{2cm}|p{1cm}|p{2.5cm}|p{2.5cm}|p{1.3cm}|p{1.25cm}|p{1cm}|p{1cm}|p{1cm}|p{1cm}}
    \caption{Unfiltered fixed and variable O\&M cost data for nuclear plants.} \label{tab:opex_unfiltered}\\
    \hline
    Reference & Page & Cost Description & Reactor & Reference Year & Capacity in kW & Fixed O\&M Cost & Unit Fixed Cost & Variable O\&M Cost & Unit Var. Cost \\ \hline
    \hline
    \endfirsthead
    \caption[]{(continued)}\\
    \hline
    Reference & Page & Cost Description & Reactor & Reference Year & Capacity in kW & Fixed O\&M Cost & Unit Fixed Cost & Variable O\&M Cost & Unit Var. Cost \\ \hline
    \endhead
    \hline
    \multicolumn{10}{l}{Continued on next page.}\\
    \hline
    \endfoot
    \hline
    \endlastfoot
    
        \citet{barkatullah_current_2017}& 132 & Asssumed parameters & n.a. & 2017 & ~ & 93.28 & US-\$(2018) per kW & 2.14 & US-\$ per MWh \\ 
        \citet{boldon_small_2014}& 19 & Input Water-Cooled SMR & SMR & 2014 & 1260 & n.a. & n.a. & 18 & US-\$(2014) per MWh \\ 
        \citet{stein_advancing_2022}& 30 & Lower Cost, High Learning (Best Case) // FOAK CAPEX & Traditional Nuclear & 2020 & n.a. & 121 & US-\$(2020) per kW-yr & 2.43 & US-\$(2020) per MWh \\ 
        \citet{stein_advancing_2022}& 30 & Lower Cost, High Learning (Best Case) // FOAK CAPEX & SMR & 2020 & n.a. & 98 & US-\$(2020) per kW-yr & 3.08 & US-\$(2020) per MWh \\ 
        \citet{stein_advancing_2022}& 30 & Lower Cost, High Learning (Best Case) // FOAK CAPEX & HTGR & 2020 & n.a. & 39 & US-\$(2020) per kW-yr & 0.35 & US-\$(2020) per MWh \\ 
        \citet{stein_advancing_2022}& 30 & Lower Cost, High Learning (Best Case) // FOAK CAPEX & Advanced reactor with thermal storage (ARTES) & 2020 & n.a. & 121 & US-\$(2020) per kW-yr & 2.43 & US-\$(2020) per MWh \\ 
        \citet{stein_advancing_2022}& 30 & Upper Cost, Low Learning (Worst Case) // FOAK CAPEX & Traditional Nuclear & 2020 & n.a. & 94 & US-\$(2020) per kW-yr & 7.31 & US-\$(2020) per MWh \\ 
        \citet{stein_advancing_2022}& 30 & Upper Cost, Low Learning (Worst Case) // FOAK CAPEX & SMR & 2020 & n.a. & 98 & US-\$(2020) per kW-yr & 3.08 & US-\$(2020) per MWh \\ 
        \citet{stein_advancing_2022}& 30 & Upper Cost, Low Learning (Worst Case) // FOAK CAPEX & HTGR & 2020 & n.a. & 189 & US-\$(2020) per kW-yr & 0 & US-\$(2020) per MWh \\ 
        \citet{stein_advancing_2022}& 30 & Upper Cost, Low Learning (Worst Case) // FOAK CAPEX & Advanced reactor with thermal storage (ARTES) & 2020 & n.a. & 40 & US-\$(2020) per kW-yr & 5 & US-\$(2020) per MWh \\ 
        \citet{budi_fuel_2019}& 3 & Annual O\&M costs & HTGR & n.a & 600 & 59780000 & US-\$(2012) & n.a. & n.a. \\ 
        \citet{budi_fuel_2019}& 3 & Annual O\&M costs & HTGR & n.a. & 350 & 56630000 & US-\$(2012) & n.a. & n.a. \\ 
        \citet{dixon_advanced_2017}& xii & Mean Costs & Thermal LWR Reactor & 2017 & n.a. & 72 & US-\$(2017) per kW & 2 & US-\$(2017) per kW \\ 
        \citet{dixon_advanced_2017}& xii & Mean Costs & Fast Reactors & 2017 & n.a. & 78 & US-\$(2017) per kW & 2.1 & US-\$(2017) per kW \\ 
        \citet{dixon_advanced_2017}& xii & Mean Costs & Gas-cooled reactors & 2017 & n.a. & 0 & US-\$(2017) per kW & 0 & US-\$(2017) per kW \\ 
        \citet{dixon_advanced_2017}& xii & Mean Costs & PHWR & 2017 & n.a. & 73 & US-\$(2017) per kW & 1.8 & US-\$(2017) per kW \\ 
        \citet{international_atomic_energy_agency_advances_2020}& 56 & SMART (FOAK) & SMR & 2020 & 30 & n.a. & n.a. & 28 &US-\$ per MWh \\ 
        \citet{iea_projected_2020}& 49 & Nuclear New Build - France & EPR & 2019 & 1650 & n.a. & n.a. & 14.26 & US-\$ per MWh \\ 
        \citet{iea_projected_2020}& 49 & Nuclear New Build - Japan & ALWR & 2019 & 1152 & n.a. & n.a. & 25.84 & US-\$ per MWh \\ 
        \citet{iea_projected_2020}& 49 & Nuclear New Build - Korea & ALWR & 2019 & 1377 & n.a. & n.a. & 18.44 & US-\$ per MWh \\ 
        \citet{iea_projected_2020}& 49 & Nuclear New Build - Russia & VVER & 2019 & 1122 & n.a. & n.a. & 10.15 & US-\$ per MWh \\ 
        \citet{iea_projected_2020}& 49 & Nuclear New Build - Slovak Republic & Other nuclear & 2019 & 1004 & n.a. & n.a. & 9.72 & US-\$ per MWh \\ 
        \citet{iea_projected_2020}& 49 & Nuclear New Build - United States & LWR & 2019 & 1100 & n.a. & n.a. & 11.6 & US-\$ per MWh \\ 
        \citet{iea_projected_2020}& 49 & Nuclear New Build - China & LWR & 2019 & 950 & n.a. & n.a. & 26.42 & US-\$ per MWh \\ 
        \citet{iea_projected_2020}& 49 & Nuclear New Build - India & LWR & 2019 & 950 & n.a. & n.a. & 23.84 & US-\$ per MWh \\ 
        \citet{ingersoll_cost_2020}& 29 & Model assumptions & Advanced Nuclear Reactor & 2020 & n.a. & 31 & US-\$(2019) per kWe-yr & n.a. & n.a. \\ 
        \citet{lazard2021}& 18 & Nuclear (New Build, Low Case) & n.a. & 2021 & 2200 & 121 & US-\$(2021) per kW-yr & 4 & US-\$ per MWh \\ 
        \citet{lazard2021}& 18 & Nuclear (New Build, High Case) & n.a. & 2021 & 2200 & 140.5 & US-\$(2021) per kW-yr & 4.5 & US-\$ per MWh \\ 
        \citet{lazard2021}& 18 & Nuclear (Operating, Low Case) & n.a. & 2021 & 2200 & 83.5 & US-\$(2021) per kW-yr & 2.6 & US-\$ per MWh \\ 
        \citet{lazard2021}& 18 & Nuclear (Operating, High Case) & n.a. & 2021 & 2200 & 119.3 & US-\$(2021) per kW-yr & 4.2 & US-\$ per MWh \\ 
        \citet{mit_future_2018}& 159, 220 & Benchmark Scenario Cost & Nuclear & 2018 & n.a. & 95000 & US-\$(2017) per MW-yr & 6.89 & US-\$(2017) per MWh \\ 
        \citet{mit_future_2018}& 149 & US (low) & Nuclear & 2018 & 1000 & 95 & US-\$(2017) per MW-yr & 6.89 & US-\$(2017) per MWh \\ 
        \citet{mit_future_2018}& 150 & China (nominal) & Nuclear & 2018 & 1000 & 59.677 & US-\$(2017) per MW-yr & 4.33 & US-\$(2017) per MWh \\ 
        \citet{mit_future_2018}& 151 & France (nominal) & Nuclear & 2018 & 1000 & 115.123 & US-\$(2017) per MW-yr & 8.35 & US-\$(2017) per MWh \\ 
        \citet{mit_future_2018}& 150 & UK (nominal) & Nuclear & 2018 & 1000 & 180.759 & US-\$(2017) per MW-yr & 13.11 & US-\$(2017) per MWh \\ 
        \citet{national_nuclear_laboratory_smr_2016}& 46 & LCOE build-up for GWe LWR & LWR & n.a. & n.a. & n.a. & n.a. & 13.55 & US-\$ per MWh \\ 
        \citet{nuclear_energy_institute_nuclear_2021}& 6 & Operating Costs & US Nuclear & 2020 & n.a. & n.a. & n.a. & 18.27 &US-\$ per MWh \\ 
        \citet{nrel_2021_2021}& Nuclear-Moderate & Nuclear & 2019 & n.a. & 144.97 & US-\$(2019) per kW-yr & 2.35 & US-\$(2019) per MWh \\ 
        \citet{timilsina_demystifying_2020}& 12 & Nuclear & n.a. & 2019 & n.a. & n.a. & n.a. & 4.6 & US-\$(2019) per MWh \\ 
        \citet{tolley_economic_2004}& 7 & No-Policy Nuclear LCOE Installations (Max OCC) & Advanced Nuclear Reactor & 2003 & 1000 & 60 & US-\$(2003) per kW & 2.1 & US-\$(2003) per MWh \\ 
        \citet{wealer_investing_2021}& 5 & Inputs for MC & n.a. & 2018 & 1600 & 93280 & US-\$(2018) per MW & 2.14 & US-\$ per MWh \\ 
    \end{longtable}
    \normalsize
\end{landscape}

\begin{table}[!ht]
    \caption{Parameters and Interim Calculation Steps to Determine O\&M Input Parameters}\label{tab:opex_combination_results}
    \centering
    \begin{tabular}{l|c|c}
    \hline
        Parameter & Value & Unit \\ \hline
        Hours per Year & 8760 & Hours \\ 
        Capacity Factor & 0.95 & \% \\ 
        Full Load Hours & 8322 & Hours \\ 
        Median Ratio Fix/Var & 2.5 & ~ \\ 
        25\% Percentile of Combined O\&M Cost & 97.43 & US-\$(2018)/kW \\ 
        25\%-Quantile Median Ratio Fixed O\&M Cost & 69.59 & US-\$(2018)/kW \\ 
        25\%-Quantile Median Ratio Variable O\&M Cost & 3.35 & US-\$(2018)/MWh \\ \hline
    \end{tabular}
\end{table}

\begin{landscape}
\footnotesize
    \begin{longtable} {p{2cm}|p{2cm}|p{1.25cm}|p{1.2cm}|p{1.25cm}|p{1.2cm}|p{1.1cm}|p{1.1cm}|p{1.1cm}|p{0.75cm}}
    \caption{Fixed and variable O\&M costs for nuclear plants in OECD countries.} \label{tab:opex_combined}\\
    \hline
    Reference & Cost Description & Fix O\&M Cost & Unit\_Fix & Var. O\&M Cost & Unit\_Var & Fix O\&M Cost in US-\$(2018) per kW & Var. O\&M Cost in US-\$(2018) per MWh & Comb. O\&M Cost in US-\$(2018) per kW & Ratio Fix/Var \\ \hline
    \hline
    \endfirsthead
    \caption[]{(continued)}\\
    \hline
    Reference & Cost Description & Fix O\&M Cost & Unit\_Fix & Var. O\&M Cost & Unit\_Var & Fix O\&M Cost in US-\$(2018) per kW & Var. O\&M Cost in US-\$(2018) per MWh & Comb. O\&M Cost in US-\$(2018) per kW & Ratio Fix/Var \\ \hline
    \endhead
    \hline
    \multicolumn{10}{l}{Continued on next page.}\\
    \hline
    \endfoot
    \hline
    \endlastfoot
        
        \citet{barkatullah_current_2017}& Asssumed parameters & 93.28 & US-\$(2018) per kW & 2.14 & US-\$ per MWh & 93.28 & 2.14 & 111.09 & 5 \\ 
        \citet{budi_fuel_2019}& Annual O\&M costs & 99.63333333 & US-\$(2012) & n.a. & n.a. & 108.97 & 0.00 & 108.97 & 0 \\ 
        \citet{budi_fuel_2019}& Annual O\&M costs & 94.38333333 & US-\$(2012) & n.a. & n.a. & 103.23 & 0.00 & 103.23 & 0 \\ 
        \citet{dixon_advanced_2017}& Mean Costs & 78 & US-\$(2017) per kW & 2.1 & US-\$(2017) per kW & 79.90 & 2.15 & 97.81 & 4 \\ 
        \citet{dixon_advanced_2017}& Mean Costs & 72 & US-\$(2017) per kW & 2 & US-\$(2017) per kW & 73.76 & 2.05 & 90.81 & 4 \\ 
        \citet{dixon_advanced_2017}& Mean Costs & 73 & US-\$(2017) per kW & 1.8 & US-\$(2017) per kW & 74.78 & 1.84 & 90.13 & 5 \\ 
        \citet{iea_projected_2020}& Nuclear New Build - Japan & n.a. & n.a. & 25.84 & US-\$ per MWh & 0.00 & 25.38 & 211.22 & 0 \\ 
        \citet{iea_projected_2020}& Nuclear New Build - Korea & n.a. & n.a. & 18.44 & US-\$ per MWh & 0.00 & 18.11 & 150.73 & 0 \\ 
        \citet{iea_projected_2020}& Nuclear New Build - France & n.a. & n.a. & 14.26 & US-\$ per MWh & 0.00 & 14.01 & 116.56 & 0 \\ 
        \citet{iea_projected_2020}& Nuclear New Build - United States & ~ & ~ & 11.6 & US-\$ per MWh & 0.00 & 11.39 & 94.82 & 0 \\ 
        \citet{iea_projected_2020}& Nuclear New Build - Slovak Republic & ~ & ~ & 9.72 & US-\$ per MWh & 0.00 & 9.55 & 79.45 & 0 \\ 
        \citet{ingersoll_cost_2020}& Model assumptions & 31 & US-\$(2019) per kW & n.a. & n.a. & 30.45 & 0.00 & 30.45 & 0 \\ 
        \citet{lazard2021}& Nuclear (New Build, High Case) & 140.5 & US-\$(2021) per kW-yr & 4.5 & US-\$ per MWh & 130.20 & 4.17 & 164.91 & 4\\
        \citet{lazard2021}& Nuclear (Operating, High Case) & 119.3 & US-\$(2021) per kW-yr & 4.2 & US-\$ per MWh & 110.56 & 3.89 & 142.95 & 3 \\ 
        \citet{lazard2021}& Nuclear (New Build, Low Case) & 121 & US-\$(2021) per kW-yr & 4 & US-\$ per MWh & 112.13 & 3.71 & 142.98 & 4 \\ 
        \citet{lazard2021}& Nuclear (Operating, Low Case) & 83.5 & US-\$(2021) per kW-yr & 2.6 & US-\$ per MWh & 77.38 & 2.41 & 97.43 & 4 \\ 
        \citet{mit_future_2018}& UK (nominal) & 180.759 & US-\$(2017) per kW-yr & 13.11 & US-\$(2017) per MWh & 185.17 & 13.43 & 296.93 & 2 \\ 
        \citet{mit_future_2018}& France (nominal) & 115.123 & US-\$(2017) per kW-yr & 8.35 & US-\$(2017) per MWh & 117.93 & 8.55 & 189.12 & 2 \\ 
        \citet{mit_future_2018}& Benchmark Scenario Cost & 95 & US-\$(2017) per MWh & 6.89 & US-\$(2017) per MWh & 97.32 & 7.06 & 156.06 & 2 \\ 
        \citet{mit_future_2018}& US (low) & 95 & US-\$(2017) per kW-yr & 6.89 & US-\$(2017) per MWh & 97.32 & 7.06 & 156.06 & 2 \\ 
        \citet{national_nuclear_laboratory_smr_2016}& LCOE build-up for GWe LWR & 212.79 & US-\$(2016) per kW & 13.55 & US-\$(2016) per MWh & 222.62 & 14.18 & 340.59 & 2 \\ 
        \citet{nrel_2021_2021}& Nuclear-Moderate & 144.97 & US-\$(2019) per kW-yr & 2.35 & US-\$(2019) per MWh & 142.39 & 2.31 & 161.6 & 7 \\ 
        \citet{timilsina_demystifying_2020}& Nuclear & n.a. & n.a. & 4.6 & US-\$(2019) per MWh & 0.00 & 4.52 & 37.6 & 0 \\ 
        \citet{tolley_economic_2004}& No-Policy Nuclear LCOE Installations (Max OCC) & 60 & US-\$(2003) per kW & 2.1 & US-\$(2003) per MWh & 81.90 & 2.87 & 105.76 & 3 \\ 
        \citet{wealer_investing_2021}& Inputs for MC & 93.28 & US-\$(2018) per kW & 2.14 & US-\$ per MWh & 93.28 & 2.14 & 111.09 & 5 \\ \hline
    \end{longtable}
    \normalsize
\end{landscape}

\begin{landscape}
\footnotesize
    \begin{longtable} {p{2cm}|p{1cm}|p{2cm}|p{1.25cm}|p{1.2cm}|p{1cm}|p{1cm}|p{1.2cm}|p{1.2cm}}
    \caption{Fuel costs for nuclear plants.} \label{tab:fuel_cost}\\
    \hline
    Reference & Page & Description & Fuel Type & Reactor & Reference Year & Value & Unit & US-\$(2018) per MWh \\ \hline
    \hline
    \endfirsthead
    \caption[]{(continued)}\\
    \hline
    Reference & Page & Description & Fuel Type & Reactor & Reference Year & Value & Unit & US-\$(2018) per MWh \\ \hline
    \endhead
    \hline
    \multicolumn{9}{l}{Continued on next page.}\\
    \hline
    \endfoot
    \hline
    \endlastfoot
 
        \citet{barkatullah_current_2017}& 132 & Inputs for MC & n.a. & n.a. & 2018 & 10.11 & US-\$ per MWe*h & 10.11 \\ 
        \citet{boldon_small_2014}& 19 & Fuel Cost & n.a. & n.a. & 2014 & 9.1 & US-\$ per MWe*h & 9.65 \\ 
        \citet{iea_projected_2020}& 39 & Assumptions for LCOE & ~ & Nuclear & 2019 & 9.33 & US-\$ per MWe*h & 9.16 \\ 
        \citet{iea_projected_2020}& 39 & Assumptions for LCOE (Japan) & ~ & Nuclear & 2019 & 13.9 & US-\$ per MWe*h & 13.65 \\ 
        \citet{iea_projected_2020}& 59 & Nuclear New Build - France & ~ & EPR & 2019 & 9.33 & US-\$ per MWe*h & 9.16 \\ 
        \citet{iea_projected_2020}& 59 & Nuclear New Build - Japan & ~ & ALWR & 2019 & 13.92 & US-\$ per MWe*h & 13.67 \\ 
        \citet{iea_projected_2020}& 59 & Nuclear New Build - Korea & ~ & ALWR & 2019 & 9.33 & US-\$ per MWe*h & 9.16 \\ 
        \citet{iea_projected_2020}& 59 & Nuclear New Build - Slovak Republic & ~ & Other nuclear & 2019 & 9.33 & US-\$ per MWe*h & 9.16 \\ 
        \citet{iea_projected_2020}& 59 & Nuclear New Build - United States & ~ & LWR & 2019 & 9.33 & US-\$ per MWe*h & 9.16 \\ 
        \citet{ingersoll_cost_2020}& 29 & Model assumptions & Advanced Nuclear Reactor & ~ & 2019 & 4.44 & US-\$ per MWe*h & 4.36 \\ 
        \citet{lazard2021}& 18 & Nuclear (New Build) & n.a. & n.a. & 2021 & 8.88 & US-\$ per MWe*h & 8.23 \\ 
        \citet{lazard2021}& 18 & Nuclear (Operational) & n.a. & n.a. & 2021 & 6.24 & US-\$ per MWe*h & 5.78 \\ 
       \citet{mit_future_2018}& 240 & Fuel Cycle Cost min & n.a. & n.a. & 2017 & 9 & US-\$ per MWe*h & 9.22 \\ 
       \citet{mit_future_2018}& 240 & Fuel Cycle Cost max & n.a. & n.a. & 2017 & 17.54 & US-\$ per MWe*h & 17.97 \\ 
       \citet{mit_future_2018}& 159 & Fuel Cost Benchmark & n.a. & n.a. & 2017 & 10.7 & US-\$ per MWe*h & 10.96 \\ 
       \citet{mit_future_2018}& 151 & US Fuel Cost & Uranium & n.a. & 2017 & 10.7 & US-\$ per MWe*h & 10.96 \\ 
       \citet{mit_future_2018}& 153 & UK Fuel Cost & Uranium & n.a. & 2017 & 10.7 & US-\$ per MWe*h & 10.96 \\ 
       \citet{mit_future_2018}& 153 & France Fuel Cost & Uranium & n.a. & 2017 & 10.7 & US-\$ per MWe*h & 10.96 \\ 
        \citet{national_nuclear_laboratory_smr_2016}& 46 & Fuel and waste cost & Uranium & LWR & 2015 & 10.25 & US-\$ per MWe*h & 10.86 \\ 
        \citet{nrel_2021_2021}& Nuclear-Moderate & Nuclear & n.a. & n.a. &2013 & 7.10 & US-\$ per MWe*h & 7.65 \\ 
        \citet{pannier_comparison_2014}& 4 & NuScale & Enriched Uranium & SMR & 2014 & 20.57 & US-\$ per MWe*h & 21.82 \\ 
        \citet{pannier_comparison_2014}& 4 & SMART & Enriched Uranium & SMR & 2014 & 18.07 & US-\$ per MWe*h & 19.17 \\ 
        \citet{pannier_comparison_2014}& 4 & mPower & Enriched Uranium & SMR & 2014 & 21.29 & US-\$ per MWe*h & 22.58 \\ 
        \citet{pannier_comparison_2014}& 4 & HI-SMUR & Enriched Uranium & SMR & 2014 & 24.10 & US-\$ per MWe*h & 25.56 \\ 
        \citet{pannier_comparison_2014}& 4 & W-SMR & Enriched Uranium & SMR & 2014 & 13.67 & US-\$ per MWe*h & 14.50 \\ 
        \citet{pannier_comparison_2014}& 4 & n.a. & Uranium & EC 6 & 2014 & 15.25 & US-\$ per MWe*h & 16.18 \\ 
        \citet{pannier_comparison_2014}& 4 & n.a. & Enriched Uranium & VVER1000 & 2014 & 11.97 & US-\$ per MWe*h & 12.70 \\
        \citet{pannier_comparison_2014}& 4 & n.a. & Enriched Uranium & AP1000 & 2014 & 11.97 & US-\$ per MWe*h & 12.70 \\ 
        \citet{pannier_comparison_2014}& 4 & n.a. & Enriched Uranium & ABWR & 2014 & 11.11 & US-\$ per MWe*h & 11.79 \\
        \citet{pannier_comparison_2014}& 4 & n.a. & Enriched Uranium & US-APWR & 2014 & 10.19 & US-\$ per MWe*h & 10.81 \\ 
        \citet{pannier_comparison_2014}& 4 & n.a. & Enriched Uranium & US-EPR & 2014 & 10.67 & US-\$ per MWe*h & 11.31 \\
        \citet{pannier_comparison_2014}& 4 & n.a. & Enriched Uranium & ESBWR & 2014 & 11.28 & US-\$ per MWe*h & 11.97 \\ 
        \cite{timilsina_demystifying_2020}& 12 & Nuclear & n.a. & n.a. & 2019 & 8.8 & US-\$ per MWe*h & 8.64 \\ 
        \citet{tolley_economic_2004}& 7 & No-Policy Nuclear LCOE Installations (Max OCC) & ~ & Advanced Nuclear Reactor & 2003 & 4.35 & US-\$ per MWe*h & 5.94 \\ 
        \citet{wealer_investing_2021}& 5 & Inputs for MC & n.a. & n.a. & 2018 & 10.11 & US-\$ per MWe*h & 10.11 \\ \hline
    \end{longtable}
    \normalsize
\end{landscape}

\begin{table}[!ht]
    \caption{Capacity factors for nuclear plants.} \label{tab:capacity_factors}
    \centering
    \begin{tabular}{p{2cm}|p{1cm}|p{2.5cm}|p{1.25cm}}
    \hline
        Reference & Page & Description & Capacity Factor \\ \hline
        \citet{barkatullah_current_2017}& 132 & Asssumed parameters & 0.9 \\
        \citet{iea_projected_2020}& 39 & Assumptions for LCOE & 0.85 \\
        \citet{ingersoll_cost_2020} & 26 & Model assumptions & 0.92 \\
        \citet{lazard2021}& 18 & Nuclear (New Build, Low Case) & 0.92 \\ 
        \citet{lazard2021}& 18 & Nuclear (New Build, High Case) & 0.89 \\ 
        \citet{lazard2021}& 18 & Nuclear (Operating, Low Case) & 0.95 \\ 
        \citet{lazard2021}& 18 & Nuclear (Operating, High Case) & 0.88 \\ 
        \citet{national_nuclear_laboratory_smr_2016}& 46 & Nationally nominated factor for modelling & 0.9 \\
        \citet{nrel_2021_2021}& Nuclear-Moderate & 0.93 \\ 
        \citet{timilsina_demystifying_2020}& 12 & Nuclear & 0.9 \\ 
        \citet{tolley_economic_2004}& 7 & No-Policy Nuclear LCOE Installations (Max OCC) & 0.85 \\ 
        \citet{wealer_ten_2021}& 58 & Historical Capacity Factor France & 0.697 \\ 
        \citet{wealer_ten_2021}& 58 & Historical Capacity Factor Germany & 0.71 \\ 
        \citet{wealer_investing_2021}& 5 & Inputs for MC & 0.85 \\ \hline
    \end{tabular}
\end{table}

\begin{table}[!ht]
    \caption{Construction times of nuclear plants.}\label{tab:construction_time}
    \centering
    \begin{tabular}{p{2cm}|p{1cm}|p{2cm}|p{1.5cm}|p{1.5cm}}
    \hline
        Reference & Page & Description & Reactor & Construction Time in Years \\ \hline
        \citet{international_atomic_energy_agency_advances_2020}& 308 & HTR-PM & SMR & 9 \\ 
        \citet{international_atomic_energy_agency_advances_2020}& 308 & CAREM25 & SMR & 9.5 \\ 
        \citet{international_atomic_energy_agency_advances_2020} & 308 & ACP100 & SMR & 4.5 \\ 
        \citet{international_atomic_energy_agency_advances_2020} & 308 & NuScale & SMR & 5.5 \\ 
        \citet{international_atomic_energy_agency_advances_2020}& 308 & RITM-200M & SMR & 4.5 \\ 
        \citet{international_atomic_energy_agency_advances_2020} & 56 & SMART (FOAK) & SMR & 4.5 \\
        \citet{iea_projected_2020}& 39 & Assumptions for LCOE & Nuclear New Build & 7 \\ 
        \citet{lazard2021}& 18 & Nuclear (New Build, High Case) & n.a. & 5.75 \\ 
        \citet{linares_economics_2013}& n.a. & n.a. & n.a. & 5 \\
        \citet{linares_economics_2013}& n.a. & n.a. & n.a. & 9 \\ 
        \citet{mit_future_2018}& 149 & US (nominal) & Nuclear & 7 \\
        \citet{mit_future_2018}& 149 & UK (nominal) & Nuclear & 7 \\ 
        \citet{mit_future_2018}& 149 & China (nominal) & Nuclear & 7 \\ 
        \citet{mit_future_2018}& 149 & France (nominal) & Nuclear & 7 \\ 
        \citet{nrel_2021_2021}& Nuclear-Moderate & Nuclear & 6 \\
        \citet{tolley_economic_2004}& 7 & No-Policy Nuclear LCOE Installations (Max OCC) & ~ & 7 \\
        \citet{wealer_investing_2021}& 5 & Inputs for MC & n.a. & 8.7 \\ 
        \citet{schneider_world_2021}& n.a. & n.a. & n.a. & 7 \\ \hline
    \end{tabular}    
\end{table}

\begin{table}[!ht]
    \caption{Operational lifetimes of nuclear plants.} \label{tab:operational_lifetime}
    \centering
    \begin{tabular}{p{2cm}|p{1cm}|p{2cm}|p{1.5cm}|p{1.5cm}}
    \hline
        Reference & Page & Description & Reactor & Operational Lifetime \\ \hline
        \citet{barkatullah_current_2017}& 132 & Asssumed parameters & Nuclear & 60 \\ 
        \citet{iea_projected_2020}& 39 & Assumptions for LCOE & Nuclear New Build & 60 \\ 
        \citet{lazard2021}& 18 & Nuclear (New Build, Low Case) & Nuclear & 40 \\ 
        \citet{lazard2021}& 18 & Nuclear (New Build, High Case) & Nuclear & 40 \\ 
        \citet{lazard2021}& 18 & Nuclear (Operating, Low Case) & Nuclear & 40 \\ 
        \citet{lazard2021}& 18 & Nuclear (Operating, High Case) & Nuclear & 40 \\ 
        \citet{mit_future_2018}& 149 & US (nominal) & Nuclear & 40 \\ 
        \citet{mit_future_2018}& 149 & UK (nominal) & Nuclear & 40 \\ 
        \citet{mit_future_2018}& 149 & China (nominal) & Nuclear & 40 \\ 
        \citet{mit_future_2018}& 149 & France (nominal) & Nuclear & 40 \\ 
        \citet{timilsina_demystifying_2020}& 12 & Nuclear & n.a. & 45 \\ 
        \citet{tolley_economic_2004}& 7 & No-Policy Nuclear LCOE Installations (Max OCC) & ~ & 40 \\ 
        \citet{wealer_investing_2021}& 5 & Inputs for MC & Nuclear & 40 \\ \hline
    \end{tabular}
\end{table}

\begin{table}[!ht]
    \centering
    \caption{Thermal efficiency of nuclear plants.} \label{tab:thermal efficiency} 
    \begin{tabular}{p{2cm}|p{1cm}|p{2cm}|p{1.5cm}|p{1.5cm}}
    \hline
        Reference & Page & Description & Reactor & Efficiency \\ \hline
        \citet{iea_projected_2020}& 49 & Nuclear New Build - France & EPR & 0.33 \\ 
        \citet{iea_projected_2020}& 49 & Nuclear New Build - Japan & ALWR & 0.33 \\ 
        \citet{iea_projected_2020}& 49 & Nuclear New Build - Korea & ALWR & 0.36 \\ 
        \citet{iea_projected_2020}& 49 & Nuclear New Build - Russia & VVER & 0.38 \\ 
        \citet{iea_projected_2020}& 49 & Nuclear New Build - Slovak Republic & Other nuclear & 0.32 \\ 
        \citet{iea_projected_2020}& 49 & Nuclear New Build - United States & LWR & 0.33 \\ 
        \citet{iea_projected_2020}& 49 & Nuclear New Build - China & LWR & 0.33 \\ 
        \citet{iea_projected_2020}& 49 & Nuclear New Build - India & LWR & 0.33 \\ 
        \citet{ingersoll_cost_2020}& 29 & Model assumptions & Advanced Nuclear Reactor & 0.4 \\ 
        \citet{pannier_comparison_2014}& 4 & NuScale & SMR & 0.28 \\ 
        \citet{pannier_comparison_2014}& 4 & SMART & SMR & 0.27 \\ 
        \citet{pannier_comparison_2014}& 4 & mPower & SMR & 0.31 \\ 
        \citet{pannier_comparison_2014}& 4 & HI-SMUR & SMR & 0.31 \\
        \citet{pannier_comparison_2014}& 4 & W-SMR & SMR & 0.33 \\ 
        \citet{pannier_comparison_2014}& 4 & n.a. & EC 6 & 0.36 \\ 
        \citet{pannier_comparison_2014}& 4 & n.a. & VVER1000 & 0.33 \\ 
        \citet{pannier_comparison_2014}& 4 & n.a. & AP1000 & 0.32 \\ 
        \citet{pannier_comparison_2014}& 4 & n.a. & ABWR & 0.35 \\ 
        \citet{pannier_comparison_2014}& 4 & n.a. & US-APWR & 0.36 \\ 
        \citet{pannier_comparison_2014}& 4 & n.a. & US-EPR & 0.36 \\ 
        \citet{pannier_comparison_2014}& 4 & n.a. & ESBWR & 0.35 \\ \hline
    \end{tabular}
\end{table}

\section{Inputs for historic LCOEs} \label{lcoeSup}

The tables here list the cost assumptions to compute the historic LCOEs in Fig. \ref{fig:lcoe}. Full-load hours are an average from the European data also used as an input for the energy system model. All other data is taken from the Lazard reports for the respective years \citep{lazard2009,lazard2010,lazard2011,lazard2012,lazard2013,lazard2014,lazard2015,lazard2016,lazard2017,lazard2018,lazard2019,lazard2020,lazard2021}. In addition, we assume a uniform interest rate of 5\%.

\begin{landscape}
\footnotesize
    \begin{longtable} {p{1.5cm}|p{2cm}|p{1cm}|p{1.2cm}|p{.5cm}|p{.5cm}|p{.5cm}|p{.5cm}|p{.5cm}|p{.5cm}|p{.5cm}|p{.5cm}|p{.5cm}|p{.5cm}|p{.5cm}|p{.5cm}|p{.5cm}}
    \caption{Inputs for historic LCOEs.} \label{tab:loce_input}\\
    technology & parameter & unit & category & 2009 & 2010 & 2011 & 2012 & 2013 & 2014 & 2015 & 2016 & 2017 & 2018 & 2019 & 2020 & 2021 \\
    \hline
    \endfirsthead
    \caption[]{(continued)}\\
    \hline
    technology & parameter & unit & category & 2009 & 2010 & 2011 & 2012 & 2013 & 2014 & 2015 & 2016 & 2017 & 2018 & 2019 & 2020 & 2021 \\
    \endhead
    \hline
    \multicolumn{9}{l}{Continued on next page.}\\
    \hline
    \endfoot
    \hline
    \endlastfoot 
        Nuclear & Total capital cost (incl. construction) &US-\$/kW & Low & 6,325 & 5,385 & 5,385 & 5,385 & 5,385 & 5,385 & 5,400 & 5,400 & 6,500 & 6,500 & 6,900 & 7,675 & 7,800 \\ 
        ~ & ~ & ~ & Up & 8,375 & 8,199 & 8,199 & 8,199 & 8,199 & 8,199 & 8,200 & 8,200 & 11,800 & 12,250 & 12,200 & 12,500 & 12,800 \\ 
        ~ & Fixed O\&M &US-\$/kW/a & Low & 12.8 & 12.8 & 12.8 & 12.8 & 60.0 & 95.0 & 135.0 & 135.0 & 135.0 & 115.0 & 108.5 & 119.0 & 121.0 \\ 
        ~ & ~ & ~ & Up & 12.8 & 12.8 & 12.8 & 12.8 & 60.0 & 115.0 & 135.0 & 135.0 & 135.0 & 135.0 & 133.0 & 133.3 & 140.5 \\ 
        ~ & Variable O\&M &US-\$/MWh & Low & 11.0 & ~ & ~ & ~ & ~ & 0.3 & 0.5 & 0.5 & 0.8 & 0.8 & 3.4 & 3.8 & 4.0 \\
        ~ & ~ & ~ & Up & 11.0 & ~ & ~ & ~ & ~ & 0.8 & 0.8 & 0.8 & 0.8 & 0.8 & 4.3 & 4.3 & 4.5 \\ 
        ~ & Heat Rate & BTU/kWh & Low & 10,450 & 10,450 & 10,450 & 10,450 & 10,450 & 10,450 & 10,450 & 10,450 & 10,450 & 10,450 & 10,450 & 10,450 & 10,450 \\
        ~ & ~ & ~ & Up & 10,450 & 10,450 & 10,450 & 10,450 & 10,450 & 10,450 & 10,450 & 10,450 & 10,450 & 10,450 & 10,450 & 10,450 & 10,450 \\
        ~ & Capacity factor & \% & Low & 90\% & 90\% & 90\% & 90\% & 90\% & 90\% & 90\% & 90\% & 90\% & 90\% & 91\% & 92\% & 92\% \\ 
        ~ & ~ & ~ & Up & 90\% & 90\% & 90\% & 90\% & 90\% & 90\% & 90\% & 90\% & 90\% & 90\% & 90\% & 89\% & 89\% \\ 
        ~ & Fuel Price &US-\$/MMBtu & Low & 0.50 & 0.50 & 0.50 & 0.50 & 0.65 & 0.70 & 0.85 & 0.85 & 0.85 & 0.85 & 0.85 & 0.85 & 0.85 \\ 
        ~ & ~ & ~ & Up & 0.50 & 0.50 & 0.50 & 0.50 & 0.65 & 0.70 & 0.85 & 0.85 & 0.85 & 0.85 & 0.85 & 0.85 & 0.85 \\ 
        ~ & Lifetime & a & Low & 20 & 40 & 40 & 40 & 40 & 40 & 40 & 40 & 40 & 40 & 40 & 40 & 40 \\ 
        ~ & ~ & ~ & Up & 20 & 40 & 40 & 40 & 40 & 40 & 40 & 40 & 40 & 40 & 40 & 40 & 40 \\
        PV, openspace & Total capital cost (incl. construction) &US-\$/kW & Low & 3,250 & 3,500 & 2,500 & 2,000 & 1,750 & 1,500 & 1,400 & 1,300 & 1,100 & 950 & 900 & 825 & 800 \\
        ~ & ~ & ~ & Up & 4,000 & 4,000 & 4,000 & 2,750 & 2,000 & 1,750 & 1,600 & 1,450 & 1,375 & 1,250 & 1,100 & 975 & 950 \\ 
        ~ & Fixed O\&M &US-\$/kW/a & Low & 25.0 & 37.5 & 15.0 & 13.0 & 13.0 & 13.0 & 10.0 & 9.0 & 9.0 & 9.0 & 9.0 & 9.5 & 9.5 \\ 
        ~ & ~ & ~ & Up & 25.0 & 37.5 & 25.0 & 25.0 & 20.0 & 20.0 & 13.0 & 12.0 & 12.0 & 12.0 & 12.0 & 13.5 & 13.0 \\ 
        ~ & Capacity factor & \% & Low & 15.4\% & 15.4\% & 15.4\% & 15.4\% & 15.4\% & 15.4\% & 15.4\% & 15.4\% & 15.4\% & 15.4\% & 15.4\% & 15.4\% & 15.4\% \\
        ~ & ~ & ~ & Up & 15.4\% & 15.4\% & 15.4\% & 15.4\% & 15.4\% & 15.4\% & 15.4\% & 15.4\% & 15.4\% & 15.4\% & 15.4\% & 15.4\% & 15.4\% \\ 
        ~ & Lifetime & a & Low & 20 & 20 & 20 & 20 & 20 & 20 & 30 & 30 & 30 & 30 & 30 & 30 & 30 \\
        ~ & ~ & ~ & Up & 20 & 20 & 20 & 20 & 20 & 20 & 30 & 30 & 30 & 30 & 30 & 30 & 30 \\ 
        PV, rooftop residential & Total capital cost (incl. construction) &US-\$/kW & Low & ~ & ~ & ~ & ~ & ~ & 3,500 & 4,100 & 2,000 & 3,125 & 2,950 & 2,800 & 2,525 & 2,475 \\ 
        ~ & ~ & ~ & Up & ~ & ~ & ~ & ~ & ~ & 4,500 & 5,300 & 2,800 & 3,560 & 3,250 & 2,950 & 2,825 & 2,850 \\
        ~ & Fixed O\&M &US-\$/kW/a & Low & ~ & ~ & ~ & ~ & ~ & 25.0 & 17.5 & 20.0 & 20.0 & 14.5 & 14.0 & 15.0 & 15.0 \\ 
        ~ & ~ & ~ & Up & ~ & ~ & ~ & ~ & ~ & 30.0 & 22.5 & 25.0 & 25.0 & 25.0 & 25.0 & 18.0 & 18.0 \\ 
        ~ & Capacity factor & \% & Low & ~ & ~ & ~ & ~ & ~ & 15.1\% & 15.1\% & 15.1\% & 15.1\% & 15.1\% & 15.1\% & 15.1\% & 15.1\% \\ 
        ~ & ~ & ~ & Up & ~ & ~ & ~ & ~ & ~ & 15.1\% & 15.1\% & 15.1\% & 15.1\% & 15.1\% & 15.1\% & 15.1\% & 15.1\% \\ 
        ~ & Lifetime & a & Low & ~ & ~ & ~ & ~ & ~ & 20 & 20 & 20 & 20 & 25 & 25 & 25 & 25 \\ 
        ~ & ~ & ~ & Up & ~ & ~ & ~ & ~ & ~ & 20 & 20 & 20 & 20 & 25 & 25 & 25 & 25 \\ 
        PV, rooftop industry & Total capital cost (incl. construction) &US-\$/kW & Low & ~ & 4,000 & 3,750 & 3,000 & 3,000 & 2,500 & 2,600 & 2,100 & 2,000 & 1,900 & 1,750 & 1,600 & 1,400 \\ 
        ~ & ~ & ~ & Up & ~ & 4,500 & 4,500 & 3,500 & 3,500 & 3,000 & 3,750 & 3,750 & 3,750 & 3,250 & 2,950 & 2,825 & 2,850 \\ 
        ~ & Fixed O\&M &US-\$/kW/a & Low & ~ & 25.0 & 15.0 & 13.0 & 13.0 & 13.0 & 15.0 & 15.0 & 15.0 & 15.0 & 15.0 & 11.8 & 11.8 \\ 
        ~ & ~ & ~ & Up & ~ & 25.0 & 25.0 & 20.0 & 20.0 & 20.0 & 20.0 & 20.0 & 20.0 & 20.0 & 20.0 & 18.0 & 18.0 \\ 
        ~ & Capacity factor & \% & Low & ~ & 15.1\% & 15.1\% & 15.1\% & 15.1\% & 15.1\% & 15.1\% & 15.1\% & 15.1\% & 15.1\% & 15.1\% & 15.1\% & 15.1\% \\ 
        ~ & ~ & ~ & Up & ~ & 15.1\% & 15.1\% & 15.1\% & 15.1\% & 15.1\% & 15.1\% & 15.1\% & 15.1\% & 15.1\% & 15.1\% & 15.1\% & 15.1\% \\ 
        ~ & Lifetime & a & Low & ~ & 20 & 20 & 20 & 20 & 20 & 25 & 25 & 25 & 25 & 25 & 25 & 25 \\
        ~ & ~ & ~ & Up & ~ & 20 & 20 & 20 & 20 & 20 & 25 & 25 & 25 & 25 & 25 & 25 & 25 \\
        Wind, onshore & Total capital cost (incl. construction) &US-\$/kW & Low & 1,900 & 2,250 & 1,300 & 1,500 & 1,500 & 1,400 & 1,250 & 1,250 & 1,200 & 1,150 & 1,100 & 1,050 & 1,025 \\
        ~ & ~ & ~ & Up & 2,500 & 2,600 & 1,900 & 2,000 & 2,000 & 1,800 & 1,700 & 1,700 & 1,650 & 1,550 & 1,500 & 1,450 & 1,350 \\ 
        ~ & Fixed O\&M &US-\$/kW/a & Low & 40.0 & 60.0 & 30.0 & 30.0 & 30.0 & 35.0 & 35.0 & 35.0 & 30.0 & 28.0 & 28.0 & 27.0 & 22.5 \\ 
        ~ & ~ & ~ & Up & 50.0 & 60.0 & 30.0 & 30.0 & 30.0 & 40.0 & 40.0 & 40.0 & 40.0 & 36.5 & 36.5 & 39.5 & 36.0 \\ 
        ~ & Variable O\&M &US-\$/MWh & Low & ~ & ~ & ~ & 6.0 & 6.0 & ~ & ~ & ~ & ~ & ~ & ~ & ~ & ~ \\ 
        ~ & ~ & ~ & Up & ~ & ~ & ~ & 10.0 & 10.0 & ~ & ~ & ~ & ~ & ~ & ~ & ~ & ~ \\ 
        ~ & Capacity factor & \% & Low & 24.8\% & 24.8\% & 24.8\% & 24.8\% & 24.8\% & 24.8\% & 24.8\% & 24.8\% & 24.8\% & 24.8\% & 24.8\% & 24.8\% & 24.8\% \\ 
        ~ & ~ & ~ & Up & 24.8\% & 24.8\% & 24.8\% & 24.8\% & 24.8\% & 24.8\% & 24.8\% & 24.8\% & 24.8\% & 24.8\% & 24.8\% & 24.8\% & 24.8\% \\ 
        ~ & Lifetime & a & Low & 20 & 20 & 20 & 20 & 20 & 20 & 20 & 20 & 20 & 20 & 20 & 20 & 20 \\
        ~ & ~ & ~ & Up & 20 & 20 & 20 & 20 & 20 & 20 & 20 & 20 & 20 & 20 & 20 & 20 & 20 \\ 
        Wind, offshore & Total capital cost (incl. construction) &US-\$/kW & Low & ~ & 3,750 & 3,100 & 3,100 & 3,100 & 3,100 & 3,100 & 2,750 & 2,360 & 2,250 & 2,350 & 2,600 & 2,500 \\
        ~ & ~ & ~ & Up & ~ & 5,000 & 5,000 & 5,000 & 5,000 & 5,500 & 5,500 & 4,500 & 4,500 & 3,800 & 3,550 & 3,775 & 2,500 \\ 
        ~ & Fixed O\&M &US-\$/kW/a & Low & ~ & 60.0 & 60.0 & 60.0 & 60.0 & 60.0 & 60.0 & 80.0 & 80.0 & 80.0 & 80.0 & 67.3 & 65.8 \\ 
        ~ & ~ & ~ & Up & ~ & 100.0 & 100.0 & 100.0 & 100.0 & 100.0 & 100.0 & 110.0 & 110.0 & 110.0 & 110.0 & 81.8 & 79.5 \\ 
        ~ & Variable O\&M &US-\$/MWh & Low & ~ & 13.0 & 13.0 & 13.0 & 13.0 & 13.0 & 13.0 & ~ & ~ & ~ & ~ & ~ & ~ \\ 
        ~ & ~ & ~ & Up & ~ & 18.0 & 18.0 & 18.0 & 18.0 & 18.0 & 18.0 & ~ & ~ & ~ & ~ & ~ & ~ \\
        ~ & Capacity factor & \% & Low & ~ & 37.9\% & 37.9\% & 37.9\% & 37.9\% & 37.9\% & 37.9\% & 37.9\% & 37.9\% & 37.9\% & 37.9\% & 37.9\% & 37.9\% \\
        ~ & ~ & ~ & Up & ~ & 37.9\% & 37.9\% & 37.9\% & 37.9\% & 37.9\% & 37.9\% & 37.9\% & 37.9\% & 37.9\% & 37.9\% & 37.9\% & 37.9\% \\ 
        ~ & Lifetime & a & Low & ~ & 20 & 20 & 20 & 20 & 20 & 20 & 20 & 20 & 20 & 20 & 20 & 20 \\ 
        ~ & ~ & ~ & Up & ~ & 20 & 20 & 20 & 20 & 20 & 20 & 20 & 20 & 20 & 20 & 20 & 20 \\ \hline
    \end{longtable}
    \normalsize
\end{landscape}

\section{Projections of LCOEs for different interest rates and lifetimes} \label{diffInter}

\begin{figure}[!htbp]
	\centering
		\includegraphics[scale=0.375]{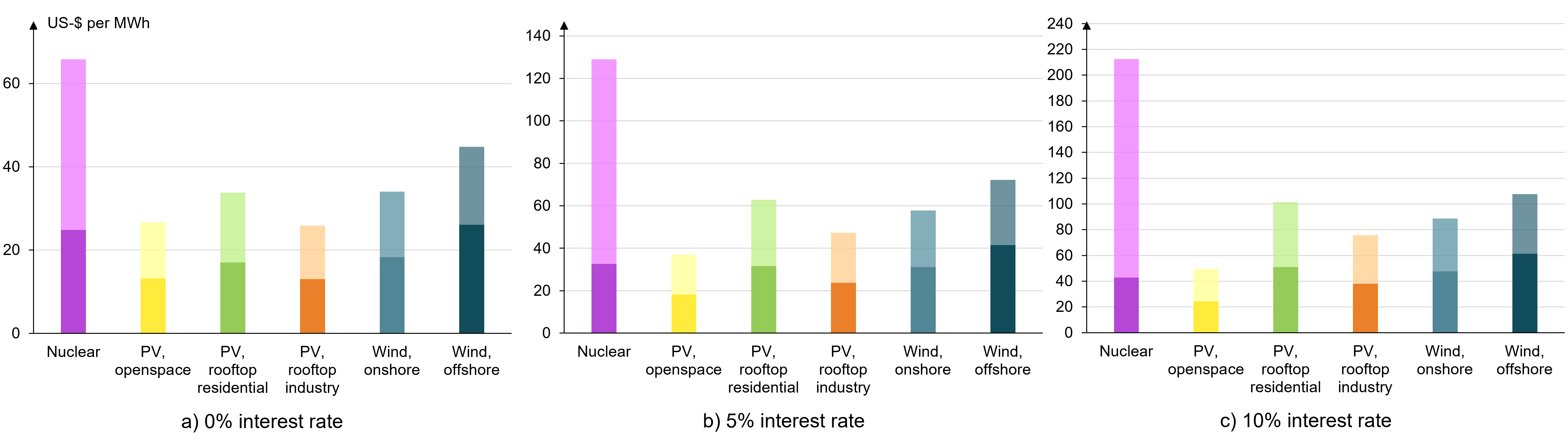}
	\caption{LCOEs for different interest rates}
	\label{fig:diffInter}
\end{figure}

Fig. \ref{fig:diffInter} shows how LCOEs for Nuclear and renewable technologies depend on the assumed interest rate. Absolute values change significantly; the overall cost ratios are relatively stable since all technologies are capital-intensive.  

\section{Structure of system model} \label{graphSec}

\begin{figure}[!htbp]
	\centering
		\includegraphics[scale=0.25]{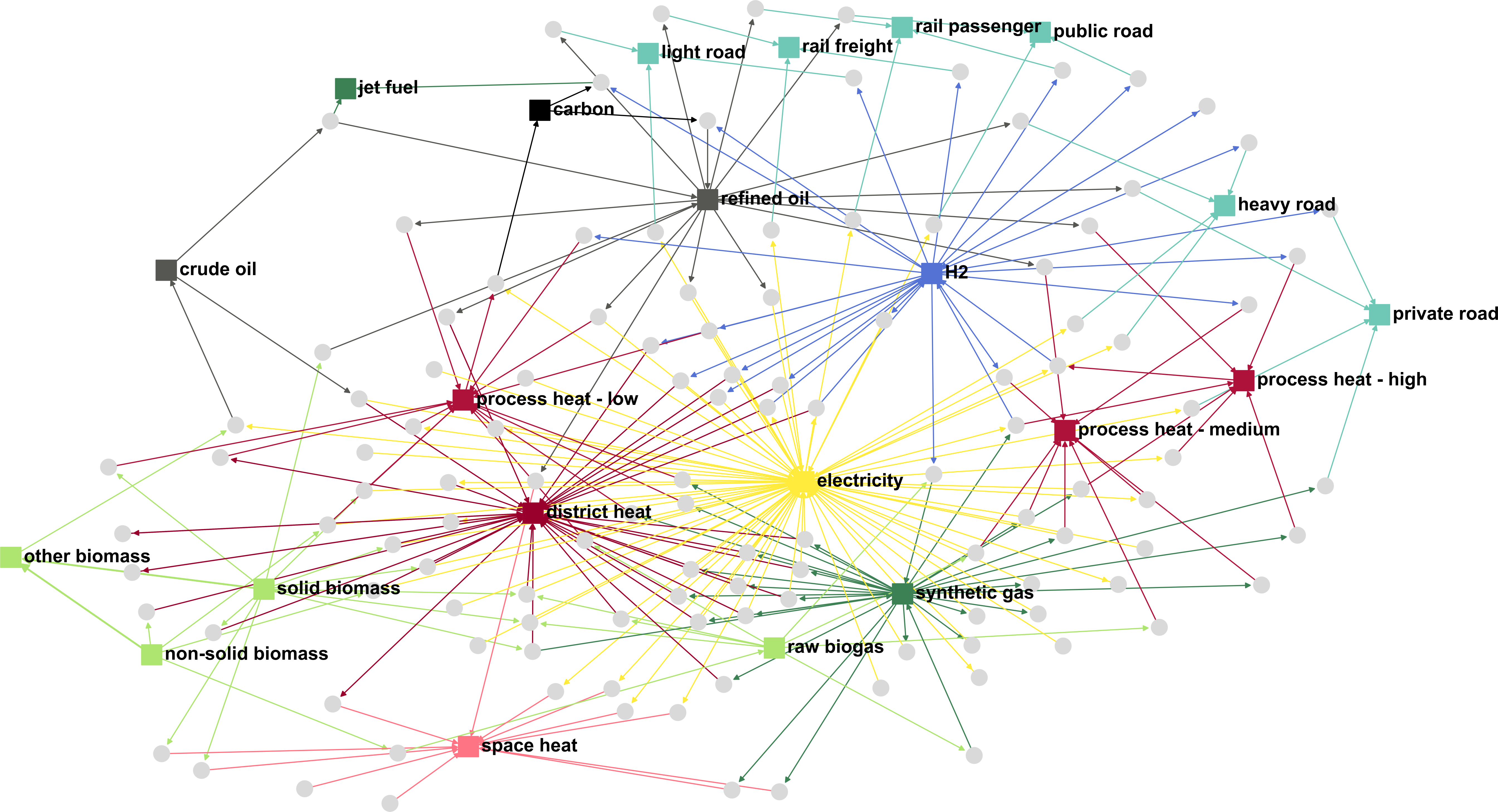}
	\caption{Full graph of model carriers and technologies}
	\label{fig:all}
\end{figure}

This section provides a comprehensive overview of all technologies and energy carriers considered in the system model based on graphs. In these graphs, vertices either represent energy carriers, depicted as colored squares, or technologies, depicted as gray circles. Entering edges of technologies refer to input carriers; outgoing edges refer to outputs. For illustrative purposes, Fig \ref{fig:all} shows the graph of all carriers and technologies. In the following, subgraphs of this graph are used to go into further detail.

\begin{figure}[!htbp]
	\centering
		\includegraphics[scale=0.25]{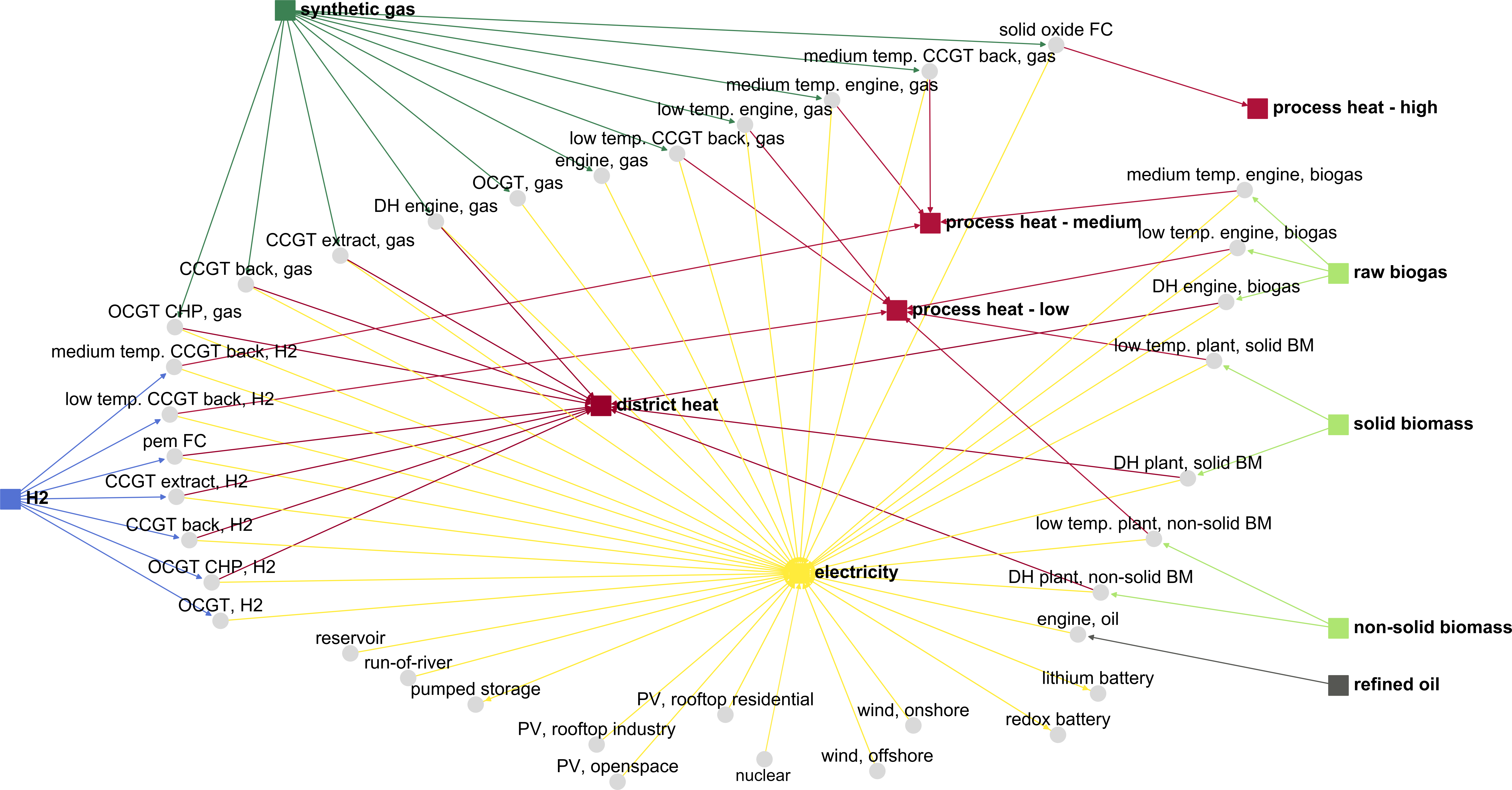}
	\caption{Subgraph for electricity supply}
	\label{fig:powerAll}
\end{figure}

\begin{figure}[!htbp]
	\centering
		\includegraphics[scale=0.25]{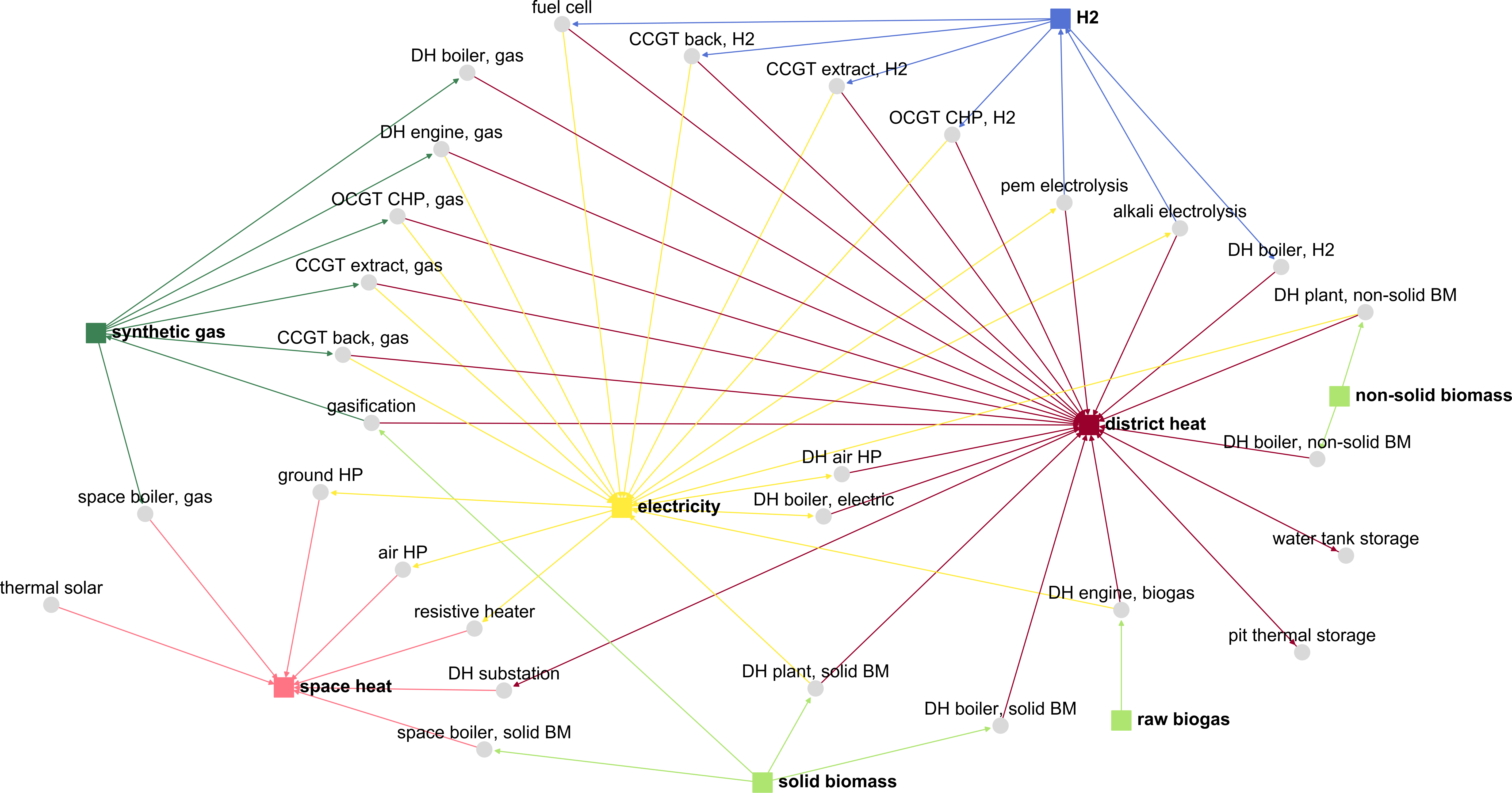}
	\caption{Subgraph for supply of space and district heating}
	\label{fig:spaceDistrictHeat}
\end{figure}

\begin{figure}[!htbp]
	\centering
		\includegraphics[scale=0.25]{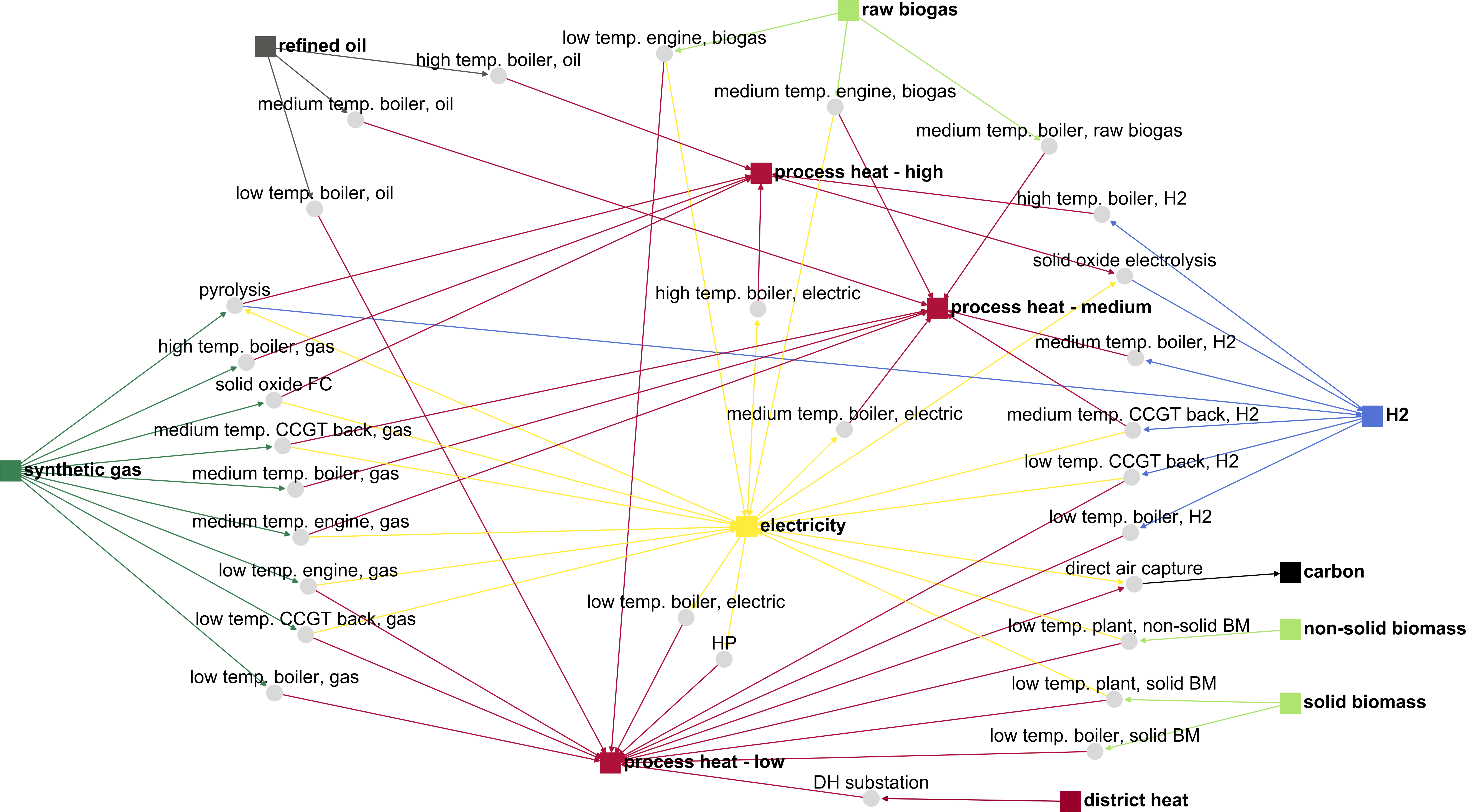}
	\caption{Subgraph for process heat}
	\label{fig:process}
\end{figure}

\begin{figure}[!htbp]
	\centering
		\includegraphics[scale=0.25]{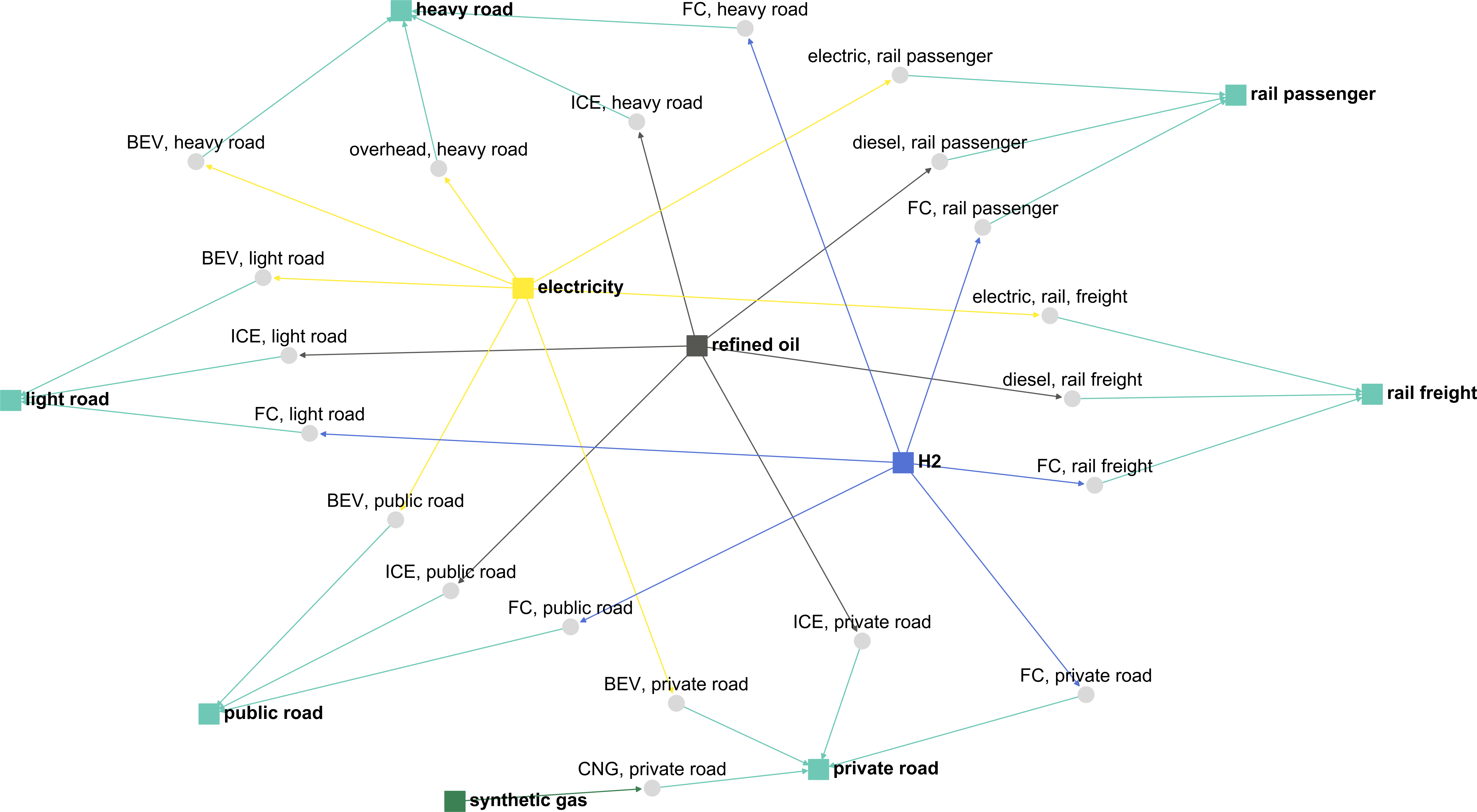}
	\caption{Subgraph of the transport sector}
	\label{fig:transport}
\end{figure}

\begin{figure}[!htbp]
	\centering
		\includegraphics[scale=0.25]{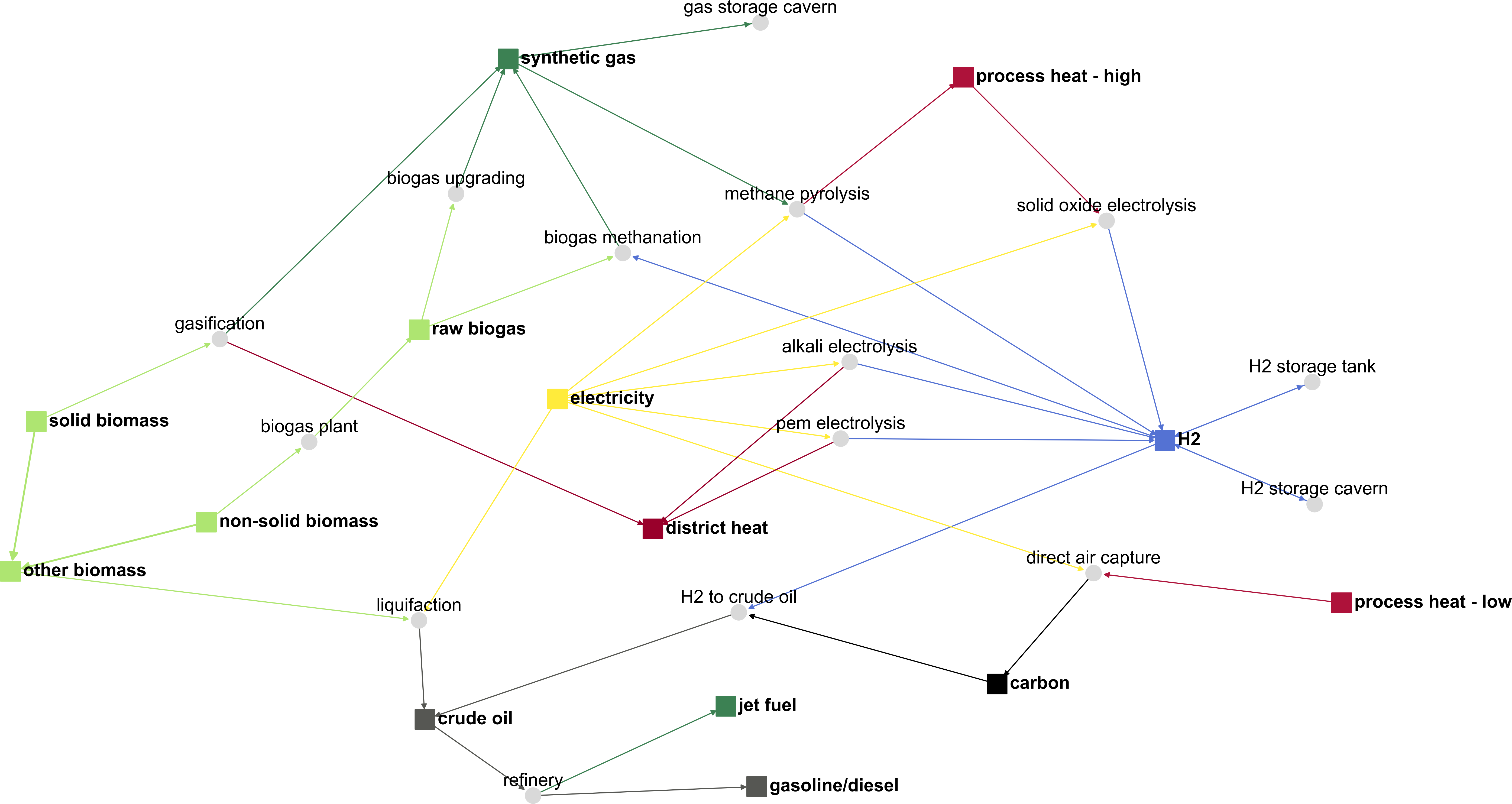}
	\caption{Subgraph for conversion of synthetic fuels}
	\label{fig:powerToX}
\end{figure}

\section{Renewable potential} \label{supEE}

\begin{figure}[!htbp]
	\centering
		\includegraphics[scale=0.1]{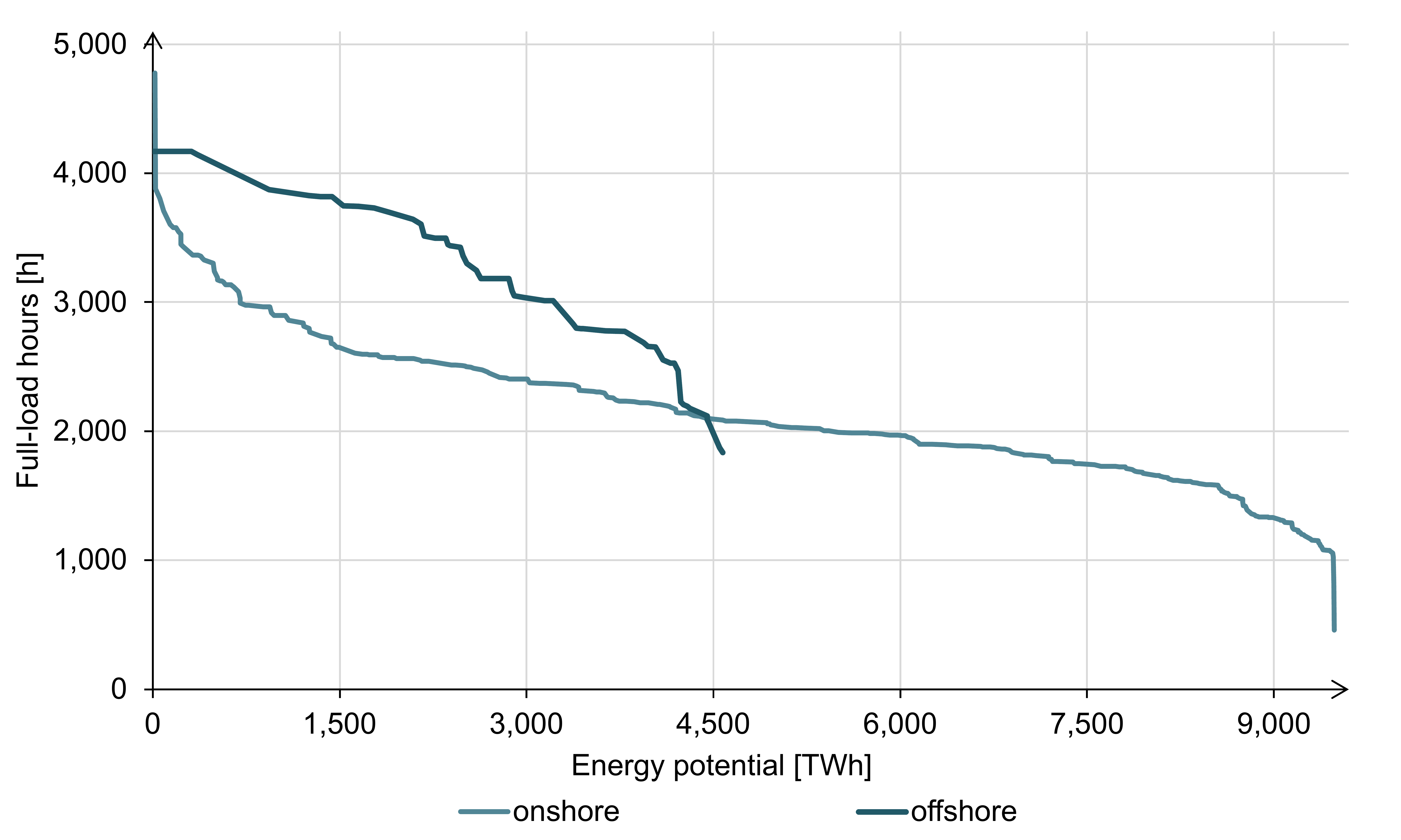}
	\caption{Full-load hours over total energy potential for wind}
	\label{fig:flhWind}
\end{figure}

\begin{figure}[!htbp]
	\centering
		\includegraphics[scale=0.1]{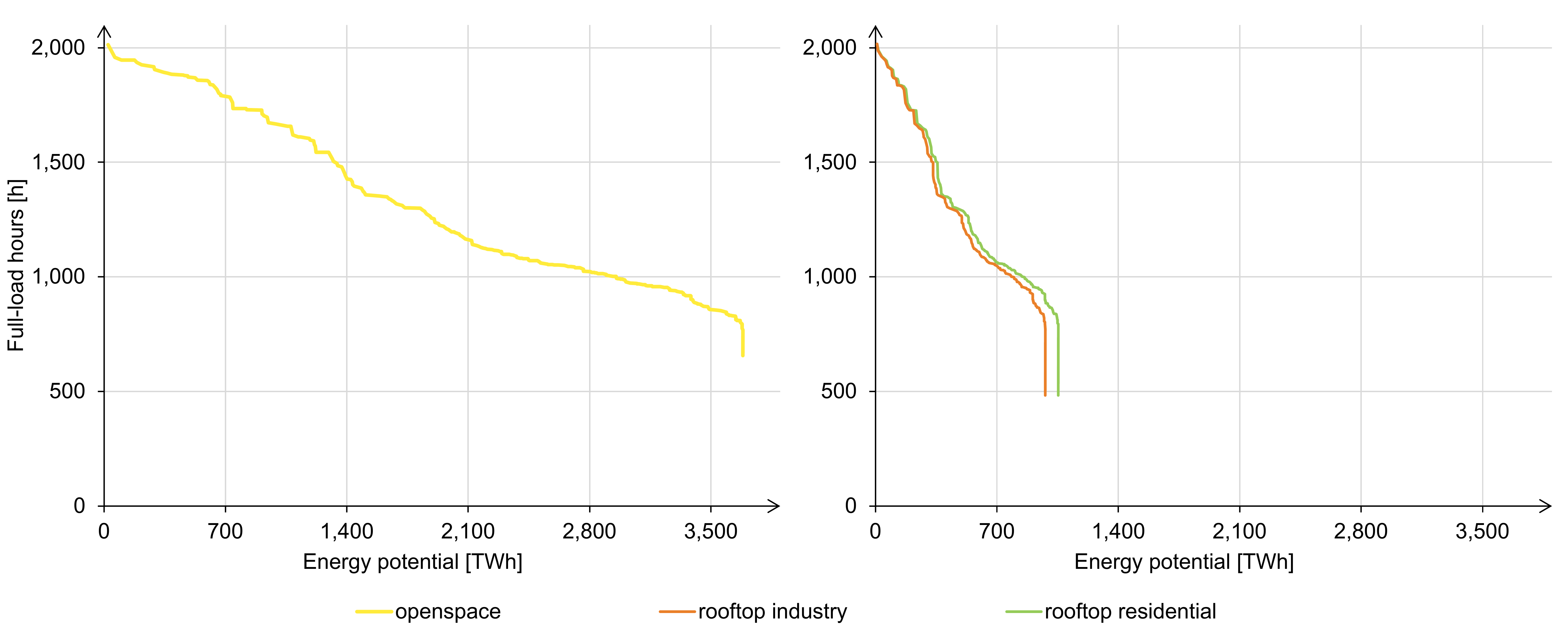}
	\caption{Full-load hours over total energy potential for PV}
	\label{fig:flhSolar}
\end{figure}

Figs. \ref{fig:flhWind} and \ref{fig:flhSolar} show how full-load hours vary across the total energy potential of renewables. The intention of these figures is to illustrate how utillizing a greater share of the renewable potential results in decreasing full-load hours and consequentially increasing incremental costs. 

The available energy potential for biomass amounts to 1,081 TWh building on \citet{jrc2015} and assuming 206 TWh of the total potential are used for the decarbonization of feedstock in the chemical industry instead of the modeled energy system \citet{dechema2017}.  

\section{Technology data for system model}

The tables here list the technology data for electricity generation in the system model from \citet{DEA}. For hydrogen fuelled plants, we assume a 15\% mark-up on the costs of the corresponding natural gas plant in line with \citet{Oberg2022}. This assumption neglects the cheaper alternative of retrofitting pre-existing gas plants instead. For all technologies we assume a uniform interest rate of 5\%.

\begin{landscape}
\footnotesize
\begin{longtable} {p{3cm}|p{2cm}|p{2cm}|p{1.5cm}|p{2cm}|p{2cm}}
    \caption{Technology data for electricity generation in system model.} \label{tab:technology_data_elec}\\
    \hline        
        technology & investment costs [m€/GW\textsubscript{in}] & operational costs [m€/GW\textsubscript{in}/a] & lifetime [a] & electrical efficiency & availability \\ \hline
    \endfirsthead
    \caption[]{(continued)}\\
    \hline
        technology & investment costs [m€/GW\textsubscript{in}] & operational costs [m€/GW\textsubscript{in}/a] & lifetime [a] & electrical efficiency & availability \\ \hline
    \endhead
    \hline
    \multicolumn{6}{l}{Continued on next page.}\\
    \hline
    \endfoot
    \hline
    \endlastfoot 
        combined cycle gas turbine, backpressure turbine & 586.0 & 13.7 & 25 & 50.3\% & 93.0\% \\ 
        combined cycle gas turbine, extraction turbine & 480.7 & 15.9 & 25 & 58.3\% & 93.0\% \\ 
        combined cycle H2 turbine, backpressure turbine & 673.9 & 13.7 & 25 & 50.3\% & 93.0\% \\ 
        combined cycle H2 turbine, extraction turbine & 552.8 & 15.9 & 25 & 58.3\% & 93.0\% \\ 
        open cycle gas turbine & 177.8 & 3.2 & 25 & 41.5\% & 97.2\% \\ 
        open cycle gas turbine CHP & 226.6 & 7.7 & 25 & 41.5\% & 93.0\% \\ 
        open cycle H2 turbine & 204.4 & 3.2 & 25 & 41.5\% & 97.2\% \\ 
        open cycle H2 turbine CHP & 260.6 & 7.7 & 25 & 41.5\% & 93.0\% \\ 
        engine, biogas & 384.8 & 3.9 & 25 & 43.3\% & 95.1\% \\ 
        engine, diesel & 118.8 & 2.9 & 25 & 35.0\% & 96.8\% \\ 
        engine, gas & 221.5 & 2.9 & 25 & 47.6\% & 98.7\% \\ 
        engine CHP, gas & 415.5 & 4.2 & 25 & 46.9\% & 95.5\% \\
        non-solid biomass plant CHP & 272.4 & 9.7 & 25 & 28.1\% & 91.2\% \\
        solid biomass plant CHP & 870.5 & 24.9 & 25 & 26.8\% & 91.2\% \\ 
        polymer electrolyte fuel cell & 475.0 & 23.8 & 10 & 50.0\% & 99.7\% \\ 
        solid oxide fuel cell & 840.0 & 42.0 & 20 & 59.6\% & 100.0\% \\
        PV, openspace & 271.2 & 5.4 & 18 & ~ & ~ \\ 
        PV, rooftop residential & 693.8 & 9.5 & 40 & ~ & ~ \\
        PV, rooftop industry & 511.7 & 7.8 & 40 & ~ & ~ \\ 
        wind, onshore & 963.1 & 11.3 & 30 & ~ & ~ \\ 
        wind, offshore, shallow water & 1,577.6 & 32.5 & 30 & ~ & ~ \\ 
        wind, offshore, deep water & 1,777.3 & 32.5 & 30 \\ \hline
\normalsize
\end{longtable}
\end{landscape}

The energy-to-power ratios for both battery technologies are constrained to be between 0.1 and 10.

\begin{table}[!ht]
    \centering
    \caption{Technology data for electricity storage in system model.}  
    \label{tab:technology_data_stor} 
    \begin{tabular}{|l|c|c|c|c|c|c|}
    \hline
        ~ & ~ & ~ & \multicolumn{2}{c|}{investment costs} & \multicolumn{2}{c|}{operational costs} \\
        technology & cycle efficiency & lifetime & power capacity & energy capacity & power capacity & energy capacity  \\ 
        ~ & ~ & [a] & [m€/GW] & [m€/GWh] & [m€/GW/a] & [m€/GWh/a] \\
        \hline
        lithium battery & 0.89 & 18 & 80.9 & 199.6 & 1.21 & 2.99 \\ 
        redox battery & 0.52 & 18 & 614 & 174.5 & 9.21 & 2.62 \\ 
        pumped storage & 0.81 & - & - & - & - & - \\ \hline
    \end{tabular}
\end{table}

\section{Supply and demand time-series} \label{timeSeries}

\begin{figure}[!htbp]
	\centering
		\includegraphics[scale=0.1]{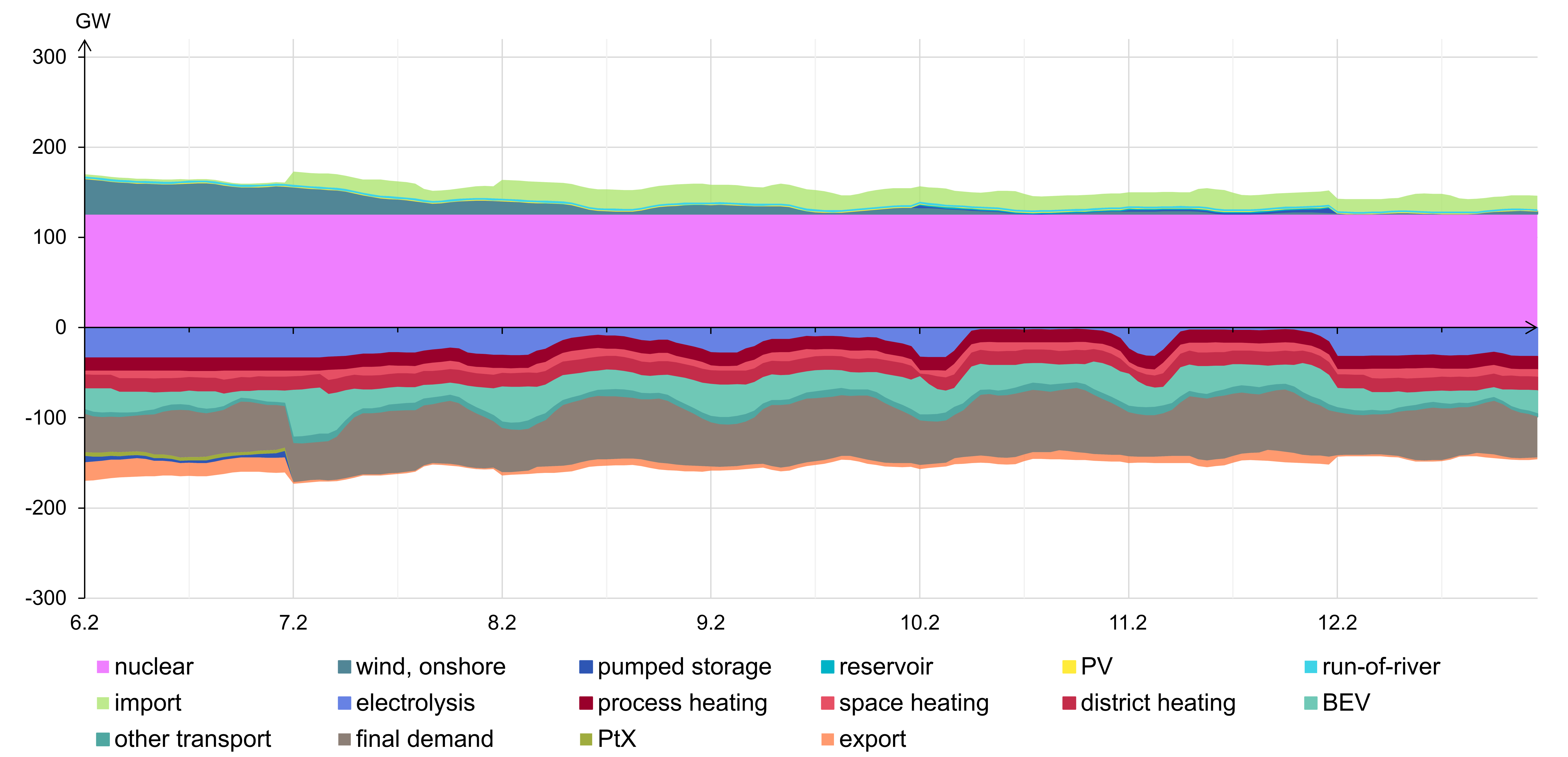}
	\caption{Supply and demand in Germany for one week and the 52\%-nuclear system}
	\label{fig:tsNu}
\end{figure}

\begin{figure}[!htbp]
	\centering
		\includegraphics[scale=0.1]{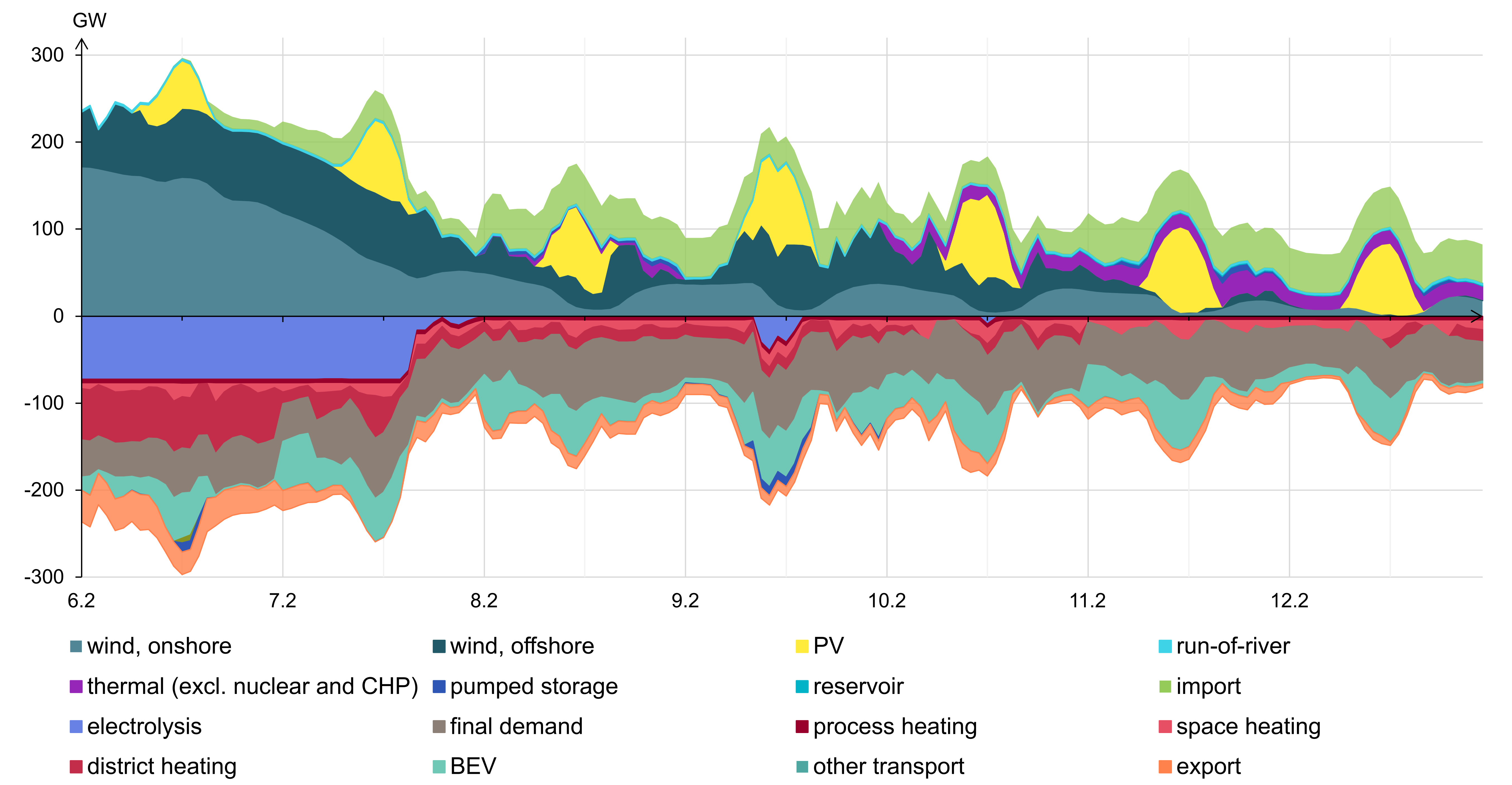}
	\caption{Supply and demand in Germany for one week and the system without nuclear}
	\label{fig:tsNoNu}
\end{figure}

Figs. \ref{fig:tsNu} and \ref{fig:tsNoNu} show hourly electricity supply and demand for an exemplary February week in Germany. The intention of these figures is to retrace the duration curves presetented in Fig. \ref{fig:durCur} of the paper. In contrast to the 52\%-nuclear system, Fig. \ref{fig:tsNoNu} shows how demand adapts to supply greatly reducing the residual demand to be met by thermal plants.

\section{Final demand}

The table below lists the final demand for energy and transport services the system model must meet in each case. 

\begin{landscape}
\footnotesize
    \begin{longtable} {p{1.5cm}|p{1.5cm}|p{1cm}|p{1.2cm}|p{.75cm}|p{.75cm}|p{.75cm}|p{.75cm}|p{.75cm}|p{.75cm}|p{.75cm}|p{.75cm}|p{.75cm}}
    \caption{Final demand for different energy carriers in the energy system model.} \label{tab:final_demand}\\
    \hline
    ~ & ~ & ~ & \multicolumn{3}{c}{process heat [TWh]} & \multicolumn{3}{c}{passenger transport [Gpkm]} & \multicolumn{3}{c}{freight transport [Gpkm]} & ~ \\
    country &	electricity [TWh] &	space heat [TWh] & low (until 100°C) & medium (100 to 500°C) &	high (above 500°C) & rail	& road private	& road public &	rail	& road heavy &	road light &	jet fuel [TWh] \\ \hline
    \endfirsthead
    \caption[]{(continued)}\\
    \hline
     ~ & ~ & ~ & \multicolumn{3}{c}{process heat [TWh]} & \multicolumn{3}{c}{passenger transport [Gpkm]} & \multicolumn{3}{c}{freight transport [Gpkm]} & ~ \\
    country	electricity [TWh] &	space heat [TWh] & low (until 100°C) & medium (100 to 500°C) &	high (above 500°C) & Rail	& road private	& road public &	rail	& road heavy &	road light &	jet fuel [TWh] \\ \hline
    \endhead
    \hline
    \multicolumn{9}{l}{Continued on next page.}\\
    \hline
    \endfoot
    \hline
    \endlastfoot 
        Albania & 4.6 & 4.2 & 2.7 & 3.8 & 4.9 & 0.1 & 8.6 & 3.0 & 2.5 & 3.8 & 0.0 & 0.0 \\ 
        Austria & 62.8 & 45.2 & 17.0 & 26.2 & 20.7 & 14.0 & 83.8 & 11.0 & 12.6 & 25.6 & 0.9 & 18.1 \\
        Bosnia and Herzegovina  & 9.6 & 4.9 & 3.2 & 4.5 & 5.7 & 0.1 & 11.5 & 4.0 & 3.4 & 5.1 & 0.1 & 0.0 \\ 
        Belgium & 73.6 & 62.5 & 17.1 & 22.6 & 28.8 & 11.7 & 117.8 & 14.9 & 5.7 & 30.9 & 3.9 & 51.7 \\
        Bulgaria & 29.2 & 11.0 & 9.1 & 4.7 & 4.8 & 1.4 & 55.9 & 7.8 & 7.1 & 20.2 & 0.3 & 5.4 \\ 
        Switzerland & 51.8 & 43.7 & 7.4 & 30.8 & 6.6 & 20.8 & 80.6 & 6.4 & 6.4 & 11.6 & 0.3 & 0.5 \\
        Czech Republic & 57.8 & 38.9 & 17.5 & 16.7 & 19.0 & 10.0 & 76.5 & 17.7 & 19.2 & 46.6 & 3.8 & 5.0 \\ 
        Germany & 480.7 & 470.9 & 112.1 & 119.0 & 153.0 & 101.7 & 950.7 & 64.8 & 84.8 & 303.3 & 8.6 & 151.0 \\ 
        Denmark & 29.5 & 30.3 & 5.6 & 7.0 & 2.6 & 6.9 & 67.1 & 7.9 & 2.0 & 14.8 & 0.2 & 4.3 \\ 
        Estonia & 7.2 & 5.3 & 1.0 & 1.9 & 0.6 & 0.4 & 11.3 & 2.4 & 4.1 & 4.7 & 0.1 & 0.0 \\ 
        Spain & 241.9 & 62.8 & 25.5 & 57.2 & 54.1 & 31.0 & 370.9 & 35.0 & 11.4 & 213.9 & 3.1 & 6.1 \\ 
        Finland & 65.8 & 30.4 & 27.7 & 42.4 & 8.7 & 4.6 & 67.8 & 8.1 & 11.8 & 28.4 & 0.5 & 16.7 \\
        France & 390.3 & 257.8 & 40.4 & 57.1 & 64.6 & 95.3 & 770.4 & 59.2 & 19.5 & 169.3 & 4.8 & 114.2 \\ 
        Greece & 45.3 & 16.0 & 5.2 & 5.9 & 5.9 & 1.2 & 106.0 & 21.0 & 0.6 & 27.0 & 1.2 & 69.5 \\
        Croatia & 16.7 & 10.3 & 1.6 & 3.9 & 2.6 & 0.8 & 25.6 & 5.3 & 3.6 & 12.2 & 0.3 & 3.3 \\ 
        Hungary & 37.1 & 36.6 & 7.1 & 4.1 & 6.9 & 6.7 & 55.1 & 16.2 & 14.5 & 35.8 & 1.1 & 6.1 \\ 
        Ireland & 24.9 & 15.9 & 4.1 & 5.8 & 3.8 & 1.1 & 28.5 & 5.2 & 0.1 & 9.0 & 0.2 & 0.0 \\
        Italy & 260.6 & 215.9 & 57.1 & 48.5 & 72.7 & 59.0 & 768.2 & 89.9 & 17.0 & 96.3 & 16.4 & 66.9 \\ 
        Lithuania & 11.7 & 8.0 & 3.7 & 2.1 & 1.5 & 0.4 & 30.3 & 2.9 & 112.4 & 52.3 & 0.8 & 29.6 \\ 
        Luxemboourg & 4.7 & 3.5 & 0.8 & 1.1 & 2.1 & 0.4 & 7.2 & 1.1 & 0.7 & 7.3 & 0.1 & 0.0 \\
        Latvia & 6.2 & 7.6 & 1.7 & 3.9 & 1.5 & 0.6 & 15.3 & 2.6 & 46.9 & 14.8 & 0.2 & 0.0 \\ 
        Montenegro & 2.8 & 0.9 & 0.5 & 0.7 & 1.1 & 0.1 & 4.1 & 0.1 & 0.9 & 1.4 & 0.0 & 0.0 \\
        Macedonia & 5.2 & 3.1 & 1.9 & 2.6 & 3.4 & 0.1 & 7.5 & 2.2 & 2.1 & 3.2 & 0.0 & 0.0 \\ 
        Netherlands & 107.7 & 62.8 & 30.0 & 28.1 & 42.5 & 24.5 & 187.4 & 6.8 & 8.7 & 66.7 & 1.6 & 215.6 \\ 
        Norway & 106.6 & 34.0 & 4.8 & 30.7 & 2.6 & 3.6 & 64.6 & 4.2 & 3.9 & 20.7 & 0.6 & 17.0 \\ 
        Poland & 145.1 & 98.0 & 23.6 & 40.8 & 42.5 & 21.8 & 218.7 & 35.6 & 127.9 & 340.0 & 9.0 & 35.1 \\ 
        Portugal & 46.5 & 9.1 & 8.4 & 16.1 & 9.1 & 4.0 & 85.2 & 7.0 & 5.1 & 30.4 & 0.6 & 29.1 \\ 
        Romania & 55.0 & 31.2 & 11.1 & 14.1 & 25.2 & 4.4 & 81.4 & 15.6 & 40.1 & 60.3 & 0.8 & 11.6 \\ 
        Serbia & 32.7 & 11.2 & 6.8 & 9.6 & 12.0 & 0.3 & 28.8 & 10.0 & 8.5 & 12.7 & 0.2 & 0.0 \\
        Sweden & 104.7 & 55.4 & 12.1 & 51.0 & 10.0 & 11.7 & 100.0 & 8.7 & 19.2 & 42.0 & 0.6 & 7.7 \\ 
        Slovenia & 12.3 & 6.5 & 1.7 & 2.6 & 2.3 & 0.5 & 26.3 & 3.6 & 13.1 & 23.7 & 0.3 & 0.0 \\ 
        Slovakia & 26.2 & 15.1 & 11.8 & 5.6 & 15.2 & 3.7 & 27.9 & 6.1 & 17.2 & 32.5 & 1.5 & 2.5 \\ 
        United Kingdom & 256.4 & 271.8 & 46.9 & 68.6 & 42.1 & 69.1 & 670.3 & 36.5 & 16.7 & 153.7 & 7.1 & 143.9 \\ \hline
    \end{longtable}
    \normalsize
\end{landscape}

\end{document}